\documentclass[a4paper,usenames,dvipsnames,11pt]{article}
\pdfoutput=1

\usepackage{jheppub}
\usepackage{slashed}
\usepackage{mathrsfs,booktabs,multirow,tabularx}
\usepackage{stmaryrd}
\usepackage{xspace}
\usepackage{fancyvrb}
\usepackage[makeroom]{cancel}
\usepackage{amsmath}    
\usepackage{amssymb}    %
\usepackage{graphicx}   
\usepackage{verbatim}   
\usepackage{lscape}
\usepackage{subfig}
\usepackage{listings}

\def\beq{\begin{equation}}
\def\beqn{\begin{eqnarray}}
\def\eeq{\end{equation}}
\def\eeqn{\end{eqnarray}}

\def\pdistr#1#2{\left(\frac{1}{#1}\right)_{\!#2}}

\def\lppdistr#1#2{\left(\frac{\log\left(#1\right)}{#1}\right)_{\!#2}}

\def\PDF#1#2{\Gamma_{\!#1/#2}}
\def\ePDF#1{\Gamma_{\!#1}}

\newcommand\sss{\scriptscriptstyle}

\newcommand\ep{\epsilon}
\newcommand\epb{\bar{\epsilon}}
\newcommand\ooepb{\frac{1}{\epb}}
\newcommand\half{\frac{1}{2}}
\newcommand\as{\alpha_{\sss S}}
\newcommand\aem{\alpha}

\newcommand\aemz{\alpha(\mu_0)}

\newcommand\aemmu{\alpha(\mu)}
\newcommand\aemR{\aem_{\!{\sss R}}}
\newcommand\aemmZ{\aem_{\mZ}}
\newcommand\aemGmu{\aem_{G_\mu}}

\newcommand\gE{\gamma_{\sss\rm E}}
\newcommand{\bq}{\bar{q}}

\newcommand{\bt}{\bar{t}}
\newcommand{\bl}{\bar{l}}
\newcommand{\bN}{\bar{N}}

\newcommand{\epem}{e^+e^-}
\newcommand{\lp}{e^+}
\newcommand{\lm}{e^-}
\newcommand{\lpm}{e^{\pm}}

\newcommand{\zp}{z_+}
\newcommand{\zm}{z_-}

\newcommand{\yp}{y_+}
\newcommand{\ym}{y_-}

\newcommand{\ord}{{\cal O}}
\newcommand\aNLO{{\sc\small MadGraph5\_aMC@NLO}}
\newcommand\aNLOs{{\sc\small MG5\_aMC}}

\newcommand\NC{N_{\sss c}}

\newcommand\Nl{N_l}
\newcommand\Nu{N_u}
\newcommand\Nd{N_d}
\newcommand\hsig{\hat{\sigma}}

\newcommand\ee{e_e}
\newcommand\el{e_l}

\newcommand\eu{e_u}
\newcommand\ed{e_d}

\newcommand\APmat{{\mathbb P}}
\newcommand\bAPmat{\bar{{\mathbb P}}}
\newcommand\Eop{{\mathbb E}}
\newcommand\Mmat{{\mathbb M}}
\newcommand\Kmat{{\mathbb K}}

\newcommand\MSb{\overline{\rm MS}}

\newcommand\bm{\overline{m}}

\newcommand\muz{\mu_0}

\newcommand{\pt}{p_{\sss T}}
\newcommand\stepf{\Theta}
\newcommand\mZ{m_{\sss Z}}
\newcommand\mW{m_{\sss W}}
\newcommand\cw{c_{\sss W}}
\newcommand\sw{s_{\sss W}}
\newcommand{\Melleq}{\stackrel{\infty}{=}}
\newcommand{\GammaLO}{\Gamma_{\rm LO}}

\newcommand\ePDFc{{\sc\small ePDF}}
\newcommand\LHAPDF{{\sc\small LHAPDF}}
\newcommand\eMELA{{\sc\small eMELA}}
\newcommand\MELA{{\sc\small MELA}}

\title{Improving methods and predictions at high-energy $e^+e^-$ 
colliders within collinear factorisation}

\author[a]{V. Bertone,}
\affiliation[a]{IRFU, CEA, Universit\'e Paris-Saclay, 
F-91191 Gif-sur-Yvette, France}

\author[b,c]{M. Cacciari,}
\affiliation[b]{Sorbonne Universit\'e, CNRS, Laboratoire de Physique 
Th\'orique et Hautes \'Energies,\\ LPTHE, F-75005 Paris, France}
\affiliation[c]{Universit\'e Paris Cit\'e, F-75006 Paris, France}

\author[d]{S. Frixione,}
\affiliation[d]{INFN, Sezione di Genova, Via Dodecaneso 33, I-16146, 
Genoa, Italy}

\author[e]{G. Stagnitto,}
\affiliation[e]{Physik-Institut, Universit\"at Z\"urich, 
Winterthurerstrasse 190, CH-8057 Z\"urich, Switzerland}

\author[f]{M. Zaro,}
\affiliation[f]{TIFLab, Universit\`a degli Studi di Milano and\\ 
INFN, Sezione di Milano, Via Celoria 16, 20133 Milano, Italy}

\author[g]{X. Zhao}
\affiliation[g]{Dipartimento di Matematica e Fisica, Universit{\`a} 
di Roma Tre and \\INFN, Sezione di Roma Tre, I-00146 Rome, Italy}

\emailAdd{valerio.bertone@cern.ch}
\emailAdd{cacciari@lpthe.jussieu.fr}
\emailAdd{Stefano.Frixione@cern.ch}
\emailAdd{giovanni.stagnitto@physik.uzh.ch}
\emailAdd{Marco.Zaro@mi.infn.it}
\emailAdd{xiaoran.zhao@uniroma3.it}

\abstract{
We illustrate how electron Parton Distribution Functions (PDFs) with 
next-to-leading collinear logarithmic accuracy must be employed in 
the context of perturbative predictions for high-energy $e^+e^-$-collision 
processes. In particular, we discuss how the renormalisation group equation
evolution of such PDFs is affected by the presence of multiple fermion
families and their respective mass thresholds, and by the dependences
on the choices of the factorisation and renormalisation schemes. 
We study the impact of the uncertainties associated with the PDFs on
physical cross sections, in order to arrive at realistic precision
estimates for observables computed with collinear-factorisation formulae.
We do so by presenting results for the production of a heavy neutral
object as well as for $t\bt$ and $W^+W^-$ pairs, including 
next-to-leading-order effects of electroweak origin.
}


\preprint{
\begin{flushright}
TIF-UNIMI-2022-10\\
ZU-TH 26/22\\
\today
\end{flushright}
}

\begin{document}
\maketitle
\flushbottom

\section{Introduction\label{sec:intro}}
In order to attain the precision goals of future $\epem$ colliders
programmes~\cite{CEPCStudyGroup:2018ghi,FCC:2018evy,
Bambade:2019fyw,CLICdp:2018cto}, computations in perturbative QED play a 
paramount role, since results obtained at increasingly large orders in the 
coupling constant $\aem$ help reduce the theoretical uncertainties that 
affect them. A key aspect of this kind of predictions is that, on top of being
inherently accurate, they give one the ability to assess in a quantitative
manner the errors that one makes by identifying them with actual
measurements; in fact, such an ability is an integral part of any
precision-physics studies.

The calculation of matrix elements of $\ord(\aem^p)$ relative to Born, with 
$p=1,2$ (i.e.~next-to-leading order, NLO, and next-to-next-to-leading order,
NNLO) and possibly even larger, is a necessary ingredient of accurate 
results, but not a sufficient one. When integrated over the phase space, such
matrix elements give rise to terms of the type \mbox{$\aem^p\log^n{\cal Q}$},
with ${\cal Q}$ a small number whose precise nature may depend on the
definition of the observable one is looking at, and/or on the characteristics
of QED. Depending on ${\cal Q}$, one typically has $n\le p$; thus, the terms 
above spoil the good behaviour of the perturbative series\footnote{In fact,
the actual behaviour of the series is worse than that, since the 
coefficient of the \mbox{$\aem^p\log^n{\cal Q}$} term may also be
logarithmically enhanced, by a mechanism different w.r.t.~the one that gives 
rise to $\log{\cal Q}$ -- the most common example of a double logarithmic 
enhancement is that due to emissions simultaneously soft and collinear.}, 
and must be resummed. A prominent example of observable-independent logarithmic
terms is that for which \mbox{${\cal Q}=m^2/E^2$}, with $m$ the electron
mass and $E$ a hard scale typical of the process (e.g.~the center-of-mass
energy). These terms are present even for fully-inclusive observables, 
and arise from the collinear emissions off initial- and final-state
particles\footnote{Strictly speaking, in the case of bare-lepton observables
such logarithms are observable dependent. However, they can be resummed with 
the same techniques as their observable-independent counterparts.}. Here,
we shall concentrate on the initial-state case, in view of the fact that the
associated logarithms (which we call collinear logarithms, a.k.a.~mass
singularities in the literature) are ubiquitous, and that those relevant to 
final-state emissions can be treated in a fully analogous manner.
The resummation is carried out in the context of collinear-factorisation
formulae~\cite{Kuraev:1985hb,Ellis:1986jba}; alternative resummation
techniques, in particular YFS~\cite{Yennie:1961ad,Jadach:2000ir} (that
addresses the resummation of soft logarithms) and parton 
shower~\cite{Anlauf:1991wr,Fujimoto:1993ge,Munehisa:1995si,
CarloniCalame:2000pz}, will not be discussed here.

By writing the collider-level cross section $d\Sigma_{\epem}$ for a 
generic $\epem\to X$ production process as follows:
\beq
d\Sigma_{\epem}(P_{\lp},P_{\lm})=\sum_{kl}\int d\yp d\ym\,
{\cal B}_{kl}(\yp,\ym)\,d\sigma_{kl}(\yp P_{\lp},\ym P_{\lm})\,,
\label{beamstr}
\eeq
collinear factorisation amounts to using the following expression:
\beqn
d\sigma_{kl}(p_k,p_l)&=&\sum_{ij}\int d\zp d\zm\,
\PDF{i}{k}(\zp,\mu,m)\,\PDF{j}{l}(\zm,\mu,m)
\nonumber\\*&&\phantom{\sum_{ij}\int}\times
d\hsig_{ij}(\zp p_k,\zm p_l,\mu)\,.
\label{master0}
\eeqn
In eq.~(\ref{beamstr}) the functions ${\cal B}_{kl}$ account for 
collective phenomena in beam dynamics, such as beamstrahlung, which give 
rise to particles $k$ and $l$, that will eventually initiate the hard 
scattering. Dominant contributions are those for which the identities
of these particles coincide with those of the respective beams (thus,
$k=\lp$ and $l=\lm$ here), but others (e.g.~$k=\gamma$ and $l=\gamma$)
are also interesting; in this paper, we shall limit ourselves to consider
only the former. In the collinear-factorisation master formula, 
eq.~(\ref{master0}), for any given $(k,l)$ pair the particle-level cross 
section $d\sigma_{kl}$ is expressed as an incoherent sum of convolutions 
between parton-level cross sections $d\hsig_{ij}$ and PDFs 
$\PDF{\alpha}{\beta}(z_\pm)$; each of the latter is the probability density 
for finding parton $\alpha$ inside particle $\beta$ with a given fraction 
$z_\pm$ of the longitudinal momentum of the particle. The PDFs are entirely
and solely responsible for resumming the initial-state radiation (ISR)
collinear logarithms, and they do so thanks to their being solutions
of appropriate renormalisation-group equations~\cite{Gribov:1972ri,
Lipatov:1974qm,Altarelli:1977zs,Dokshitzer:1977sg}. As far as the
partonic cross sections $d\hsig_{ij}$ are concerned, as the notation of 
eq.~(\ref{master0}) understands we regard them as computed with massless 
electrons, all logarithmically dominant mass effects being included in the 
PDFs; massive-electron results could be employed too (after eliminating
double-counting terms), but the differences with the former 
are only\footnote{With the exception of Yukawa-induced processes.}
of power-suppressed type, \mbox{$\ord(m^{2q}/E^{2q})$} 
for some $q\ge 1$. The indices $i$ and $j$ assume values equal to the 
identities of the partons that emerge from branching processes initiated 
by the respective particles $k$ and $l$, compatible with the perturbative 
order at which the PDFs are computed; in particular, beyond leading order 
(LO) these include leptons, quarks, and photons, with the dominant 
contributions being due to $i=k$ and $j=l$.

A common misconception is that eq.~(\ref{master0}), being based on PDFs
obtained by integrating out all non-collinear degrees of freedom, does not
give an adequate description of transverse degrees of freedom. We stress that
this is certainly not the case, but some clarifications are in order. In the 
context of a fixed-order computation of $d\hsig_{ij}$, the kinematics of 
the system $X$ and of any recoil partons (i.e.~light fermions and photons) 
that accompany it in the final state is taken into account exactly at the
relative order $\aem^p$ at which the computation is carried out; the higher $p$,
the better the kinematical description (of any degrees of freedom). However, 
in certain corners of the phase space, e.g.~where \mbox{$\pt(X)\ll E$},
this is not relevant, since large logarithmic terms render the order-by-order
accurate kinematical description ultimately irrelevant; this is nothing
but the emergence of observable-dependent logarithms discussed before
(in this example, \mbox{${\cal Q}=\pt(X)/E$}). The solution is to embed
the resummation of such logarithms into $d\hsig_{ij}$; the crucial point
is that this is done {\em within} a collinear-factorisation approach,
not {\em instead} of it. If this procedure is carried out analytically
it usually gives a highly accurate result for the degrees of freedom 
associated with $X$, but integrates out those associated with the recoil
products. While this is in fact a desirable property in the context of
a theory-to-data comparison, it may prevent one from adopting directly at 
the theoretical level the same cuts as on data, which then must be corrected
for that. Again in the context of collinear factorisation, this situation
can be further addressed by turning to fully-exclusive predictions, such
as those obtained from the aforementioned parton-shower approach or with
beyond-LO matching methods that have proven to be extremely effective in 
hadronic collisions, such as MC@NLO~\cite{Frixione:2002ik} and 
Powheg~\cite{Nason:2004rx}.

The bottom line is that, while different strategies exist that one can 
employ at the short-distance level, they are all underpinned by a collinear
{\em factorisation} picture. Thus, if one is interested in the analysis of 
the implications of the choice of the PDFs on sufficiently inclusive cross 
sections, one can essentially adopt the strategy that is most convenient
from the computational viewpoint. The obvious candidate is therefore that
where cross sections are computed at fixed order; for the purposes of this
paper, next-to-leading order accuracy is sufficient, which has the additional
benefit of being full automated.

In particular, our goal is that of assessing the impact of the effects 
due to increasing the accuracy of the PDFs, from LO+leading logarithm 
(LO+LL)~\cite{Skrzypek:1990qs,Skrzypek:1992vk,Cacciari:1992pz} to 
NLO+next-to-leading logarithm (NLO+NLL)~\cite{Frixione:2019lga,
Bertone:2019hks,Frixione:2021wzh}. We point out that all predictions
based on collinear factorisation for $\epem$ cross sections obtained 
thus far in the literature have employed LO+LL PDFs. In view of
the accuracy necessary at future $\epem$ machines this is problematic
on at least two counts: it does not match the precision typically available
at the matrix-element level and, in keeping with what has been already 
mentioned at the beginning of this introduction, it does not allow one to
properly define a theoretical systematics associated with PDF choices.

While the NLO+NLL results of refs.~\cite{Frixione:2019lga,Bertone:2019hks,
Frixione:2021wzh} are technically complete, for the goal stated above to
be phenomenologically sensible they must be supplemented by a careful
treatment of the evolution in the presence of multiple fermion families
and their mass thresholds, as well as of the $W$ boson, which was beyond
the scope of the original papers. A further interesting aspect is the
dependence on the choice of the renormalisation scheme -- in
refs.~\cite{Frixione:2019lga,Bertone:2019hks,Frixione:2021wzh} only
$\MSb$ has been considered. We shall discuss these items in the first
part of this paper. Finally, we use this work as an opportunity to
upgrade the treatment of $\epem$ ISR effects in the automated framework of
\aNLO~\cite{Alwall:2014hca,Frederix:2018nkq} (called \aNLOs\ henceforth)
from the LO+LL accuracy~\cite{Frixione:2021zdp} to the NLO+NLL one.
We stress that this implies that \aNLOs\ is now capable of computing
NLO EW corrections also for processes with massless initial-state
leptons.

This paper is organised as follows. In sect.~\ref{sec:elem} we review
refs.~\cite{Frixione:2019lga,Bertone:2019hks,Frixione:2021wzh} (with 
additional material reported in appendix~\ref{sec:pap123}), and
discuss the two aspects that must be improved in order to carry out 
simulations  that are phenomenologically viable at high-energy $\epem$ 
colliders. We deal with them in turn, in sect.~\ref{sec:evol} (evolution
with multiple fermion families) and sect.~\ref{sec:UV} (UV-renormalisation
scheme dependence). These sections also present the resulting $z\simeq 1$
analytical forms for the PDFs; the $z<1$ numerical solutions are discussed
in sect.~\ref{sec:numsolevol} instead. In sect.~\ref{sec:res} we then use 
the NLO+NLL PDFs thus obtained to predict observables relevant to the 
production in $\epem$ collisions of a heavy neutral system, of a $W^+W^-$ pair, 
and of a $t\bt$ pair. We finally draw our conclusions in sect.~\ref{sec:conc}.
In appendices~\ref{sec:NLOxLL} and~\ref{sec:LOxNLL} we give prescriptions
relevant to the cases where the perturbative accuracies of the PDFs and
the short-distance cross sections are not the same.  
In appendix~\ref{sec:NLOxsec} we briefly explain how \aNLOs\ has been
upgraded for the computations of NLO EW corrections with massless
initial-state leptons. Concurrently with this paper, we release
a new public version of \aNLOs, and a code (\eMELA) that implements
the NLO+NLL PDFs derived here.
\enlargethispage*{20pt}

\section{Theoretical ingredients and phenomenological issues\label{sec:elem}}
In ref.~\cite{Frixione:2019lga} the NLO-accurate initial conditions
for all possible combinations of partons (indices $i$ and $j$ in
eq.~(\ref{master0})) and particles (indices $k$ and $l$ in
eq.~(\ref{master0})) have been derived. These are meant to be
imposed at a mass scale $\mu_0\sim m$ and, at variance with their 
trivial LO counterparts, depend on $\aem(\mu_0)$ and contain a
\mbox{$\log\mu_0/m$} term. The initial conditions relevant to
electrons/positrons (i.e.~$k,l=e^\pm$) have then been used in
refs.~\cite{Bertone:2019hks,Frixione:2021wzh} to obtain NLL-accurate
PDFs, which is all one needs\footnote{We stress that the same methods can
be applied to the evolution of other particles, such as photons, should
the reason become compelling for also considering these particles as 
emerging from beam-beam interactions.} to deal with the largely dominant
case where beam dynamics results in particles whose identities are the same 
as those of the corresponding beams. In the notation of eq.~(\ref{beamstr}),
this is equivalent to setting:
\beq
{\cal B}_{kl}(\yp,\ym)=\delta_{k\lp}\delta_{l\lm}{\cal B}_{\epem}(\yp,\ym)\,.
\label{Bee}
\eeq
In view of eq.~(\ref{Bee}), we shall adopt the following
simpler notation for the PDFs:
\beq
\PDF{\alpha}{\lm}=\PDF{\bar{\alpha}}{\lp}\equiv\ePDF{\alpha}\,,
\label{simplnot}
\eeq
where the first equality follows from charge-conjugation invariance, and
by $\bar{\alpha}$ we have denoted the antiparticle of $\alpha$.

While both ref.~\cite{Bertone:2019hks} and ref.~\cite{Frixione:2021wzh} 
work in the $\MSb$ UV-renormalisation scheme, for the PDFs they adopt different 
{\em factorisation} schemes, equal to $\MSb$ and the so-called $\Delta$,
respectively (see ref.~\cite{Frixione:2021wzh} for more details).
The $\MSb$- and $\Delta$-defined PDFs have significantly different 
behaviours at $z\to 1$, in spite of both having an integrable divergence 
there\footnote{More precisely, $\PDF{\lpm}{\lpm}$ has a power-like integrable
divergence at $z\to 1$ (accompanied by logarithms in $\MSb$ but 
not in $\Delta$), whereas $\PDF{\gamma}{\lpm}$ is logarithmically
divergent in $\MSb$; the other PDFs do not diverge.}, whereby this region 
gives by far the dominant contribution to the cross section independently of
the factorisation scheme adopted. The difference due to the factorisation
scheme choice in the PDFs is unphysical, and is compensated by its analogue
in the partonic cross sections; this compensation can never be exact in
physical observables, and thus the residual dependence is typically 
regarded as a theoretical systematics, which we shall investigate in
this paper. Finally, motivated by the dominance of the $z\to 1$ region,
one important feature of refs.~\cite{Bertone:2019hks,Frixione:2021wzh} 
is the availability of the analytical result for the PDFs in this region:
this is crucial in order to obtain numerically-stable integrated predictions.

The results of refs.~\cite{Bertone:2019hks,Frixione:2021wzh} can immediately
be used in eq.~(\ref{master0}) to predict observables. While from a technical
point of view this poses no problem, it is not expected to give a good 
phenomenological description. The reason is that those PDFs have been
derived by considering only the electron, positron, and photon. This is
not a issue in itself, and it actually has a clear physical motivation
(the effects of other partons being suppressed by at least a relative
$\ord(\aem^2)$), but its implication is that, for consistency, $\aem$ must also
be run with a single lepton family; it is known that by doing so one obtains 
values that lead to a poor description of the data. The solution, which 
we shall present in sect.~\ref{sec:evol}, is that of evolving the PDFs by 
including all of the fermion families in the relevant energy ranges: this 
will allow one to employ a phenomenologically-sound coupling constant, 
as well as to take automatically into account the $\ord(\aem^2)$ 
perturbative suppressions mentioned above. 

A second item which has not been discussed in refs.~\cite{Bertone:2019hks,
Frixione:2021wzh} is that of the renormalisation-scheme dependence of
the PDFs: the results of those papers are relevant to the $\MSb$ scheme.
While this is the natural choice in the context of a renormalisation-group 
evolution (RGE), the existence of phenomenologically appealing 
renormalisation schemes in the Standard Model alternative to
(and much more frequently used than) $\MSb$ motivates the definition
of the PDFs in such schemes too. We shall address this point in
sect.~\ref{sec:UV}, both in a general way and by presenting explicit
results for the so-called $\aem(\mZ)$~\cite{Dittmaier:2001ay} and 
$G_\mu$~\cite{Dittmaier:2001ay,Denner:1991kt} schemes.

We note that while both of the aspects discussed above are relevant 
to LO+LL PDFs as well, they have been largely ignored in the literature.
There are a few reasons for that. Firstly, owing to the LO initial conditions
being independent of a small scale ($\mu_0$) and of the coupling 
constant ($\aem(\mu_0)$).
Secondly, at the LO+LL the running of the coupling constant can formally 
be neglected. And thirdly, at the LO+LL one simply does not
define a theoretical systematics: a choice of parameters is made 
with a specific application in mind, for which an alternative choice
is deemed less suitable, and is thus ignored. In the present work,
as far as the choices of settings are concerned, we shall treat the
LO+LL and NLO+NLL cases on equal footing, and explore the consequences
of this strategy.

\section{PDF evolution with multiple fermion families\label{sec:evol}}
In order to generalise the results of refs.~\cite{Bertone:2019hks,
Frixione:2021wzh} to the case of several fermions, for a given fermion
type $f$ we denote its electric charge (in units of the positron charge)
by $e_f$, its mass by $m_f$, and its number of colours by $\NC^{(f)}$. 
The total number of families is denoted by \mbox{$M=\Nl+\Nu+\Nd$}, with 
$\Nl$ leptons, $\Nu$ up quarks, and $\Nd$ down quarks. With abuse of 
notation, $f$ can assume both numerical and alpha-numerical values,
with the assignments in the former case determined by the hierarchy of
the fermion masses:
\beq
m_1<m_2<\ldots<m_M<m_{M+1}\equiv\infty\,.
\label{thrslist}
\eeq
Thus, for an electron, $f=e$ or $f=1$ (and $m_f\equiv m$), the electron being 
the lightest fermion. The definition of $m_{M+1}$ in eq.~(\ref{thrslist}) is 
a matter of convenience, in that it will simplify some of the formulae to be
given below. Our definition for the one-loop charge-renormalisation constant 
in the $\MSb$ scheme is the following\footnote{The resulting running coupling
constant at scale $\mu$ is denoted by $\aem(\mu)$. It is always clear from
the context when one refers to the $\MSb$ coupling, or to the $\aem(\mZ)$
scheme. Note that in the latter scheme the coupling constant is {\em not} 
denoted by $\aem(\mZ)$, and its value is not equal to $\aem(\mZ)$ 
-- see sect.~\ref{sec:Rsch}.}:
\beqn
Z_{\aem_{\MSb}}\!\!&=&\!\!
\frac{\aem}{\pi}\,\Bigg\{
\frac{1}{3}\sum_{f=light}\NC^{(f)}e_f^2\,\ooepb+
\frac{1}{3}\sum_{f=heavy}\NC^{(f)}e_f^2
\left(\ooepb-\log\frac{m_f^2}{\mu^2}\right)
\nonumber
\\*&&\phantom{\frac{\aem}{\pi}}
-\frac{7}{4}\left(\ooepb-\log\frac{\mW^2}{\mu^2}\right)\stepf(\mu\le\mW)
-\frac{7}{4}\,\ooepb\,\stepf(\mu>\mW)\Bigg\},
\label{ZaMSbdef}
\eeqn
where:
\beq
\ooepb=\frac{1}{\ep}-\gE+\log(4\pi)\,.
\label{Deltadef}
\eeq
For future reference (see sect.~\ref{sec:res}) we note that other SM 
parameters, in particular the masses of the heavy particles, are 
renormalised on-shell.
In the rest of this section, the role of the $W$ will mostly be ignored;
we shall discuss it more fully in sect.~\ref{sec:res0}. The ``light'' and 
``heavy'' fermions (that are summed over in the first and second terms on 
the r.h.s.~of eq.~(\ref{ZaMSbdef})) are determined by whether $m_f<\mu$ 
or $m_f\ge\mu$, respectively, except in the case of the top quark, that is 
always considered heavy. Equation~(\ref{ZaMSbdef}) implies that light 
fermions contribute to the running of $\aem$, while heavy fermions decouple. 
Likewise, the $W$ contributes to the running of $\aem$ only for scales
larger than its mass. More precisely, by defining
\beq
C^{(a,k)}=\sum_f \NC^{(f)}e_f^a\,\stepf\left(m_f<m_{k+1}\right)\,,
\;\;\;\;\;\;\;\;1\le k\le M\,,
\label{Cakdef}
\eeq
and by using the conventions of ref.~\cite{Bertone:2019hks} for the
coefficients of the $\beta$ function
\beq
\frac{\partial\aem(\mu)}{\partial\log\mu^2}=\beta(\aem)=
b_0\aem^2+b_1\aem^3+\ldots\,,
\label{betaQED}
\eeq
in the range
\beq
m_k\le\mu<m_{k+1}
\label{krange}
\eeq
one has $b_0=b_0^{(k)}$ and $b_1=b_1^{(k)}$, where:
\beq
b_0^{(k)}=\frac{C^{(2,k)}}{3\pi}-\frac{7}{4\pi}\,\stepf(\mu>\mW)\,,
\;\;\;\;\;\;\;\;
b_1^{(k)}=\frac{C^{(4,k)}}{4\pi^2}\,,
\label{b0b1k}
\eeq
with the coefficient $b_1^{(k)}$ obtained, in pure QED, by imposing at two 
loops the same decoupling conditions as at one loop. Equation~(\ref{b0b1k}) 
formalises the fact that when crossing the mass threshold of fermion $k$ 
e.g.~downwards (i.e.~when moving from $\mu=m_k+\varepsilon$ to 
$\mu=m_k-\varepsilon$) the role of $k$ changes from light to heavy. 
In keeping with eq.~(\ref{ZaMSbdef}), $\aem(\mu)$ is continuous at the 
thresholds (as well as at the $W$ mass). We shall discuss in 
sect.~\ref{sec:UV} how we determine the value of $\aem$ at a reference
scale, which we shall set equal to the $Z$ mass.

\subsection{Structure of the evolution equations\label{sec:strevol}}
The distinction between light and heavy fermions made above mirrors the usual 
treatment of these objects in hadronic PDFs, where the latter participate
in the evolution only at scales larger than their thresholds; furthermore, 
PDFs are continuous at such thresholds. It is natural to use the
very same approach for electron PDFs. We refrain from repeating here most of 
the details of the multi-flavour treatment in the context of RGEs, and 
rather refer the reader to ref.~\cite{deFlorian:2016gvk}, which is 
particularly convenient because it also presents the explicit results for
the Altarelli-Parisi (AP) kernels for any coupling-constant combination 
$\as^q\aem^p$ with \mbox{$q+p\le 2$}. Clearly, here we are interested in 
the case \mbox{$(q,p)=(0,2)$}. 

The structure of the evolution equations simplifies considerably if one
expresses them in terms of suitable linear combinations of PDFs, as 
opposed to individual PDFs. There is a certain freedom in their choices; 
most of the definitions of ref.~\cite{deFlorian:2016gvk} are well-suited 
to the computation of the analytical solution in the $z\to 1$ 
region\footnote{Numerically, we solve a marginally different system
of equations -- see sect.~\ref{sec:numsolevol}.}. 
In particular, we make use of the non-singlet combinations:
\beq
\ePDF{f,{\rm\sss NS}}=\ePDF{f}-\ePDF{\bar{f}}\,,
\label{defNS}
\eeq
for any fermion $f$. The other relevant combinations are the 
following\footnote{These are given for the maximal number of light
fermions. When crossing downwards the mass threshold of a given fermion,
the corresponding PDF is eliminated from the system, and some PDF combinations
become degenerate.\label{ft:Nflav}}:
\beqn
\ePDF{l2}&=&\ePDF{\lm}+\ePDF{\lp}-\left(\ePDF{\mu^-}+\ePDF{\mu^+}\right),
\label{defl2}
\\
\ePDF{l3}&=&\ePDF{\lm}+\ePDF{\lp}+\ePDF{\mu^-}+\ePDF{\mu^+}
-2\left(\ePDF{\tau^-}+\ePDF{\tau^+}\right)\,,
\label{defl3}
\\
\ePDF{uc}&=&\ePDF{u}+\ePDF{\bar{u}}-\left(\ePDF{c}+\ePDF{\bar{c}}\right),
\label{defuc}
\\
\ePDF{ds}&=&\ePDF{d}+\ePDF{\bar{d}}-\left(\ePDF{s}+\ePDF{\bar{s}}\right),
\label{defds}
\\
\ePDF{sb}&=&\ePDF{s}+\ePDF{\bar{s}}-\left(\ePDF{b}+\ePDF{\bar{b}}\right),
\label{defsb}
\\
\ePDF{\Sigma^l}&=&\sum_l^{\Nl}\left(\ePDF{l^-}+\ePDF{l^+}\right)\,,
\label{defSigl}
\\
\ePDF{\Sigma^u}&=&\sum_u^{\Nu}\left(\ePDF{u}+\ePDF{\bar{u}}\right)\,,
\label{defSigu}
\\
\ePDF{\Sigma^d}&=&\sum_d^{\Nd}\left(\ePDF{d}+\ePDF{\bar{d}}\right)\,.
\label{defSigd}
\eeqn
In addition to these, the photon PDF must be included, whereas the 
gluon one is ignored.

With the conventions of ref.~\cite{Bertone:2019hks}, the perturbative 
coefficients of the AP kernels relevant to the evolution in the range 
of eq.~(\ref{krange}) (i.e.~with $k$ light fermions) are defined as 
follows (in the $\MSb$ renormalisation scheme):
\beq
\APmat^{[k]}(x,\mu)=\sum_{j=0}^\infty\left(\frac{\aem(\mu)}{2\pi}\right)^j
\APmat^{[j,k]}(x)\,,
\label{APmatex}
\eeq
where $\APmat^{[k]}$ and $\APmat^{[j,k]}$ are matrices whose elements 
are either the individual AP kernels ($P_{ab}^{[k]}$ and $P_{ab}^{[j,k]}$,
if the evolution is carried out in terms of individual PDFs), or
some linear combination of them (if the evolution is carried out in terms
of singlet and non-singlet combinations of PDFs, as is always the case in 
practice). Equation~(\ref{APmatex}) implies that a given $P_{ab}^{[j,k]}$ 
in this paper coincides with $P_{ab}^{(0,j+1)}$ of 
ref.~\cite{deFlorian:2016gvk} (in that paper, the dependence
on the number of light fermions is implicit). In the following, we shall
add a lower index $N$ to denote a quantity in Mellin space, which
we compute with the standard definition of the Mellin transform:
\beq
M[f]\equiv f_N=\int_0^1 dz\,z^{N-1} f(z)\,,
\;\;\;\;\;\;\;\;
\bN=N\,e^{\gE}\,.
\eeq
Notation-wise, the upper index $[k]$ on the 
l.h.s.~of eq.~(\ref{APmatex}) may be omitted when the corresponding 
expression is understood to be valid for any $k$ 
(see e.g.~eqs.~(\ref{Pll})--(\ref{Plplus})).

In the lepton sector we employ the customary decomposition:
\beqn
P_{l_il_j}&=&\delta_{ij}P_{ll}^{\rm V}+P_{ll}^{\rm S}\,,
\label{Pll}
\\
P_{l_i\bl_j}&=&\delta_{ij}P_{l\bl}^{\rm V}+P_{ll}^{\rm S}\,,
\label{Plbl}
\\
P_l^\pm&=&P_{ll}^{\rm V}\pm P_{l\bl}^{\rm V}\,,
\label{Plplus}
\eeqn
having already used the fact that, up to $\ord(\aem^2)$,
\mbox{$P^{\rm S}_{ll}=P^{\rm S}_{l\bl}$}. Equations~(\ref{Pll})--(\ref{Plplus})
are symbolic both for the full expressions of the kernels, and for each
of their perturbative coefficients $P^{[j,k]}$. At the LO there are neither 
flavour-changing nor singlet contributions, and therefore:
\beqn
P_{l_il_j,N}^{[0,k]}&=&\delta_{ij}P_{ll,N}^{{\rm V}[0,k]}\,\Melleq\,
\delta_{ij}\el^2\left(-2\log\bN+\frac{3}{2}-\frac{1}{N}\right)\,,
\label{Pll0}
\\
P_{l_i\bl_j,N}^{[0,k]}&=&0\,,
\label{Plbl0}
\\
P_{l,N}^{\pm[0,k]}&=&P_{ll,N}^{{\rm V}[0,k]}\,.
\label{Plplus0}
\eeqn
Here, the symbol $\Melleq$ indicates the fact that, in Mellin space, terms 
subleading for $N\to\infty$ are neglected. We point out that this operation 
is only relevant to the analytical computations in the $z\to 1$ region:
elsewhere, numerical results that employ the complete expressions of
the evolution kernels are used instead. At the NLO, explicit computations
that make use of kernels reported in ref.~\cite{deFlorian:2016gvk} 
lead to the following expressions:
\beqn
P_{{ll},N}^{{\rm V}[1,k]}&\Melleq&
\el^2\left[
\frac{20}{9}C^{(2,k)}\log\bN+
\el^2\left(\frac{3}{8}-\frac{\pi^2}{2}+6\zeta_3\right)-
\frac{C^{(2,k)}}{18}\left(3+4\pi^2\right)\right]
\nonumber\\*&&
+\frac{\el^2}{N}\left[-4\el^2\log\bN+\frac{27\el^2+22C^{(2,k)}}{9}
\right],
\label{PllVN1asy}
\\
P_{{l\bl},N}^{{\rm V}[1,k]}&=&
\ord\left(\frac{1}{N^2}\right)\,,
\label{PlblVN1asy}
\\
P_{{ll},N}^{{\rm S}[1,k]}&=&
\ord\left(\frac{1}{N^3}\right)\,,
\label{PllSN1asy}
\eeqn
whence:
\beq
P_{l,N}^{\pm[1,k]}\,\Melleq\,P_{{ll},N}^{{\rm V}[1,k]}\,.
\label{Plplus1}
\eeq
From ref.~\cite{deFlorian:2016gvk} we also obtain the analogous kernels
relevant to the quark sector, namely:
\beqn
P_{u,N}^{\pm[j,k]}&\Melleq&P_{l,N}^{\pm[j,k]}(\ee\to\eu)\,,
\label{Pupm}
\\
P_{d,N}^{\pm[j,k]}&\Melleq&P_{l,N}^{\pm[j,k]}(\ee\to\ed)\,,
\label{Pdpm}
\eeqn
for both the LO ($j=0$) and the NLO ($j=1$) contributions. 

A simple algebra leads to the fact that the non-singlet combinations
of eq.~(\ref{defNS}) evolve independently, with a kernel equal to $P_f^-$.
Likewise, the PDF combinations of eqs.~(\ref{defl2})--(\ref{defsb})
also evolve independently, with a kernel equal to $P_l^+$ ($\ePDF{l2}$
and $\ePDF{l3}$), $P_u^+$ ($\ePDF{uc}$), and $P_d^+$ ($\ePDF{ds}$
and $\ePDF{sb}$). There thus remain the PDF combinations of
eqs.~(\ref{defSigl})--(\ref{defSigd}), plus the photon PDF; these four
quantities evolve together, as follows:
\beq
\frac{d}{d\log\mu^2}
\begin{pmatrix}
\Gamma_{\Sigma^u} \\
\Gamma_{\Sigma^d} \\
\Gamma_{\Sigma^\ell} \\
\Gamma_{\gamma}
\end{pmatrix} =
\begin{pmatrix}
  P_u^+ + 2 N_u P_{uu}^{\rm S} &
  2 N_u P_{ud}^{\rm S} &
  2 N_u P_{u\ell}^{\rm S} &
  2 N_u P_{u \gamma} \\
  2 N_d P_{du}^{\rm S} &
  P_d^+ + 2 N_d P_{dd}^{\rm S} &
  2 N_d P_{d\ell}^{\rm S} &
  2 N_d P_{d \gamma} \\
  2 N_l P_{\ell u}^{\rm S} &
  2 N_l P_{\ell d}^{\rm S} &
  P_l^+ + 2 N_l {P}_{\ell\ell}^{\rm S} &
  2 N_l P_{\ell \gamma} \\
  P_{\gamma u} &
  P_{\gamma d} &
  P_{\gamma \ell} &
  P_{\gamma\gamma} 
\end{pmatrix}
\otimes
\begin{pmatrix}
\Gamma_{\Sigma^u} \\
\Gamma_{\Sigma^d} \\
\Gamma_{\Sigma^\ell} \\
\Gamma_{\gamma}
\end{pmatrix}.
\label{APQEDfull}
\eeq
Equation~(\ref{APQEDfull}) simplifies considerably in the $z\simeq 1$
region, owing to the $N\to\infty$ behaviour of the relevant kernels
(see eq.~(\ref{PllSN1asy}) -- analogous ones holds for quarks).
By taking that into account, one arrives at:
\beq
\frac{d}{d\log\mu^2}
\left(
\begin{array}{c}
\ePDF{\Sigma^u,N}\\
\ePDF{\Sigma^d,N}\\
\ePDF{\Sigma^l,N}\\
\ePDF{\gamma,N}
\end{array}
\right)
\,\Melleq\,
\left(
\begin{array}{cccc}
P_{u,N}^+ & 0 & 0 & 2\Nu P_{u\gamma,N}
\\
0 & P_{d,N}^+ & 0 & 2\Nd P_{d\gamma,N}
\\
0 & 0 & P_{l,N}^+ & 2\Nl P_{l\gamma,N}
\\
P_{\gamma u,N} & P_{\gamma d,N} &P_{\gamma l,N} & P_{\gamma\gamma,N}
\end{array}
\right)
\left(
\begin{array}{c}
\ePDF{\Sigma^u,N}\\
\ePDF{\Sigma^d,N}\\
\ePDF{\Sigma^l,N}\\
\ePDF{\gamma,N}
\end{array}
\right).
\label{APQEDfullasy}
\eeq
At the LO:
\beqn
P_{l\gamma,N}^{[0,k]}&\Melleq&\frac{\el^2}{N}\,,
\label{P0lga}
\\
P_{\gamma l,N}^{[0,k]}&\Melleq&\frac{\el^2}{N}\,,
\label{P0gal}
\\
P_{\gamma\gamma,N}^{[0,k]}&\Melleq&-\frac{2}{3}\,C^{(2,k)}\,,
\label{P0gaga}
\eeqn
while at the NLO:
\beqn
P_{l\gamma,N}^{[1,k]}&\Melleq&\frac{\el^4}{N}\left[
\log^2\bN+\frac{15-\pi^2}{6}\right],
\label{P1lga}
\\
P_{\gamma l,N}^{[1,k]}&\Melleq&\frac{\el^2}{N}\left[
-\el^2\log^2\bN+\frac{15\el^2+4C^{(2,k)}}{3}\,\log\bN
-\frac{3\el^2(36+\pi^2)+64C^{(2,k)}}{18}\right],
\label{P1gal}
\\
P_{\gamma\gamma,N}^{[1,k]}&\Melleq&-C^{(4,k)}\,.
\label{P1gaga}
\eeqn
The remaining kernels are computed by observing that, at any order:
\beqn
&&P_{u\gamma}=P_{l\gamma}\left(\el^a\to\NC\eu^a\right)\,,\;\;\;\;\;\;
P_{d\gamma}=P_{l\gamma}\left(\el^a\to\NC\ed^a\right)\,,
\label{Pudga}
\\
&&P_{\gamma u}=P_{\gamma l}\left(\el\to\eu\right)\,,\;\;\;\;\;\;
\phantom{\NC}
P_{\gamma d}=P_{\gamma l}\left(\el\to\ed\right)\,.
\label{Pgadu}
\eeqn
Equations~(\ref{APQEDfullasy})--(\ref{Pgadu}) show that, at $\ord(N^{-1})$, 
the singlets couple to each other only through the photon. One is therefore 
in the same situation as in the single-fermion-family evolution described
in detail in app.~B of ref.~\cite{Bertone:2019hks}; we shall exploit 
this fact in sect.~\ref{sec:ansolevol}.

\subsection{Solution of the evolution equations\label{sec:solevol}}
In this section, we work with the $\MSb$ renormalisation scheme;
this condition will be relaxed later (see sect.~\ref{sec:UV}).
However, our treatment applies to any factorisation scheme,
although explicit results will be given only for $\MSb$ and $\Delta$.

We remind the reader that QED PDFs have a meaningful perturbative
expansion at any scale, for whose coefficients we use the same 
conventions as in eq.~(\ref{APmatex}), namely:
\beq
\ePDF{i}(x,\mu)=\sum_{j=0}^\infty\left(\frac{\aem(\mu)}{2\pi}\right)^j
\ePDF{i}^{[j]}(x,\mu)\,.
\label{PDFex}
\eeq
By computing the terms on the r.h.s.~of eq.~(\ref{PDFex}) from first
principles, and by setting $\mu=\muz$, with $\muz$ a scale of the order
of the electron mass, one obtains the initial conditions for the RGE
evolution of the PDFs. Up to the NLO~\cite{Frixione:2019lga}:
\beqn
\ePDF{i}^{[0]}(z,\muz)&=&\delta_{i\lm}\delta(1-z)\,,
\label{G0sol}
\\
\ePDF{\lm}^{[1]}(z,\muz)&=&\ee^2\left[\frac{1+z^2}{1-z}\left(
\log\frac{\muz^2}{m^2}-2\log(1-z)-1\right)\right]_+ +\ee^2 K_{ee}(z)\,,
\label{G1sol2}
\\
\ePDF{\gamma}^{[1]}(z,\muz)&=&\ee^2\frac{1+(1-z)^2}{z}\left(
\log\frac{\muz^2}{m^2}-2\log z-1\right) +\ee^2 K_{\gamma e}(z)\,,
\label{Ggesol2}
\\
\ePDF{i}^{[1]}(z,\muz)&=&0\,,
\;\;\;\;\;\;\;\;i\,\notin\,\{\lm,\gamma\}\,.
\label{Gpossol2}
\eeqn
We point out that while the computation of ref.~\cite{Frixione:2019lga}
has been carried out for a single fermion family, its results apply here
as well, since at $\ord(\aem)$ the only possible elementary branching
is $e\to e\gamma$, whence eq.~(\ref{Gpossol2}) follows trivially. 
The functions $K_{ij}$ that appear on the r.h.s.~of
eqs.~(\ref{G1sol2}) and~(\ref{Ggesol2}) are responsible for defining
the factorisation scheme. Technically, they are the finite parts of
the subtraction terms for initial-state collinear singularities, 
the residues of whose poles in $1/\bar{\epsilon}$ are the 
Altarelli-Parisi kernels; as such, if one works in the $\MSb$
factorisation scheme one must set them equal to zero. Apart from having 
to fulfill certain conditions that stem from momentum and charge conservation, 
these functions are completely arbitrary. This arbitrariness is compensated
by an analogous one in the partonic cross sections $d\hsig_{ij}$, so that
the l.h.s.~of eq.~(\ref{master0}) is independent (at the perturbative
accuracy at which one is working) of the choice of the factorisation
scheme. At the NLO, the FKS subtraction formalism~\cite{Frixione:1995ms,
Frixione:1997np} includes explicitly the functions $K_{ij}$ in the 
expressions of $d\hsig_{ij}$ (in the context of \aNLOs, see
ref.~\cite{Frederix:2009yq}; see also appendix~\ref{sec:LOxNLL}).

We finally mention the fact that, in keeping with refs.~\cite{Bertone:2019hks,
Frixione:2021wzh}, the evolution equations for the PDFs are more conveniently 
re-expressed as an evolution equation for an evolution operator in Mellin 
space, $\Eop_N^{(K)}$. The upper index $(K)$ here reminds one that the 
evolution operator is dependent on the choice of the factorisation scheme.

\subsection{Analytical solution\label{sec:ansolevol}}
In order to deal with the running of the coupling constant in an
easier manner, in refs.~\cite{Bertone:2019hks,Frixione:2021wzh} it was 
shown that in the evolution equation it is convenient to use the variable
\beq
t=\frac{1}{2\pi b_0}\log\frac{\aem(\mu)}{\aem(\mu_0)}\,,
\label{tdef}
\eeq
rather than the scale $\mu$ (the two are in one-to-one correspondence).
It is immediately clear that eq.~(\ref{tdef}) cannot be used in the presence 
of mass thresholds, since the value of $b_0$ depends on the range where the 
evolution takes place. However, one can easily generalise
eq.~(\ref{tdef}). Specifically, by taking eqs.~(\ref{krange}) 
and~(\ref{b0b1k}) into account, we introduce the quantities:
\beq
\bm_k=\min\Big(m_{k+1},\max\big(\mu_0,m_k\big)\Big)=
\left\{
\begin{array}{ll}
m_k &~~~\mu_0<m_k\\
\mu_0 &~~~m_k\le\mu_0<m_{k+1}\\
m_{k+1} &~~~\mu_0\ge m_{k+1}\\
\end{array}
\right.
,
\label{bmkdef}
\eeq
and we define the analogues of the variable $t$ of eq.~(\ref{tdef}),
namely:
\beq
t_k=\frac{1}{2\pi b_0^{(k)}}\log\frac{\aem(\mu)}{\aem(\bm_k)}\,.
\label{tkdef}
\eeq
For any $k$, the variable $t_k$ is the evolution variable to be used instead 
of $\mu$ in the range:
\beq
m_k\le\mu<m_{k+1}\;\bigcap\;\mu\ge\mu_0\,.
\label{krange2}
\eeq
We also need to define:
\beq
\bt_k=\frac{1}{2\pi b_0^{(k)}}\log\frac{\aem(m_{k+1})}{\aem(\bm_k)}\,,
\label{tkbardef}
\eeq
so that in terms of $t_k$ the range of eq.~(\ref{krange2}) is:
\beq
0\le t_k\le\bt_k\,.
\label{kranget}
\eeq
Note that for any $k$ such that $m_{k+1}\le\mu_0$ eq.~(\ref{tkbardef})
implies that $\bt_k=0$, and therefore that the range of eq.~(\ref{kranget})
is a zero-measure set where $t_k=0$. Indeed, in such a case the intersection
in eq.~(\ref{krange2}) is the empty set, in keeping with the fact that
for scales smaller than $\mu_0$ it is not useful to consider the
evolution of the PDFs. In practice, for electron PDFs with $\muz=m$
the range in eq.~(\ref{krange2}) coincides with that in eq.~(\ref{krange}),
and $\bt_k\ne 0$ for any $k$. Still, eq.~(\ref{kranget}) is fully general,
and can be used when $\muz>m$, as well as for PDFs relevant to particles
different from the electron, in particular for muons.

In each of the ranges given in eq.~(\ref{kranget}) the evolution 
operator obeys the same evolution equation as that derived in 
ref.~\cite{Frixione:2021wzh}\footnote{That equation, being valid
for a generic factorisation scheme, encompasses and generalises the one
introduced in ref.~\cite{Bertone:2019hks}, that is relevant to $\MSb$.}, 
with the only formal differences due to the coefficients of the $\beta$ 
function and to the AP kernels, that here depend on the number of active 
fermion families. Thus, from eq.~(2.25) of ref.~\cite{Frixione:2021wzh}:
\beqn
\frac{\partial \Eop_{N}^{(K)}(t_k)}{\partial t_k}&=&
b_0^{(k)}\aem(\mu)\Kmat_N
\left(I+\frac{\aem(\mu)}{2\pi}\,\Kmat_N\right)^{-1}
\Eop_{N}^{(K)}(t_k)
\label{EKopevtk}
\\*&+&
\frac{b_0^{(k)}\aem^2(\mu)}{\beta(\aem(\mu))}
\sum_{j=0}^\infty\left(\frac{\aem(\mu)}{2\pi}\right)^j
\nonumber
\\*&&\!\!\!\!\!\phantom{\frac{b_0\aem^2(\mu)}{\beta(\aem(\mu))}}\times
\left(I+\frac{\aem(\mu)}{2\pi}\,\Kmat_N\right)\APmat_N^{[j,k]}
\left(I+\frac{\aem(\mu)}{2\pi}\,\Kmat_N\right)^{-1}\Eop_{N}^{(K)}(t_k)\,.
\nonumber
\eeqn
Here, $\Kmat$ is the matrix whose entries are the $K_{ij}$ functions, defined 
according to the conventions introduced in ref.~\cite{Frixione:2021wzh}.
As is obvious from eqs.~(\ref{G1sol2}) and~(\ref{Ggesol2}), here this matrix
does not depend on the number of active flavours.

As was discussed in sect.~\ref{sec:strevol}, the non-singlet combinations 
of eq.~(\ref{defNS}) evolve independently, with kernels $P_f^-$. The 
relevant evolution operator is therefore a scalar, and thus we re-write 
eq.~(\ref{EKopevtk}) with a simplified notation and by 
expanding\footnote{Note that the inverse operator in the first term on the 
r.h.s.~of eq.~(\ref{EKopevtk}) is not expanded; this is crucial in order to 
obtain a solution, called $\Delta_1$ in ref.~\cite{Frixione:2021wzh}, which
is sensible at $z\to 1$ -- see that paper for more details.} in $\aem$:
\beqn
\frac{\partial E_{N}^{(K)}(t_k)}{\partial t_k}&=&
b_0^{(k)}\aem(\mu)K_N
\left(1+\frac{\aem(\mu)}{2\pi}\,K_N\right)^{-1}
E_{N}^{(K)}(t_k)
\label{EKopevtkNS}
\\*&+&
\left[P_N^{-[0,k]}+\frac{\aem(\mu)}{2\pi}\left(
P_N^{-[1,k]}-\frac{2\pi b_1^{(k)}}{b_0^{(k)}}\,P_N^{-[0,k]}\right)
\right]E_{N}^{(K)}(t_k)+\ord(\aem^2)\,.
\nonumber
\eeqn
We point out that, owing to the initial conditions of 
eqs.~(\ref{G0sol})--(\ref{Gpossol2}), the only non trivial case
for eq.~(\ref{EKopevtkNS}) is that relevant to the electron non-singlet:
we thus understand \mbox{$P_N^{-[j,k]}\equiv P_{l,N}^{-[j,k]}$} and
\mbox{$K_N\equiv K_{ee,N}$} (incidentally, for all of the other non-singlets
the scheme-change kernel is equal to zero). By proceeding as was done
in ref.~\cite{Frixione:2021wzh}, one readily arrives at the solution
of eq.~(\ref{EKopevtkNS}):
\beqn
\log E_{N}^{(K)}(t_k)&=&P_N^{-[0,k]}\,t_k+\frac{1}{4\pi^2 b_0^{(k)}}
\big(\aem(\mu)-\aem(\bm_k)\big)
\left(P_N^{-[1,k]}-\frac{2\pi b_1^{(k)}}{b_0^{(k)}}\,P_N^{-[0,k]}\right)
\nonumber\\*&+&
\log\frac{1+\frac{\aem(\mu)}{2\pi}K_N}
{1+\frac{\aem(\bm_k)}{2\pi}K_N}+A_k\,,
\label{Esol1p3tk}
\eeqn
with 
\beq
\aem(\bm_k)=\aem(\mu)\,e^{-2\pi b_0^{(k)}t_k}\,,
\label{tmp15}
\eeq
and $A_k$ an integration constant, to be determined by imposing suitable
initial conditions that we shall soon discuss. We point out that $\bm_k$ 
is the value of the scale $\mu$ when $t_k=0$, according to eq.~(\ref{tkdef}), 
i.e.~of the lower bound on the integration range of eq.~(\ref{krange2}), when
the latter is non-trivial. 

The determination of the integration constants $A_k$ is done recursively.
Firstly, after choosing the scale $\mu_0$ let us denote by $\rho$ an
index such that:
\beq
m_{\rho}\le\mu_0<m_{\rho+1}\,.
\label{rhodef}
\eeq
Thus, for any $k<\rho$ we have $t_k=0$ and eq.~(\ref{tmp15}) implies
that the first three terms on the r.h.s.~of eq.~(\ref{Esol1p3tk}) are 
equal to zero. Also, below $\mu_0$ there is no evolution, and therefore 
we must have \mbox{$E_{N}^{(K)}(t_k)=I$}. Therefore:
\beq
A_k=0\;\;\;\;\;\;\forall k<\rho\,.
\label{Akeqz}
\eeq
Exactly the same arguments apply to the case $k=\rho$, since also
in this case (owing to the fact that \mbox{$t_\rho=0\Rightarrow\mu=\muz$})
the initial condition is:
\beq
E_{N}^{(K)}(t_\rho=0)=I\,,
\eeq
and therefore one also has $A_{\rho}=0$. 
Obviously, the fundamental difference between the 
solutions for $E_{N}^{(K)}(t_k)$ when $k<\rho$ and $k=\rho$ is that in the 
former case the entire r.h.s.~of eq.~(\ref{Esol1p3tk}) is equal to zero,
while in the latter case only the integration constant (for any $t_\rho>0$)
is equal to zero. 

We now consider $k=\rho+1$. By construction, the lower bound of the 
integration range in $t_{\rho+1}$, i.e.~$t_{\rho+1}=0$, represents the
same scale ($\mu=m_{\rho+1}$) as the upper bound of the integration range 
in $t_{\rho}$, i.e.~$t_{\rho}=\bt_\rho$. At such a scale value, which
corresponds to a mass threshold, the PDFs are continuous; in order to 
achieve this, we impose:
\beq
\log E_{N}^{(K)}(t_{\rho+1}=0)=\log E_{N}^{(K)}(t_{\rho}=\bt_{\rho})\,.
\label{Econt}
\eeq
By solving eq.~(\ref{Econt}) for $A_{\rho+1}$ we obtain:
\beqn
A_{\rho+1}&=&P_N^{-[0,\rho]}\,\bt_\rho+
\frac{\aem(m_{\rho+1})}{4\pi^2 b_0^{(\rho)}}
\left(1-e^{-2\pi b_0^{(\rho)}\bt_\rho}\right)
\left(P_N^{-[1,\rho]}-\frac{2\pi b_1^{(\rho)}}{b_0^{(\rho)}}\,
P_N^{-[0,\rho]}\right)
\nonumber\\*&+&
\log\frac{1+\frac{\aem(m_{\rho+1})}{2\pi}K_N}
{1+\frac{\aem(\bm_\rho)}{2\pi}K_N}\,.
\label{Arhop1sol}
\eeqn
which must be replaced in eq.~(\ref{Esol1p3tk}) in order to obtain the
complete solution for \mbox{$E_{N}^{(K)}(t_{\rho+1})$}. By doing so, it becomes
clear how the procedure can be iterated. The result for a generic $k$ 
(including $k<\rho$) thus reads as follows:
\beqn
\log E_{N}^{(K)}(t_k)&=&
\sum_{i=1}^{k-1}P_N^{-[0,i]}\,\bt_i+
P_N^{-[0,k]}\,t_k
\nonumber\\*&+&
\sum_{i=1}^{k-1}\frac{\aem(m_{i+1})}{4\pi^2 b_0^{(i)}}
\left(1-e^{-2\pi b_0^{(i)}\bt_i}\right)
\left(P_N^{-[1,i]}-\frac{2\pi b_1^{(i)}}{b_0^{(i)}}\,
P_N^{-[0,i]}\right)
\nonumber\\*&+&
\frac{\aem(\mu)}{4\pi^2 b_0^{(k)}}
\left(1-e^{-2\pi b_0^{(k)}t_k}\right)
\left(P_N^{-[1,k]}-\frac{2\pi b_1^{(k)}}{b_0^{(k)}}\,P_N^{-[0,k]}\right)
\nonumber\\*&+&
\log\left(
\frac{1+\frac{\aem(\mu)}{2\pi}K_N}
{1+\frac{\aem(\bm_k)}{2\pi}K_N}
\prod_{i=1}^{k-1}\frac{1+\frac{\aem(m_{i+1})}{2\pi}K_N}
{1+\frac{\aem(\bm_i)}{2\pi}K_N}\right)\,.
\label{Esol1p3tkfull}
\eeqn
By using the definition of $\bm_i$ given in eq.~(\ref{bmkdef}), it is 
immediate to see that:
\beq
\frac{1+\frac{\aem(\mu)}{2\pi}K_N}
{1+\frac{\aem(\bm_k)}{2\pi}K_N}
\prod_{i=1}^{k-1}\frac{1+\frac{\aem(m_{i+1})}{2\pi}K_N}
{1+\frac{\aem(\bm_i)}{2\pi}K_N}=
\frac{1+\frac{\aem(\mu)}{2\pi}K_N}
{1+\frac{\aem(\mu_0)}{2\pi}K_N}\,.
\eeq
Following refs.~\cite{Bertone:2019hks,Frixione:2021wzh}, in the region
of interest ($N\to\infty$) we re-express eq.~(\ref{Esol1p3tkfull}) as follows:
\beq
\log E_{N}^{(K)}(t_k)\Melleq
-\xi_1^{(k)}\log\bN+\hat{\xi}_1^{(k)}+
\log\left(\frac{1+\frac{\aem(\mu)}{2\pi}K_N}
{1+\frac{\aem(\mu_0)}{2\pi}K_N}\right)\,,
\label{ENxik1}
\eeq
where we have used eqs.~(\ref{Pll0}), (\ref{Plplus0}), (\ref{PllVN1asy}),
and~(\ref{Plplus1}) at $\ord(N^0)$ to define:
\beqn
\xi_1^{(k)}/\ee^2&=&
2\left(\sum_{i=1}^{k-1}\bt_i+t_k\right)
\nonumber\\*&-&
\sum_{i=1}^{k-1}\frac{\aem(m_{i+1})}{4\pi^2 b_0^{(i)}}
\left(1-e^{-2\pi b_0^{(i)}\bt_i}\right)
\left(\frac{20}{9}C^{(2,i)}+\frac{4\pi b_1^{(i)}}{b_0^{(i)}}\right)
\nonumber\\*&-&
\frac{\aem(\mu)}{4\pi^2 b_0^{(k)}}
\left(1-e^{-2\pi b_0^{(k)}t_k}\right)
\left(\frac{20}{9}C^{(2,k)}+\frac{4\pi b_1^{(k)}}{b_0^{(k)}}\right)\,,
\label{xikR1def}
\\
\hat{\xi}_1^{(k)}/\ee^2&=&
\frac{3}{2}\left(\sum_{i=1}^{k-1}\bt_i+t_k\right)
\nonumber\\*&+&
\sum_{i=1}^{k-1}\frac{\aem(m_{i+1})}{4\pi^2 b_0^{(i)}}
\left(1-e^{-2\pi b_0^{(i)}\bt_i}\right)
\left(\lambda_1^{(i)}-\frac{3\pi b_1^{(i)}}{b_0^{(i)}}\right)
\nonumber\\*&+&
\frac{\aem(\mu)}{4\pi^2 b_0^{(k)}}
\left(1-e^{-2\pi b_0^{(k)}t_k}\right)
\left(\lambda_1^{(k)}-\frac{3\pi b_1^{(k)}}{b_0^{(k)}}\right)\,,
\label{chik1Rdef}
\eeqn
with:
\beq
\lambda_1^{(k)}=\ee^2\left(
\frac{3}{8}-\frac{\pi^2}{2}+6\zeta_3\right)-
\frac{C^{(2,k)}}{18}\left(3+4\pi^2\right)\,.
\label{hlam1kdef}
\eeq
Noticing that eq.~(\ref{ENxik1}) has the same functional form as its 
analogue in the single-fermion-family case of refs.~\cite{Bertone:2019hks,
Frixione:2021wzh}, and recalling that the NLO initial conditions are
also unchanged, it follows that the $z\to 1$ solution for the 
electron non-singlet component is obtained from eq.~(5.63) of 
ref.~\cite{Bertone:2019hks} for $\MSb$, and from eq.~(4.40) of
ref.~\cite{Frixione:2021wzh} for $\Delta$ (both are also reported
in app.~\ref{sec:pap123}), simply with the replacements:
\beq
\xi_1\;\longrightarrow\;\xi_1^{(k)}\,,
\;\;\;\;\;\;\;\;
\hat{\xi}_1\;\longrightarrow\;\hat{\xi}_1^{(k)}\,.
\label{replxi}
\eeq
In addition, following the observation about eq.~(\ref{APQEDfullasy})
that implies that the singlet-photon sector evolution is formally 
identical to that of refs.~\cite{Bertone:2019hks,Frixione:2021wzh}, 
we obtain that in the $z\to 1$ region the electron PDF coincide
with its non-singlet counterpart, and that the photon PDF can be
obtained from eq.~(B.87) of ref.~\cite{Bertone:2019hks} for $\MSb$, 
and from eq.~(5.50) of ref.~\cite{Frixione:2021wzh} for $\Delta$ 
(both are also reported in app.~\ref{sec:pap123}), again with the 
replacements of eq.~(\ref{replxi}) and $t\to t_k$.

The results above show explicitly how the PDF evolution with
multiple fermion families, although technically more complicated
than its single-family counterpart, is qualitatively very similar
to the latter, which then provides one with a sensible physical
interpretation that is intuitively easier to understand.

\section{Alternative UV-renormalisation schemes\label{sec:UV}}
In order to obtain the analogues of the results of sect.~\ref{sec:evol}
in an UV-renormalisation scheme different from $\MSb$, one starts by
observing that the UV-scheme dependence of the PDFs may stem from
two different sources: the initial conditions, and the evolution 
equations. 

As far as the former are concerned, the NLO initial conditions are
in fact UV-scheme independent. This is straightforward to show by
using the calculation technique outlined in sect.~5 of 
ref.~\cite{Frixione:2019lga}, whereby the initial conditions are 
entirely determined by collinear and quasi-collinear contributions. 
Therefore, the only possible UV-scheme dependence enters through the 
coupling constant. Since the first non-null contribution to the initial 
conditions is of $\ord(\aem^0)$, and since the difference between the coupling 
constants defined in any two UV schemes is of $\ord(\aem^2)$, if follows
that the PDF initial conditions might have a UV-scheme dependence only
at the NNLO and beyond\footnote{Note, however, that 
it is not mandatory to expand perturbatively the $\aem$ 
factor that appears in the NLO contribution to the PDF initial conditions 
in terms of a given reference coupling; for the sake of numerical accuracy, 
it is actually better to employ the value dictated by the specific UV scheme 
chosen. By doing so, the {\em numerical} values of the PDF initial conditions
at the NLO are dependent on the UV scheme, but it remains true that, from
a perturbative standpoint, such a dependence is of higher order.}.

Coming to the evolution equations, the UV-scheme dependence enters
through both the coupling constant and the AP kernels; that of the
latter, which starts at the NLO, is in turn also driven by the 
coupling constant. More specifically, denoting by $R$ a generic
non-$\MSb$ UV-renormalisation scheme\footnote{Since in practice we
shall employ either the $\aem(\mZ)$ or the $G_\mu$ scheme as alternatives
to $\MSb$, the discussion that follows assumes that the coupling constant
defined in $R$ does not run. It is trivial to generalise the results to the 
cases where this condition is relaxed.\label{ft:NR}}, by $\aemR$ its coupling 
constant, and by $\bAPmat^{[j,k]}$ the matrices of the AP kernels in the 
$R$-scheme (i.e.~the analogues of the $\MSb$-defined $\APmat^{[j,k]}$),
we have:
\beq
\frac{\aem(\mu)}{2\pi}\sum_{j=0}^n
\left(\frac{\aem(\mu)}{2\pi}\right)^j \APmat^{[j,k]}(x)=
\frac{\aemR}{2\pi}\sum_{j=0}^n
\left(\frac{\aemR}{2\pi}\right)^j \bAPmat^{[j,k]}(x,\mu)
+\ord\left(\aem^{n+2}\right)\,,
\label{APijMSbvsaMZ}
\eeq
at any given order $\aem^{n+1}$. 
Here, we shall employ eq.~(\ref{APijMSbvsaMZ}) at the NLO, i.e.~with $n=1$,
to obtain $\bAPmat^{[j,k]}$ in terms of their $\MSb$ counterparts. In
order to do so, we need to express $\aemR$ in terms of $\aem(\mu)$.
This is most easily done by expressing $\aemR$ in terms of $\aem(\mZ)$,
and then use the RGE of the $\MSb$ coupling constant to relate $\aem(\mZ)$
to $\aem(\mu)$. Note that $\mZ$ is a convenient choice for a
``large'' scale (i.e.~larger than all of the fermion thresholds), given
its prominent role in the $\aem(\mZ)$ and $G_\mu$ schemes; if need be,
it can be replaced by any other fixed scale. By introducing the
coefficient $\Delta_{\MSb\to R}$ (which is perturbatively calculable)
thus:
\beq
\aemR=\aem(\mZ)-\Delta_{\MSb\to R}\,\aem^2(\mZ)+\ord(\aem^3)\,,
\label{aGmuvsaMSb}
\eeq
we obtain:
\beqn
\bAPmat^{[0,k]}(x,\mu)&=&\APmat^{[0,k]}(x)\,,
\label{P0aemZthrs2}
\\*
\bAPmat^{[1,k]}(x,\mu)&=&\APmat^{[1,k]}(x)
+\left(2\pi b_0^{(k)}\log\frac{\mu^2}{m_{k+1}^2}+
D^{(k)}\right)\APmat^{[0,k]}(x)\,,
\label{P1aemZthrs2}
\eeqn
having now defined
\beq
m_{M+1}=\mZ
\label{mMpoZ}
\eeq
and
\beq
D^{(k)}=2\pi\sum_{i=k+1}^M b_0^{(i)}\log\frac{m_i^2}{m_{i+1}^2}
+2\pi\Delta_{\MSb\to R}\,.
\label{Dkdef}
\eeq
In eq.~(\ref{Dkdef}) we understand that the sum in the first term
on the r.h.s.~gives no contribution when 
\mbox{$k=M$}, i.e.~\mbox{$D^{(M)}=2\pi\Delta_{\MSb\to R}$}. 

In view of the fact that we are considering here a scheme $R$ where
the coupling constant does not run (see footnote~\ref{ft:NR}), the
evolution equation for the evolution operator is best expressed directly
in terms of the scale $\mu$. Following ref.~\cite{Frixione:2021wzh},
it is straightforward to arrive at the following result:
\beq
\frac{\partial \Eop_{N}^{(K)}(\mu)}{\partial\log\mu^2}=
\sum_{j=0}^\infty\left(\frac{\aemR}{2\pi}\right)^{j+1}
\left(I+\frac{\aemR}{2\pi}\,\Kmat_N\right)
\bAPmat_N^{[j,k]}
\left(I+\frac{\aemR}{2\pi}\,\Kmat_N\right)^{-1}
\Eop_{N}^{(K)}(\mu)\,,
\label{matAPmell3thrs}
\eeq
which is the counterpart of eq.~(\ref{EKopevtk}), understood to be relevant
in the following range\footnote{Equation~(\ref{krange3}) is identical
to eq.~(\ref{krange2}); they seemingly differ because of the different
definitions of $m_{M+1}$ in eqs.~(\ref{thrslist}) and~(\ref{mMpoZ}).}:
\beq
\begin{array}{ll}
m_k\le\mu<m_{k+1}\;\bigcap\;\mu\ge\mu_0\,,\;\;\;\;\;\;&k\le M-1\,,\\
\mu\ge m_M\;\bigcap\;\mu\ge\mu_0\,,\;\;\;\;\;\;&k=M\,.
\end{array}
\label{krange3}
\eeq
As we have seen in sect.~\ref{sec:ansolevol}, if one is interested in
the $z\to 1$ behaviour the only non-trivial content of
eq.~(\ref{matAPmell3thrs}) is that relevant to the non-singlet electron
component. In that case, by applying the same iterative procedure
as in sect.~\ref{sec:ansolevol}, we arrive at the following solution
\beq
\log E_{N}^{(K)}(\mu)\Melleq
-\xi_1^{(k)}\log\bN+\hat{\xi}_1^{(k)}\,,
\label{ENxik1mu}
\eeq
where:
\beqn
\xi_1^{(k)}/\ee^2&=&\frac{\aemR}{\pi}
\left(\log\frac{\mu^2}{\bm_k^2}+
\sum_{i=1}^{k-1}\log\frac{m_{i+1}^2}{\bm_i^2}\right)
\label{ximukR1def}
\\*&+&
\frac{\aemR^2}{2\pi}\Bigg\{
b_0^{(k)}\left(\log^2\frac{\mu^2}{m_{k+1}^2}-
\log^2\frac{\bm_k^2}{m_{k+1}^2}\right)-
\sum_{i=1}^{k-1}b_0^{(i)}\log^2\frac{\bm_i^2}{m_{i+1}^2}
\nonumber
\\*&&\phantom{\frac{\aemR^2}{2\pi}}
-\frac{1}{\pi}\left(\frac{10}{9}C^{(2,k)}-D^{(k)}\right)
\log\frac{\mu^2}{\bm_k^2}
\nonumber
\\*&&\phantom{\frac{\aemR^2}{2\pi}}
-\frac{1}{\pi}\sum_{i=1}^{k-1}\left(\frac{10}{9}C^{(2,i)}-D^{(i)}\right)
\log\frac{m_{i+1}^2}{\bm_i^2}
\Bigg\}\,,
\nonumber
\\
\hat{\xi}_1^{(k)}/\ee^2&=&
\frac{3\aemR}{4\pi}
\left(\log\frac{\mu^2}{\bm_k^2}+
\sum_{i=1}^{k-1}\log\frac{m_{i+1}^2}{\bm_i^2}\right)
\label{chimuk1Rdef}
\\*&+&
\frac{3\aemR^2}{8\pi}\Bigg\{
b_0^{(k)}\left(\log^2\frac{\mu^2}{m_{k+1}^2}-
\log^2\frac{\bm_k^2}{m_{k+1}^2}\right)-
\sum_{i=1}^{k-1}b_0^{(i)}\log^2\frac{\bm_i^2}{m_{i+1}^2}\Bigg\}
\nonumber
\\*&+&\left(\frac{\aemR}{2\pi}\right)^2\Bigg\{
\lambda_1^{(k)}\log\frac{\mu^2}{\bm_k^2}+
\sum_{i=1}^{k-1}\lambda_1^{(i)}\log\frac{m_{i+1}^2}{\bm_i^2}
\nonumber
\\*&&\phantom{\left(\frac{\aemR}{2\pi}\right)^2}
+\frac{3}{2}\left(D^{(k)}\log\frac{\mu^2}{\bm_k^2}+
\sum_{i=1}^{k-1}D^{(i)}\log\frac{m_{i+1}^2}{\bm_i^2}\right)
\Bigg\}\,.
\nonumber
\eeqn
Equations~(\ref{ximukR1def}) and~(\ref{chimuk1Rdef}) must be used
on the r.h.s.'s of the replacements of eq.~(\ref{replxi}), together 
with the replacement
\beq
\aem(\kappa)\;\longrightarrow\;\aemR
\label{replaem}
\eeq
for any scale $\kappa$, in order to obtain the $z\to 1$ solution for 
the electron PDF in the $R$ UV-renormalisation scheme from eq.~(5.63) 
of ref.~\cite{Bertone:2019hks} (in the $\MSb$ factorisation scheme), 
and from eq.~(4.40) of ref.~\cite{Frixione:2021wzh} (in the $\Delta$
factorisation scheme) -- these equations are also reported in 
app.~\ref{sec:pap123}.

By means of a straightforward algebra\footnote{This entails expanding 
in series of $\aem$ the parameters $t_k$ and $\bt_k$, and employing
eq.~(\ref{aGmuvsaMSb}).}, one can verify that the expansions in series 
of $\aem$ of eqs.~(\ref{ximukR1def}) and~(\ref{chimuk1Rdef})
differ from those of eqs.~(\ref{xikR1def}) and~(\ref{chik1Rdef}), 
respectively, by terms of $\ord(\aem^3)$. Conversely, these differences 
are of $\ord(\aem^2)$ if in the $\MSb$ results of eqs.~(\ref{xikR1def}) 
and~(\ref{chik1Rdef}) one naively neglects the running of $\aem$
(by fixing the value of $\aem(\mu)$ and by setting the $\beta$-function
coefficients equal to zero in the analytical expressions). Because
of this, in order to study the behaviour of the NLL PDFs at fixed 
$\aem$, one must choose a renormalisation scheme 
where $\aem$ does not run, as opposed to simply
switching the running of $\aem$ off in $\MSb$ (as one can do at the LL).
Having said that, we note that the $\ord(\aem^3)$ terms mentioned above
are logarithmically enhanced (by $\log\mu^2$); therefore, we expect these
differences to become more relevant with increasing energies. 
We shall briefly return to this matter in sect.~\ref{sect:resfacren}.

\subsection{The $\aem(\mZ)$ and $G_\mu$ schemes\label{sec:Rsch}}
In this section we give the relevant definitions for the $\aem$
renormalisation factors in the $\aem(\mZ)$ and $G_\mu$ schemes,
whose associated coupling constants we denote by $\aemmZ$
and $\aemGmu$, respectively. We also compute their numerical
values, as well as that of the $\MSb$ coupling at $\mu=\mZ$, 
i.e.~$\aem(\mZ)$, which we shall eventually use in our numerical 
simulations of sect.~\ref{sec:res}. We work at the one-loop level, 
and we employ the notation of ref.~\cite{Denner:1991kt} for the 
two-point functions.

We start from the so-called $\aem(0)$ scheme, for which:
\beq
Z_{\aem(0)}=\Pi^{AA}(0)-2\frac{\sw}{\cw}\frac{\Sigma_T^{AZ}(0)}{\mZ^2}\,,
\label{Zaem0}
\eeq
where
\beq
\sw=\sqrt{1-\cw^2}\,,\;\;\;\;\;\;\;\;
\cw=\frac{\mW}{\mZ}\,,
\label{Weinangle}
\eeq
and
\beq
\Pi^{AA}(0)=\left.\frac{\partial\Sigma_T^{AA}(k^2)}{\partial k^2}
\right|_{k^2=0}\,,
\;\;\;\;\;\;\;\;
\Pi^{AA}(k^2)=\frac{\Sigma_T^{AA}(k^2)}{k^2}\,.
\label{PiAA0mZdef}
\eeq
The corresponding coupling constant is equal to the Thomson value:
\beq
\aem(0)=1/137.035999084\,.
\label{Thomson}
\eeq
The two-point $\gamma\gamma$ function is conveniently decomposed into
its fermion and $W$ contributions:
\beq
\Pi^{AA}(0)=\sum_f\Pi_f^{AA}(0)+\Pi_W^{AA}(0)\,.
\label{PiAA0cont}
\eeq
The renormalisation constant for $\aem$ in the $\aem(\mZ)$ scheme is
defined as follows
\beq
Z_{\aem(\mZ)}=Z_{\aem(0)}-\Delta\aem\,,
\label{ZamZ}
\eeq
with
\beq
\Delta\aem=\sum_{f=light}\Big(\Pi_f^{AA}(0)-\Pi_f^{AA}(\mZ^2)\Big)\equiv
\Delta\aem_L^{(3)}+\Delta\aem_H^{(5)}\,.
\label{Dadef}
\eeq
In the rightmost side of eq.~(\ref{Dadef}) we have separated the leptons
and the light-quark contributions, thus:
\beqn
\Delta\aem_L^{(3)}&=&
\sum_{\ell}\Big(\Pi_\ell^{AA}(0)-\Pi_\ell^{AA}(\mZ^2)\Big)
\,\stackrel{{\rm pert}}{=}\sum_{\ell}P_f\left(m_\ell,\mZ\right)\,\,,
\label{DaemL3}
\\
\Delta\aem_H^{(5)}&=&
\sum_{q=light}\Big(\Pi_q^{AA}(0)-\Pi_q^{AA}(\mZ^2)\Big)
\,\stackrel{{\rm pert}}{=}\sum_{q=light}P_f\left(m_q,\mZ\right)\,,
\label{DaemH5}
\eeqn
with:
\beq
P_f\left(m,Q\right)=
\frac{\aem}{3\pi}\,\NC^{(f)}e_f^2\left(\log\frac{Q^2}{m^2}-\frac{5}{3}\right)
+\ord\left(\frac{m^2}{Q^2}\right).
\label{Pfres}
\eeq
As the notation of eqs.~(\ref{DaemL3}) and~(\ref{DaemH5}) suggests, all 
three leptons and all the quarks except the top are considered to be light, 
which is in keeping with the fact that we work at the $Z$ mass. The symbol
$\stackrel{{\rm pert}}{=}$ indicates that the result on its right is 
obtained by means of a perturbative computation of the relevant
two-point functions; at one loop, this gives eq.~(\ref{Pfres}) -- the
power-suppressed terms on the r.h.s.~are known exactly, and are omitted
here only for the sake of brevity. From eqs.~(\ref{Zaem0}), (\ref{ZamZ}),
(\ref{DaemL3}), and~(\ref{DaemH5}) we obtain:
\beq
Z_{\aem(\mZ)}=\sum_{f=heavy}\Pi_f^{AA}(0)+\sum_{f=light}\Pi_f^{AA}(\mZ^2)+
\Pi_W^{AA}(0)-2\frac{\sw}{\cw}\frac{\Sigma_T^{AZ}(0)}{\mZ^2}\,.
\label{ZamZexp}
\eeq
With an explicit computation, this leads to the following expression:
\beqn
Z_{\aem(\mZ)}\!\!&=&\!\!
\frac{\aem}{\pi}\,\Bigg\{
\frac{1}{3}\sum_{f=light}\NC^{(f)}e_f^2
\left(\ooepb+\frac{5}{3}-\log\frac{\mZ^2}{\mu^2}\right)
\label{ZamZdef}
\\*&&\phantom{\frac{\aem}{\pi}}
+\frac{1}{3}\sum_{f=heavy}\NC^{(f)}e_f^2
\left(\ooepb-\log\frac{m_f^2}{\mu^2}\right)
-\frac{7}{4}\left(\ooepb-\log\frac{\mW^2}{\mu^2}\right)-\frac{1}{6}
\Bigg\},
\nonumber
\eeqn
which we shall use by setting $\mu=\mZ$. While eqs.~(\ref{ZamZexp})
and~(\ref{ZamZdef}) are fully general, we point out that in the SM
the ``heavy'' tag applies only to the top quark.

In order to compute the numerical value of $\aemmZ$, we start by
observing that from eq.~(\ref{ZamZ}) we have:
\beq
\aemmZ=\frac{\aem(0)}{1-\Delta\aem}\,.
\label{amZfa0}
\eeq
The r.h.s.~of this equation can be evaluated by using the Thomson
value of eq.~(\ref{Thomson}) and the lepton and quark contributions
of eqs.~(\ref{DaemL3}) and~(\ref{DaemH5}), respectively, obtained from
the perturbative result of eq.~(\ref{Pfres}).
While this procedure is perfectly fine in the case of the leptons,
for the quarks it introduces a significant source of uncertainty
that stems from the fact that light-quark masses are unphysical 
parameters. As is well known, this problem can be avoided by using
a value of $\Delta\aem_H^{(5)}$ that is obtained from a dispersion
relation fitted to $\epem\to hadrons$ data; from the PDG~\cite{Zyla:2020zbs}: 
\beq
\Delta\aem_H^{(5)}=0.02766\pm 0.00007\,.
\label{DaemH5disp}
\eeq
We observe that the perturbative result for $\Delta\aem_H^{(5)}$ stemming
from eq.~(\ref{Pfres}) is equal to about $0.036$ if the masses of 
eq.~(\ref{fermmass}) (i.e.~the PDG's central values)
are employed. Equation~(\ref{Pfres}) would lead to 
the phenomenologically-sensible result of eq.~(\ref{DaemH5disp}) only if 
the masses of the quarks, and in particular those of the three lightest
ones, were assigned much larger values (and generally way outside of the 
respective uncertainty ranges) w.r.t.~to the central ones reported on 
the PDG; we shall briefly return to this point in sect.~\ref{sec:res0}. 
By using eq.~(\ref{DaemH5disp}) we finally obtain:
\beq
\aemmZ=1/128.940\,.
\label{amZval}
\eeq
Moreover, eqs.~(\ref{ZaMSbdef}) and~(\ref{ZamZdef}) can be exploited
with the definition of eq.~(\ref{aGmuvsaMSb}) to arrive at the following
result:
\beq
\Delta_{\MSb\to\aem(\mZ)}=\frac{5}{9\pi}\,C^{(2,M)}
+\frac{1}{\pi}\left(\frac{7}{4}\log\frac{\mW^2}{\mZ^2}-\frac{1}{6}\right).
\label{DaMZ}
\eeq
With this, we can evaluate the $\MSb$ coupling constant at $\mu=\mZ$.
The analogue of eq.~(\ref{amZfa0}) reads:
\beq
\aem(\mZ)=
\frac{\aem(0)}{1-\Delta\aem-\Delta_{\MSb\to\aem(\mZ)}}\,,
\label{aMSbfa03}
\eeq
whence\footnote{\label{ft:aMSbmZ}If the $W$ contribution is neglected in 
eq.~(\ref{DaMZ}), one obtains a value of $\aem(\mZ)$ which is $0.15$\% 
larger than that in eq.~(\ref{aMSbval}).}:
\beq
\aem(\mZ)=1/127.955\,.
\label{aMSbval}
\eeq
We have explicitly verified that, by defining the $\MSb$ scheme without
the decoupling of the top quark, we obtain a value for $\aem(\mZ)$ which
is in excellent agreement with that of ref.~\cite{Degrassi:2003rw} (whose
definition is identical to that stemming from eq.~(\ref{ZaMSbdef}) except
from the role of the top) -- the very small residual differences are due
to our neglecting two-loop effects.

In the $G_\mu$ scheme, we define the renormalisation constant for $\aem$ 
thus:
\beq
Z_{\aem_{G_\mu}}=Z_{\aem(0)}-\Delta r\,,
\label{ZaGmu}
\eeq
with (see e.g.~ref.~\cite{Denner:1991kt}):
\beqn
\Delta r&=&\Pi^{AA}(0)+\Delta r^{(1)}\,,
\label{Drdef}
\\
\Delta r^{(1)}&=&-\frac{\cw^2}{\sw^2}\left(
\frac{\Sigma_T^{ZZ}(\mZ^2)}{\mZ^2}-\frac{\Sigma_T^{W}(\mW^2)}{\mW^2}
\right)-\frac{\Sigma_T^{W}(0)-\Sigma_T^{W}(\mW^2)}{\mW^2}
\nonumber\\*&&
+2\frac{\cw}{\sw}\frac{\Sigma_T^{AZ}(0)}{\mZ^2}+
\frac{\aem}{4\pi\sw^2}\left(6-\frac{7-4\sw^2}{2\sw^2}\log\cw^2\right),
\label{Dr1def}
\eeqn
where we understand that only the real parts of the two-point functions
have to be taken into account. From eq.~(\ref{Zaem0}) we then obtain:
\beq
Z_{\aem_{G_\mu}}=-2\frac{\sw}{\cw}\frac{\Sigma_T^{AZ}(0)}{\mZ^2}
+\Delta r^{(1)}\,.
\label{ZaGmu2}
\eeq
The numerical value of $\aemGmu$ is determined by means of the
following relationship:
\beq
\aemGmu=\frac{\sqrt{2}G_\mu\mW^2\left(\mZ^2-\mW^2\right)}
{\pi\mZ^2}\equiv 1/132.183\,,
\label{a550}
\eeq
from which, by employing eq.~(\ref{aGmuvsaMSb}) and by making use
of eq.~(\ref{aMSbval}), we obtain:
\beq
\Delta_{\MSb\to G_\mu}=4.092936\,.
\label{DMSbaGua550num}
\eeq
We remind the reader that $\Delta r$ is known to receive large contributions
beyond one loop. We take this effectively into account by using 
eqs.~(\ref{a550}) and~(\ref{DMSbaGua550num}), rather than their one-loop
expressions stemming from eqs.~(\ref{ZaGmu})--(\ref{Dr1def}). This is
conceptually equivalent to using eq.~(\ref{DaemH5disp}) in the
$\aem(\mZ)$ scheme rather than its perturbatively-computed counterpart.

We finally note that the definitions given above are fully consistent
with those employed~\cite{Frederix:2018nkq} in the $\aem(\mZ)$ and 
$G_\mu$ UFO models in \aNLOs.

\section{Numerical solution\label{sec:numsolevol}}
The numerical approach adopted here is conceptually identical to the one 
presented in ref.~\cite{Bertone:2019hks}: the evolution of the PDFs is 
performed by solving the evolution equations in Mellin space, in terms of 
the evolution operator introduced in sect.~\ref{sec:ansolevol}. The latter, 
applied to the PDFs evaluated at $\mu_0$ (i.e.~to the PDF initial conditions) 
returns the PDFs at the desired final scale $\mu$.

From the practical viewpoint, however, several changes are required
w.r.t.~the implementation of ref.~\cite{Bertone:2019hks}, owing to the
presence of multiple fermion families, as well as to the possibility to
adopt different factorisation and renormalisation schemes; more details
on this are given in sect.~\ref{sec:evnum}.

Moreover, the integrable divergence of the electron PDF at $z=1$ must
be handled with care in the convolution with short-distance cross sections
of eq.~(\ref{master0}), lest the numerical accuracy be degraded. In order 
to do so, in the region $z\simeq 1$ one switches from the numerical solution 
to the analytical one found in sect.~\ref{sec:ansolevol}; such a switching 
is detailed in sect.~\ref{sec:switching}.

The code that implements the numerical solution of the evolution equations 
and the switching to the analytical solution is described in 
sect.~\ref{sec:emela}.

\subsection{Partonic content\label{sec:basis}}
In order to perform the PDF evolution with multiple fermion families
according to the discussion at the beginning of sect.~\ref{sec:evol}
we have employed the so-called variable-flavour-number scheme 
(VFNS)~\cite{Collins:1978wz}, in which the PDF of each fermion, 
be it a lepton or a quark, is generated radiatively starting from 
the corresponding threshold, that we assume to be equal to the mass of the 
fermion itself. In this scheme, the entire evolution range is subdivided 
into sub-ranges (see eq.~(\ref{krange})) characterised by a specific number 
of leptons, \mbox{$n_l\le\Nl$}, and quarks, \mbox{$n_f\le \Nu+\Nd$}. 
In each such sub-range, the evolution operator is evaluated as is explained 
in ref.~\cite{Bertone:2019hks} including only $n_l$ leptons and $n_f$
quarks. Finally, all of the evolution operators thus obtained are consistently 
matched at thresholds\footnote{Up to NLL accuracy and by setting the fermion 
thresholds equal to the respective masses, the matching implies that 
radiatively-generated fermions have a vanishing PDF at the respective 
thresholds. We point out that this is generally not true, and that at orders
higher than NLL, or by choosing thresholds different from mass values,
radiatively-generated PDFs at threshold are different from zero but
perturbatively computable.}. We have chosen the fermion masses equal
to the corresponding central values quoted in the PDG~\cite{Zyla:2020zbs},
namely: 
\begin{equation}
\begin{array}{l}
m_e = 0.5109989461\cdot 10^{-3} \mbox{ GeV}\,,\\ 
m_u = 2.16\cdot 10^{-3} \mbox{ GeV}\,,\\
m_d = 4.67\cdot 10^{-3} \mbox{ GeV}\,,\\ 
m_s = 0.093 \mbox{ GeV}\,,\\
m_\mu = 0.1056583745 \mbox{ GeV}\,,\\
m_c = 1.27\mbox{ GeV}\,,\\
m_\tau = 1.77686 \mbox{ GeV}\,,\\
m_b = 4.18 \mbox{ GeV}\,,
\end{array}
\label{fermmass}
\end{equation}
which are strictly ordered as was assumed in eq.~(\ref{thrslist}). We note
that, in the case of quark masses, other criteria can be adopted to set
their values -- one explicit example is given in sect.~\ref{sec:Rsch}
in the context of the definition of the $\aem(\mZ)$ UV-renormalisation 
scheme; we shall also briefly return to this point in sect.~\ref{sec:res0}.
Finally, in this paper we set the initial scale for the evolution equal
to the electron mass, $\mu_0=m_e$.

As was already anticipated, we perform the numerical evolution using
a slightly different functional basis w.r.t.~that introduced in
sect.~\ref{sec:strevol}: while eqs.~(\ref{defl2})--(\ref{defSigd}) 
are employed as such, the eight non-singlet functions of eq.~(\ref{defNS}) 
are re-combined as follows~\cite{Bertone:2015lqa}:
\beqn
\ePDF{V_l}&=&\sum_{l=1}^{\Nl}\left(\ePDF{l}-\ePDF{\bar{l}}\right),
\label{defVl}
\\
\ePDF{V_u}&=&\sum_{u=1}^{\Nu}\left(\ePDF{u}-\ePDF{\bar{u}}\right),
\label{defVu}
\\
\ePDF{V_d}&=&\sum_{d=1}^{\Nd}\left(\ePDF{d}-\ePDF{\bar{d}}\right),
\label{defVd}
\\
\ePDF{V_1^l}&=&\ePDF{\lm}-\ePDF{\lp}-\left(\ePDF{\mu^-}-\ePDF{\mu^+}\right),
\label{defV1l}
\\
\ePDF{V_2^l}&=&\ePDF{\lm}-\ePDF{\lp}+\ePDF{\mu^-}-\ePDF{\mu^+}
-2\left(\ePDF{\tau^-}-\ePDF{\tau^+}\right)\,,
\label{defV2l}
\\
\ePDF{V_1^u}&=&\ePDF{u}-\ePDF{\bar{u}}-\left(\ePDF{c}-\ePDF{\bar{c}}\right),
\label{defV1u}
\\
\ePDF{V_1^d}&=&\ePDF{d}-\ePDF{\bar{d}}-\left(\ePDF{s}-\ePDF{\bar{s}}\right),
\label{defV1d}
\\
\ePDF{V_2^d}&=&\ePDF{d}-\ePDF{\bar{d}}+\ePDF{s}-\ePDF{\bar{s}}
-2\left(\ePDF{b}-\ePDF{\bar{b}}\,,\right)\,.
\label{defV2d}
\eeqn
As was done in eqs.~(\ref{defNS})--(\ref{defSigd}), these are given
for a maximal number of light fermions, and the same comment as in
footnote~\ref{ft:Nflav} applies here. The reason for using the basis
of eqs.~(\ref{defVl})--(\ref{defV2d}) rather than that of eq.~(\ref{defNS})
is a practical one: our numerical code is built upon an existing one
relevant to QCD~\cite{Bertone:2015cwa}, that can handle NNLL evolution 
as well: at that order eqs.~(\ref{defVl})--(\ref{defV2d}) are more 
convenient than eq.~(\ref{defNS})\footnote{This is because while 
eqs.~(\ref{defV1l})--(\ref{defV2d}) evolve with $P_f^-$, 
eqs.~(\ref{defVl})--(\ref{defVd}) evolve with 
\mbox{$P_f^-+n_f(P^{\rm S}_{ff}-P^{\rm S}_{f\bar{f}})$}, and these two
kernels coincide only up to $\ord(\aem^2)$.}.
Thus, while at the LL and NLL the two bases are equally sensible,
the choice made in this section allows us to minimise the changes
to the original computer code.

Needless to say, in the numerical code the exact forms of the AP kernels
are used, as opposed to the $N\to\infty$ ones employed for the analytical
solutions in the $z\to 1$ region. This implies that, in the singlet sector,
we solve eq.~(\ref{APQEDfull}) rather than eq.~(\ref{APQEDfullasy}).
Conversely, as in the case of the analytical solutions, when the evolution 
is carried out at the LL only the first term of the perturbative expansion 
of each AP kernel, eq.~(\ref{APmatex}), is retained, while the flavour 
structure is exactly the same as for the NLO+NLL solutions. Therefore,
contrary to what is typically done in the literature, also at the LO+LL
we have non-vanishing photon, positron, and quark PDFs.

\subsection{Evolution in Mellin space\label{sec:evnum}}
The differential equation we solve has the following structure:
\beq
\frac{\partial \Eop_{N}^{(K)}(\mu)}{\partial\log\mu^2} =
\Mmat_{N}^{(K)}(\aem(\mu),\mu)\,\Eop_{N}^{(K)}(\mu)\,,\quad\quad
\Eop_{N}^{(K)}(\mu_I) = \Eop_{I,N}^{(K)}\,,
\label{eq:eveq}
\eeq
where $\Mmat_{N}^{(K)}$ and $\Eop_{N}^{(K)}$ are either $4\times 4$-dimensional
matrices in the singlet sector (eq.~(\ref{APQEDfull})) or scalar functions
(all of the other cases).
Equation~(\ref{eq:eveq}) is understood to be relevant in the range of
eq.~(\ref{krange2}), and $\mu_I$ denotes the lower end of that range 
(i.e.~$\mu_0$ or $m_k$ when non-empty), where the evolution operator assumes 
the value $\Eop_{I,N}^{(K)}$. We solve eq.~(\ref{eq:eveq}) recursively, 
starting from the lowest $k$ value for which the range of eq.~(\ref{krange2}) 
is non-empty, and $\Eop_{I,N}^{(K)}=I$: in the case we are considering 
in this paper (the electron PDFs with all thresholds set equal to the 
respective masses), this implies $k=1$ and $\mu_0=m_1\equiv m$. Then,
the value assumed by the evolution operator at the upper end of the
range, $\Eop_{N}^{(K)}(m_{k+1})$, coincides, by continuity, with the
starting value $\Eop_{I,N}^{(K)}$ relevant to the next range
\mbox{$\mu\in(m_{k+1},m_{k+2})$}. The procedure is thus iterated,
which corresponds to the successive determination of the integration
constants $A_k$ that appear in eq.~(\ref{Esol1p3tk}) in the explicit 
example given for the non-singlet evolution of sect.~\ref{sec:ansolevol}.

The most trivial of cases is that where $\Mmat_{N}^{(K)}$ does not depend 
on $\mu$, neither directly nor through $\aem(\mu)$. This happens in the
$\MSb$ UV-renormalisation scheme when the running of $\aem$ is neglected.
We hasten to stress that this is an unphysical situation (since the 
coupling constant {\em does} run in $\MSb$), that we use solely for
testing purposes -- we shall comment briefly on this in 
sect.~\ref{sect:resfacren}. The solution is given by the following expression:
\beq
\Eop_{N}^{(K)}(\mu) = 
\exp\left(\Mmat_{N}^{(K)}\log\frac{\mu^2}{\mu_I^2}\right) 
\Eop_{I,N}^{(K)}\,.
\label{eq:nosdep}
\eeq
In the singlet sector ($4\times 4$ matrices) the r.h.s.~of 
eq.~(\ref{eq:nosdep}) cannot be given in a closed form, and we use a 
series expansion instead, by keeping as many terms as are deemed 
necessary for targeting a numerical accuracy of a relative
$10^{-8}$ precision.

When $\Mmat_{N}^{(K)}$ depends on $\mu$ only through $\aem(\mu)$
(in the cases we consider, this corresponds to the $\MSb$ UV-renormalisation 
scheme), we first trade the evolution variable $\mu$ for $t_k$, in keeping
with sect.~\ref{sec:ansolevol}, and then we adopt a discretised 
path-ordered product~\cite{Bonvini:2012sh}, as has been done in
ref.~\cite{Bertone:2019hks}. The range of eq.~(\ref{kranget}) is split
into ``small'' $n_k$ sub-intervals, \mbox{$(t_{0,k},t_{1,k})$}, $\dots$
\mbox{$(t_{n_k-1,k},t_{n_k,k})$}, which for simplicity are evenly 
spaced:
\beqn
&&t_{i,k}=t_{0,k} + i \Delta_k t\,,\;\;\;\;\;\;\;\;
i=0,\dots,n_k\,,\;\;\;\;\;\;\;\;
\Delta_k t = \frac{t_{n_k,k} -t_{0,k}}{n_k}\,,
\label{eq:steps}
\\*
&&t_{0,k}=0\,,
\label{t0kdef}
\\*
&&t_{n_k,k}=\bt_k\,.
\label{tnkdef}
\eeqn
Then, if $\Delta_k t\ll 1$, the following discretised expression for
the evolution operator works fairly well:
\beq
\Eop_{N}^{(K)}(\mu) \simeq \left[ \prod_{i=0}^{n_k-1}
\exp\left(\frac{\Delta_k t}{2}
\left(\Mmat_{N}^{(K)}\left(t_{i,k}\right)+
\Mmat_{N}^{(K)}\left(t_{i+1,k}\right)\right)\right)
\right] \Eop_{I,N}^{(K)}\,,
\label{eq:pathord}
\eeq
and we adopt it in our computations. We point out that, more in general,
the $\Delta_k t\ll 1$ condition allows the usage, as the argument of the 
exponential in eq.~(\ref{eq:pathord}), of any function of $\Mmat_{N}^{(K)}$
that depends only on the endpoints $t_{i,k}$ and $t_{i+1,k}$; the
differences induced by different choices for this function vanish
for $\Delta_k t\to 0$.

It remains to determine how to set the $n_k$ parameters so that the
sub-intervals used here are small enough to guarantee that the discretised
approach of eq.~(\ref{eq:pathord}) does not degrade the numerical accuracy 
of the solution. In order to do that, we notice that the larger the number of
fermions in the evolution, the slower the convergence of the path-ordered 
product. This ultimately happens because the evolution is predominantly
driven by the running of $\aem$, which in turn is controlled by the
$\beta$-function coefficients, whose values scale linearly with the number 
of fermions (see eq.~(\ref{b0b1k})). As a consequence of that, the total 
number of sub-intervals $\sum_k n_k$ required to achieve comparably-accurate 
results in different ranges is expected to scale roughly with the number of 
fermions, \mbox{$M=\Nl+\Nu+\Nd$}. In view of the fact that it has been 
heuristically established in ref.~\cite{Bertone:2019hks} that in the case 
of a single-fermion evolution (i.e.~for $\Nl=1$ and $\Nu+\Nd=0$), and for the 
largest scales of the order of the TeV, an appropriate total number of 
sub-intervals is equal to 20, by making a conservative choice we set
\beq
\sum_{k=1}^M n_k\ge n\equiv 300\,.
\label{n300}
\eeq
While the individual values of $n_k$ could also be assigned by taking
into account the scaling with the number of active fermions, this would
ignore the fact that mass thresholds are not evenly spaced. We can take
both effects into account with the following settings:
\beq
n_k=n\,\,\frac{v(\min(\mu,m_{k+1}))-v(\bm_k)}{v(\mu)-v(\mu_0)}\,,
\label{nkdef0}
\eeq
with $v(x)=\aem(x)$. As a final refinement, since eq.~(\ref{nkdef0}) might
lead to the undersampling of the ranges defined by two thresholds particularly
close to each other, we actually use:
\beq
n_k\;\longrightarrow\;
\stepf(n_k)\max\left(n_{\min},n_k\right)\,,
\;\;\;\;\;\;\;\;
n_{\min}=50\,. 
\label{nkdef}
\eeq
Equations~(\ref{n300})--(\ref{nkdef}) have been validated by varying 
the numerical values that appear therein. We have found that the PDFs are 
unchanged within the target relative accuracy of $10^{-8}$ for 
\mbox{$200\le n\le 1000$} and \mbox{$20\le n_{\min}\le 300$}.

We finally must consider the case where $\Mmat_{N}^{(K)}$ has a 
direct dependence on $\mu$, but $\aem$ does not run -- this happens
in the $\aem(\mZ)$ and $G_\mu$ schemes. In the non-singlet case, we 
exploit a closed-form analytical solution, the fixed-$\aem$ counterpart 
of eq.~(\ref{Esol1p3tk}) that can be easily obtained by solving
eq.~(\ref{matAPmell3thrs}). In the singlet case, we adopt two different 
strategies, which are used to validate each other. The first strategy is 
a path-ordered product analogous to that in eq.~(\ref{eq:pathord}), but 
with evolution variable 
\beq
L(\mu) = \frac{\aem}{2\pi} \log\frac{\mu^2}{\mu^2_I}
\eeq
instead of $t_k$, so that $v(x)=L(x)$ in eq.~(\ref{nkdef0}), with the 
analogues of the $t_{i,k}$ parameters of eq.~(\ref{eq:steps}) defined in
the space spanned by the values of $L$.
The second strategy is based on a Magnus expansion~\cite{Magnus:1954zz}:
\beq
\Eop_{N}^{(K)}(\mu) = 
\exp\left(\sum_{i=1}^{\infty} \Omega_i(\mu,\mu_I)\right)
\Eop_{I,N}^{(K)}\,,
\label{MagExp}
\eeq
where for instance the first two terms are given by:
\beqn
\Omega_1(\mu,\mu_I)&=&\int_{\log\mu_I^2}^{\log\mu^2} d\log\kappa^2\,
\Mmat_{N}^{(K)}(\kappa)\,,
\\
\Omega_2(\mu,\mu_I)&=&\half\int_{\log\mu_I^2}^{\log\mu^2} d\log\kappa_1^2\,
\int_{\log\mu_I^2}^{\log\kappa^2_1} d\log\kappa_2^2\,
\left[\Mmat_{N}^{(K)}(\kappa_1),\Mmat_{N}^{(K)}(\kappa_2)\right]\,.
\eeqn
We have limited ourselves to considering the first four terms in the 
series at the exponent in eq.~(\ref{MagExp}). Since the running time of the
Magnus-expansion-based implementation is shorter than that based on
the path-ordered approach, we have adopted the former as our default
strategy for the $\aem(\mZ)$ and $G_\mu$ schemes, after having verified
that the two strategies give the same PDFs within the target relative 
accuracy of $10^{-8}$ (the parameters $n$ and $n_{\min}$ of eqs.~(\ref{n300}) 
and~(\ref{nkdef}) have been used in the path-ordered case).

Finally, in order to invert the PDFs from the Mellin to the $z$ space
we employ the same strategy already used in ref.~\cite{Bertone:2019hks},
namely a numerical algorithm based on the so-called Talbot
path~\cite{DelDebbio:2007ee}, with a trapezoidal integration.

\subsection{Switching to the analytical solution\label{sec:switching}}
The numerical approach described so far cannot possibly work in
the $z\to 1$ limit, owing to the fact that the electron PDF has a very 
steep, power-like integrable singularity.
In order to address this issue we make use of the analytical solution
of sect.~\ref{sec:ansolevol} to define a PDF which is accurate and 
well-behaved in the whole $z$ range. Denoting by $\Gamma_{\rm num}$ and 
$\Gamma_{\rm an}$ the numerical solution and its associated analytical one,
respectively, we introduce a switching point $z_0$, and define the complete 
solution as follows (generalising what was done in 
ref.~\cite{Frixione:2022ofv}):
\beq
\Gamma(z) =
\begin{cases}
 \Gamma_{\rm num}(z)\,, &\phantom{aaa} z\le z_0\,, \\
 \Gamma_{\rm an}(z)\left(
s(z;z_0)\dfrac{\Gamma_{\rm num}(z_0)}{\Gamma_{\rm an}(z_0)}+
1-s(z;z_0)\right), &\phantom{aaa} z > z_0\,, \\
\end{cases}
\label{eq:msol}
\eeq
with:
\beq
s(z;z_0)=\left(\frac{1-z}{1-z_0}\right)^k\,,\;\;\;\;\;\;\;\;
k\ge 1\,.
\label{sfdef}
\eeq
The factor in round brackets on the r.h.s.~of eq.~(\ref{eq:msol}) is
chosen so that $\Gamma(z)$ is continuous at $z=z_0$ and coincides with
the analytical solution at $z=1$. For all practical purposes, this is
essentially academic. In fact, the switching point $z_0$ is chosen in such
a way that $\Gamma_{\rm num}(z_0)$ and $\Gamma_{\rm an}(z_0)$ are virtually
identical\footnote{Which constitutes a powerful consistency check of 
the numerical and analytical solutions.}, and the numerical solution 
$\Gamma_{\rm num}(z_0)$ has not yet lost accuracy. Typical values for $z_0$ 
are between $1-10^{-6}$ and  $1-10^{-7}$; for these, we observe that
\mbox{$\Gamma_{\rm num}(z_0)/\Gamma_{\rm an}(z_0)\simeq z_0$}. 
This implies that in practice the choice of $k$ in eq.~(\ref{sfdef})
has a negligible impact\footnote{More precisely, the relative differences 
between cross sections obtained with $k=0$, $k=1$, and $k=2$ are at the 
level of $10^{-6}$, i.e.~within our MC integration errors.}; in 
phenomenological applications, we employ $k=2$.

We remind the reader that in the $\MSb$ renormalisation and factorisation 
schemes an $\ord(\aem^3)$-accurate analytical solution exists also for 
$z<1$~\cite{Bertone:2019hks} that, matched with the $z\to 1$ one, offers 
an alternative to the form defined in eq.~(\ref{eq:msol}). However, such a 
solution has been derived only for single-lepton evolution, and because 
of that is not employed here.

\subsection{Code \eMELA\label{sec:emela}}
The strategy described in this section has been implemented in a code 
that we call \eMELA, which we release together with the current paper.
Such a code supersedes the one developed in ref.~\cite{Bertone:2019hks}
(\ePDFc{}), that was limited to the evolution with a single lepton in
the $\MSb$ renormalisation and factorisation schemes.

The new code is available at the link:
\begin{center}
  {\centering \texttt{https://github.com/gstagnit/eMELA}}\,,
\end{center}
where documentation and examples about its usage can also be found.
\eMELA{} is a standalone code, and can be linked to any external program.

More in detail, \eMELA{} is an improved QED-version of 
\MELA~\cite{Bertone:2015cwa}. It consists of a Fortran code responsible 
for the numerical evolution of the PDFs, and a C++ wrapper that provides 
one with the analytical solutions and the switching described in 
sect.~\ref{sec:switching}. Moreover, since a runtime evaluation of 
the numerical solution is likely too slow for phenomenological applications,
the possibility is given to the user to output the PDFs as grids compliant
with the \LHAPDF~\cite{Buckley:2014ana} format, that can be employed at
a later stage. 

We stress that, regardless of whether the numerical solution is computed 
at runtime or read from the grids, \eMELA\ always switches to the analytical 
solution at $z=z_0$. Furthermore, in order for the output to be handled
within machine-precision double arithmetic for values of $z$ arbitrarily
close to one, the code returns the PDFs multiplied by a user-defined
damping factor that vanishes at $z\to 1$, following the procedure
introduced in sect.~3 of ref.~\cite{Frixione:2021zdp}.

The interested reader can find all of the necessary technical information
by visiting the link given above. There, one can also find pre-computed 
grids with different choices of renormalisation scheme, factorisation scheme, 
and so forth. Such grids are validated, and have been used for the cross 
section calculations of sect.~\ref{sec:res}.

\section{Results\label{sec:res}}
In this section we employ the PDFs we have derived as was explained
in sects.~\ref{sec:evol}--\ref{sec:numsolevol} to compute cross sections
according to eqs.~(\ref{beamstr}) and~(\ref{master0}). We do so in
the automated \aNLOs\ framework, thereby extending the simulation
of ISR effects in $\epem$ collisions of ref.~\cite{Frixione:2021zdp}
from LO+LL to NLO+NLL accuracy. As was already mentioned in
sect.~\ref{sec:intro}, this means that \aNLOs\ can now also compute
NLO EW corrections for processes with massless initial-state leptons;
this has required some changes in the implementation of the FKS subtractions 
relevant to this case -- more details can be found in
appendix~\ref{sec:NLOxsec}. We study the processes:
\beqn
\epem&\longrightarrow&q\bar{q}(\gamma)\,,
\label{proctoy}
\\
\epem&\longrightarrow&W^+W^-(X)\,,
\label{procWW}
\\
\epem&\longrightarrow&t\bt(X)\,,
\label{proctt}
\eeqn
at the NLO accuracy (i.e.~at $\ord(\aem^3)$), where the symbols in round 
brackets denote any particle that may be present in the final state beyond 
the LO. In eq.~(\ref{proctoy}) $q$ is a massless fermion of charge $e_q$, 
and in the corresponding short-distance cross sections we retain only
the contributions proportional to $e_q^2$ -- this is therefore the
process already used in ref.~\cite{Frixione:2019lga} (the constraint
on the quark charge being specified in eq.~(4.3) there; this limits 
the real and virtual radiation to the initial state, and thus
the process is effectively equivalent to that for the production of a
heavy neutral object of variable mass), whose simple analytical 
cross sections we have used as a cross-check of the corresponding 
automated computation carried out by \aNLOs. The results
for the processes of eqs.~(\ref{procWW}) and~(\ref{proctt}) include EW 
contributions to the short-distance cross sections. The latter case is
also computed as a pure-QED process, i.e.~by ignoring EW effects. 
The process of eq.~(\ref{proctoy}) is always dealt with in QED.

Beam-dynamics effects, parametrised by the functions ${\cal B}_{kl}(\yp,\ym)$ 
of eqs.~(\ref{beamstr}) and~(\ref{Bee}), are generally ignored; when included, 
we restrict ourselves to considering a single illustrative case (a $500$-GeV 
collider with an ILC-type configuration), where such effects are implemented 
as is detailed in ref.~\cite{Frixione:2021zdp}; we discuss it in
sect.~\ref{sect:beamstr}.

The aim of this section is to document the effects of the theoretical
novelties introduced here and in refs.~\cite{Frixione:2019lga,
Bertone:2019hks,Frixione:2021wzh} on actual observables; in order
to keep the number of plots at a manageable level, we present 
results for the cumulative cross section:
\beq
\sigma(\tau_{min})=\int d\sigma\,
\stepf\!\left(\tau_{min}\le\frac{M_{p\bar{p}}^2}{s}\right)\,,
\;\;\;\;\;\;\;\;
p=q\,,t\,,W^+\,,
\label{Ixsec}
\eeq
where $M_{p\bar{p}}^2$ is the invariant mass squared of the $p\bar{p}$ pair,
and $s$ the collider c.m.~energy squared.
We employ \aNLOs\ to compute this observable at fixed order, either
leading or next-to-leading; in other words, soft logarithms that
appear at $\tau_{min}\to 1$ are not resummed.
We stress that \aNLOs\ is capable of computing simultaneously any
number of observables, subject to arbitrary final-state cuts.
Our primary interest is the assessment of the impact of NLL contributions
to the PDFs, and of the factorisation- and renormalisation-scheme
dependencies, which we shall discuss in sects.~\ref{sect:resNLL}
and~\ref{sect:resfacren}, respectively. In order to do so in a manner 
conceptually analogous to what is typically done in the literature, in those
sections we shall limit ourselves to including only the $\epem$-initiated 
{\em partonic} channel results. The contributions of other partonic channels 
that enter eqs.~(\ref{procWW}) and~(\ref{proctt}), and in particular the
$\gamma\gamma$ one, will be discussed in sect.~\ref{sect:gammaproc}
(see also sect.~\ref{sec:res0}). We typically consider all of the six
possible combinations of factorisation ($\Delta$, $\MSb$) and
renormalisation ($\MSb$, $\aem(\mZ)$, $G_\mu$) schemes, except for
$q\bq$ and $t\bt$ production in QED, in which cases no results are 
given for the $G_\mu$ renormalisation scheme.

We set the hard scale as follows:
\beq
\mu=\sqrt{s}\,,
\eeq
and employ
\beqn
\mW&=&80.379~{\rm GeV}\,,
\\
\mZ&=&91.1876~{\rm GeV}\,,
\\
m_t&=&173.3~{\rm GeV}\,.
\eeqn
We present predictions obtained with a $\sqrt{s}=500$~GeV c.m.~energy, 
but we stress that we have considered (if above the respective 
pair-production thresholds) several other cases in the range
\mbox{$50~{\rm GeV}\le\sqrt{s}<500~{\rm GeV}$}, finding quantitatively
similar results. In the legends of the plots, we shall typically employ
the following naming conventions:
\beq
{\rm xsec}\,,\;{\rm PDF}\;\left[{\rm fact~sch}\,,{\rm ren~sch}\right]\,,
\label{plotnames}
\eeq
where ``xsec'' denotes the perturbative accuracy of the short-distance
cross sections, ``PDF'' the logarithmic accuracy of the PDFs, and
``fact~sch'' and ``ren~sch'' the factorisation and renormalisation schemes,
respectively, used in the latter. Thus:
\beqn
{\rm xsec}&\in&\{{\rm LO}\,,{\rm NLO}\}\,,
\\
{\rm PDF}&\in&\{{\rm LL}\,,{\rm NLL}\}\,,
\\
{\rm fact~sch}&\in&\{\Delta\,,\MSb\}\,,
\\
{\rm ren~sch}&\in&\{\MSb\,,\aem(\mZ)\,,G_\mu\}\,.
\eeqn
The factorisation-scheme tag is absent when the corresponding PDFs
are evaluated at the LL accuracy.

Before going into the details, in sect.~\ref{sec:res0} we
discuss some general features that inform the predictions shown
later, and that may be used in the future for further phenomenological
refinements.

\subsection{General considerations and the role of the $W$
\label{sec:res0}}
We start by reminding the reader that the requirement that the
evolution of $\aem$ be phenomenologically sensible is the main
motivation for employing all fermion families, and their respective
mass thresholds, as opposed to limiting oneself to consider only 
the electron (see sect.~\ref{sec:elem}). The very same requirement,
then, demands that the contribution of the $W$ to the running of $\aem$
be included as well, since its effect is numerically significant at
large scales. For example, the values of $\aem(\mu)$ at $\mu=500$~GeV
obtained by including or by ignoring the contribution of the $W$, and
computed according to the procedure outlined in sect.~\ref{sec:Rsch} with 
consistent initial conditions at $\mu=\mZ$ (see footnote~\ref{ft:aMSbmZ}),
differ from one another by $1.7$\% (at one loop); the evolution that includes 
the $W$ is slower w.r.t.~that which excludes it, owing to the negative sign
in front of the second term on the r.h.s.~of eq.~(\ref{b0b1k}).

In view of what was said above, the results presented here are obtained
by including the $W$ contribution to the running of $\aem$; we stress,
however, that if one were interested in treating such a contribution
as a systematics, \eMELA\ allows one to switch it on and off easily.
We note that from the technical viewpoint the inclusion of the $W$ does 
not pose any problems -- the $W$ mass can simply be regarded as 
an additional threshold, and treated in
the same manner as its fermionic counterparts as is explained in 
sects.~\ref{sec:evol} and~\ref{sec:numsolevol}. However, we also remark
that by doing so we introduce two sources of inconsistency in our
framework. Firstly, the $W$ contribution to the running of $\aem$ is
included only at the one-loop level, while pure-QED effects are included
up to two loops (see eq.~(\ref{b0b1k})). Secondly, by considering weak
contributions to the running of $\aem$ one would need to do the same
in the PDF evolution equations, that in turn would entail the necessity
of considering branching processes that involve weak bosons, as well
as treating weak bosons as partons. We believe that in practice we can 
safely neglect both of these items. As far as the first one is concerned,
heuristic evidence is given by the fact that the changes in the PDFs
stemming from very large changes to the $b_1^{(M)}$ coefficient w.r.t.~to
its value of eq.~(\ref{b0b1k}) -- i.e.~setting it equal to zero or doubling
it -- are very small in absolute value, and completely negligible w.r.t.~other 
theoretical systematics considered here. 
For what concerns the second item, it should be clear that the
large masses of the weak bosons prevent them from giving noticeable
contributions even at energies of the order of 1~TeV -- the case of
{\em hadronic} PDFs, where analogous kinematic considerations apply,
has been explicitly considered, see e.g.~ref.~\cite{Bauer:2017isx,
Fornal:2018znf,Bauer:2018arx}. Having said that, we stress that the language 
adopted here gives one a blueprint for the inclusion of the weak bosons in 
PDF evolution that does not necessitate any conceptual changes, should
the need for doing so arise in the distant future.

A potential source of systematics that is ignored as such in this paper 
is the dependence on the starting scale for the PDF evolution, $\muz$, 
that we always set equal to the electron mass. Still, we point out that 
the NLO initial conditions do depend, explicitly and implicitly through the 
coupling constant, on $\muz$ (see eqs.~(\ref{PDFex})--(\ref{G1sol2})).
In particular, we remark that the presence of \mbox{$\log\muz$} terms
in the NLO initial conditions is such that it compensates an analogous
dependence stemming from the evolution, so that at the NLO any physical
cross section is $\muz$-independent. This property does {\em not} hold
true if an NLO short-distance cross section is convoluted with LO+LL
PDFs, since the latter have $\muz$-independent initial conditions, but
$\muz$-dependent evolution. There are two solutions for this issue:
either one simply understands $\muz=m$ in an LL evolution, as is typically
done in the literature; or one compensates for the $\muz$ dependence by 
including a contribution to the NLO short-distance cross section that is 
engineered to cancel it -- this is explained in more detail in 
appendix~\ref{sec:NLOxLL}. Irrespective of which of these two solutions
one adopts, it is true that a $\log\muz$ dependence remains at the NNLO and 
beyond; this is the ultimate reason that informs the setting 
$\muz=m$\footnote{We remark that analogous considerations hold
for mass thresholds.}.

Whether the choice of the light-quark masses should be considered
a systematics is debatable; these are not physical parameters, and
can also be regarded as quantities to be fitted in order to optimise
predictions for a given set of observables. For the phenomenological
results of this paper we set them equal to the central values reported 
by the PDG, as is shown in eq.~(\ref{fermmass}); here, we limit ourselves
to a few comments on the role they play in the PDFs. Firstly, we stress that 
the in coupling-constant evolution that we carry out as is explained in
sect.~\ref{sec:evol} the presence of multiple fermion families and 
their respective mass thresholds set according to eq.~(\ref{fermmass})
is such that \mbox{$\aem(m)=1/138.290$}. This value differs by only $0.9$\%
from the Thomson constant of eq.~(\ref{Thomson}). While there is no
reason why $\aem(m)$ and the Thomson value should coincide, it appears
to be phenomenologically sensible that they are close to each
other\footnote{For comparison, the evolution with all fermions kept 
active down to the electron mass leads to \mbox{$\aem(m)=1/144.983$}.}.
Secondly, it has been observed in sect.~\ref{sec:Rsch} that a value
of $\Delta\aem_H^{(5)}$ close to that of eq.~(\ref{DaemH5disp}) can
be obtained by means of a perturbative calculation by employing relatively
large values of the light-quark masses. For example, by setting the
masses of the three lightest quarks equal to the muon mass, we obtain
\mbox{$\aem(m)=1/136.956$} (and \mbox{$\Delta\aem_H^{(5)}=0.0266$}),
which differs from the Thomson value by a mere $0.06$\%. Therefore, one
can argue that such a choice is better motivated from a phenomenological
viewpoint than that of eq.~(\ref{fermmass}). While this remains to be
seen, and will not be investigated any further in this paper, we observe
the following: the $\ord(1\%)$ difference in the values of $\aem(m)$ 
obtained with eq.~(\ref{fermmass}) and with the choice discussed above
does {\em not} lead to a difference of $\ord(1\%)$ between the PDFs evolved
with these two choices of quark masses -- e.g.~for the electron PDFs at the 
NLL, and for scales of $\ord(10~{\rm GeV})$ or larger, the difference is in 
fact below the permille level\footnote{And totally negligible for $\aem(\mu)$
itself, which for large scales is primarily constrained by eq.~(\ref{aMSbval})
and by the fact that there all fermions are active in the evolution.}. 
In other words, the $\ord(1\%)$ difference in the initial condition is 
compensated by the different ``speeds'' in the evolution in the relevant 
mass ranges.

\begin{figure}[t!]
  \begin{center}
  \includegraphics[width=0.47\textwidth]{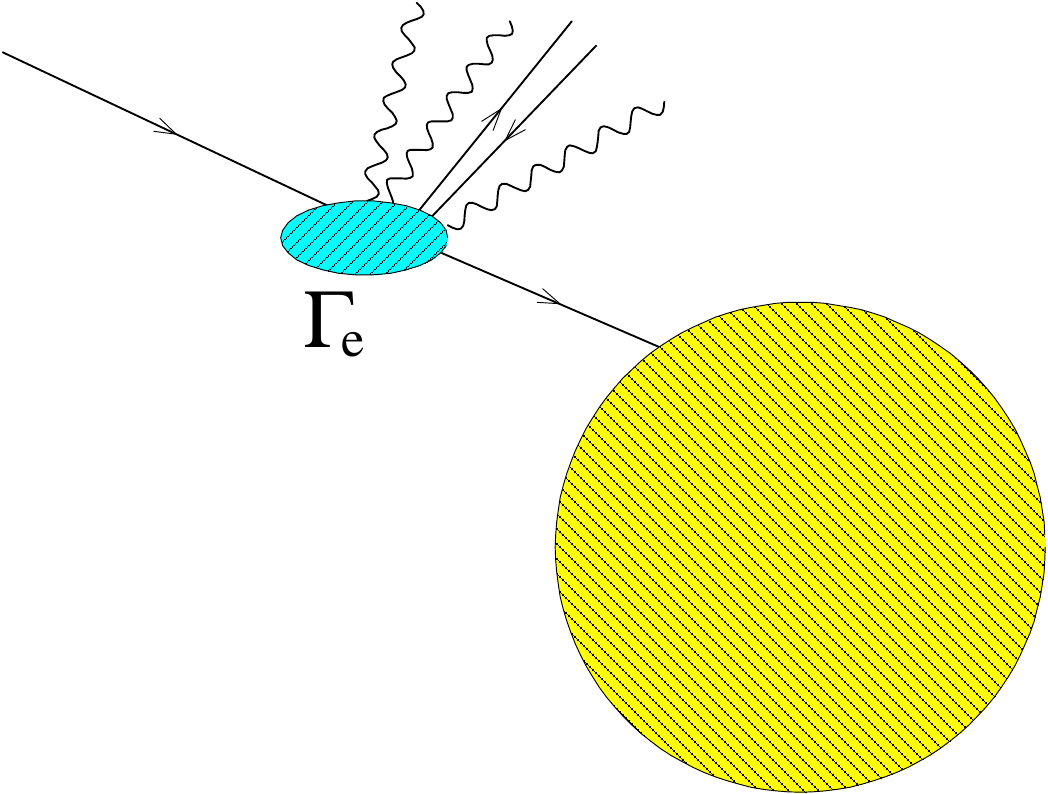}
$\phantom{a}$
  \includegraphics[width=0.47\textwidth]{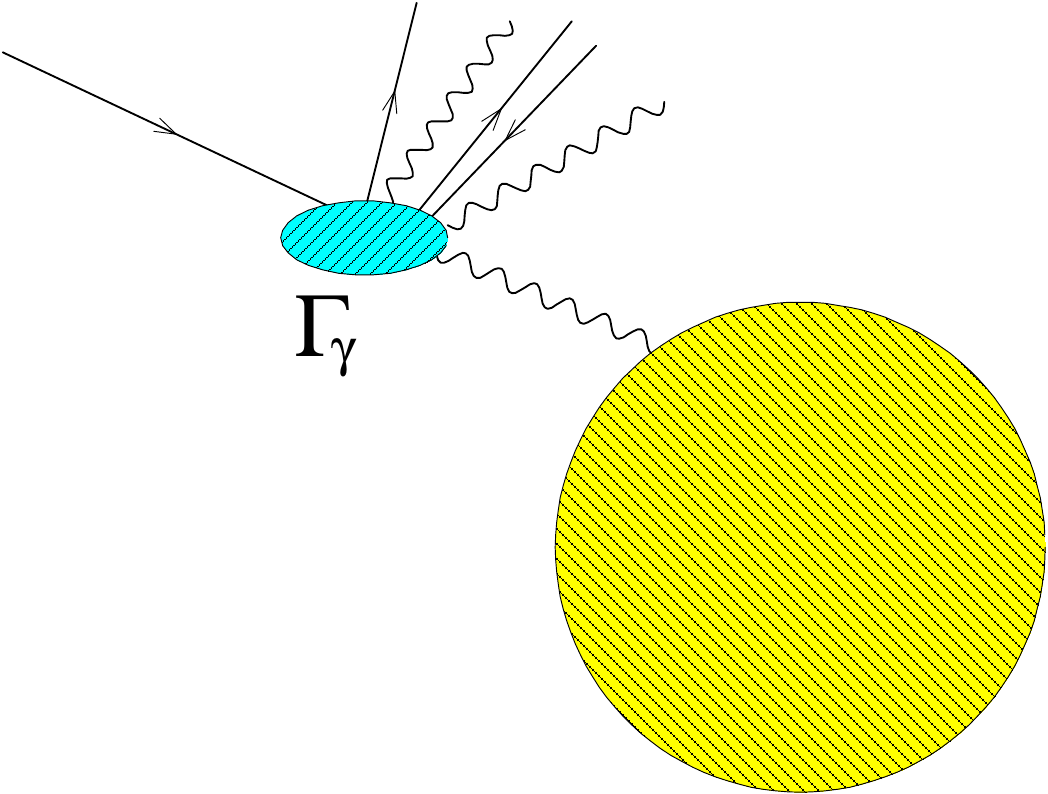}
\caption{\label{fig:Gphot} 
Electron (left panel) and photon (right panel) PDFs (blue ellipses) 
in an electron. The full yellow circle represents a generic
short-distance process.
}
  \end{center}
\end{figure}

We also remind the reader that for PDFs beyond the LL the identity of
the parton (index $i$ or $j$ in eq.~(\ref{master0})) and of the particle 
(index $k$ or $l$ in eq.~(\ref{master0})) need not coincide\footnote{At the
LO+LL whether non-zero PDFs with $i\ne k$ exist is a matter of conventions.
While we do consider them, as opposed to the standard approach in the 
literature that ignores them, in practice their contributions are minuscule.
See also sect.~\ref{sect:gammaproc}.}. When this
happens (as is schematically depicted on in the l.h.s.~panel of 
fig.~\ref{fig:Gphot} for an electron particle) one generally obtains
a contribution which is larger than those relevant to all of the other cases
(in the r.h.s.~panel of fig.~\ref{fig:Gphot} we depict the most relevant
among them, namely that of the photon PDF).
Whether this dominance is also seen at the level of physical cross
sections depends ultimately on the short-distance cross sections
($d\hsig_{ij}$ in eq.~(\ref{master0})), and on the observables one
is interested in. As an example which involves the photon PDF we
remark that already at the LO (i.e.~at $\ord(\aem^2)$) the processes 
of eqs.~(\ref{procWW}) and~(\ref{proctt}) receive contributions from
both of the following partonic production channels:
\beq
\epem\;\longrightarrow\;W^+W^-\,,\;\;\;\;\;\;\;\;
\epem\;\longrightarrow\;t\bt\,,
\label{LOeeprocWWtt}
\eeq
and 
\beq
\gamma\gamma\;\longrightarrow\;W^+W^-\,,\;\;\;\;\;\;\;\;
\gamma\gamma\;\longrightarrow\;t\bt\,.
\label{LOggprocWWtt}
\eeq
Since the photon PDF is of relative $\ord(\aem)$ w.r.t.~the electron PDF,
one may be tempted to conclude that for physical observables the partonic 
processes of eq.~(\ref{LOggprocWWtt}) will contribute to NNLO, and can 
therefore be discarded given that our results are NLO accurate. However, this 
argument, based strictly on a perturbative expansion of the PDFs, is plain 
wrong from a phenomenological viewpoint. In fact, we point out that the photon 
PDF becomes larger than the electron PDF\footnote{In spite of it still being
perturbatively suppressed by an $\aem$ factor w.r.t.~the electron PDF. 
In other words, higher-order contributions, that stem from the evolution, 
do matter.} as one moves towards small $z$ values (see 
ref.~\cite{Bertone:2019hks}), which is a region that can be accessed when 
the invariant mass of the system produced in the hard collision is much 
smaller than the available c.m.~energy. We shall return to this point 
in sect.~\ref{sect:gammaproc}. We also observe that quark-initiated
contributions to $W^+W^-$ and $t\bt$ production exist as well in the
context of a treatment that features multiple fermion families, as the
one presented in this paper; we neglect them here.

We finally point out that a soft non-collinear logarithm is exponentiated
in the fixed-$\aem$ LO+LL PDFs which have often been used in the 
literature\footnote{In eqs.~(\ref{bschdef})--(\ref{mschdef}) neglecting
such a logarithm corresponds to setting $\beta_E=\ee^2\eta$.}. While it is 
not difficult to arrive at a similar exponentiation in the case of either 
running-$\aem$ LO+LL PDFs~\cite{Gribov:1972ri,Dokshitzer:1977sg,
Nicrosini:1986sm} or (in a naive manner) NLO+NLL PDFs, we have not 
done it in this paper, since this matter deserves a more thorough analysis.
Therefore, for consistency reasons and because our present priority is
that of assessing various aspects of collinear physics, such an exponentiated
logarithmic term is not included in the LO+LL PDFs we use in the following; 
thus, for these PDFs we employ either the running scheme
(eq.~(\ref{rschdef}), in the case of $\MSb$) or the collinear scheme
(eq.~(\ref{collschdef}), in the cases of $\aem(\mZ)$ or $G_\mu$).

\subsection{Impact of NLL effects\label{sect:resNLL}}
\begin{figure}[t!]
  \begin{center}
  \includegraphics[width=0.47\textwidth]{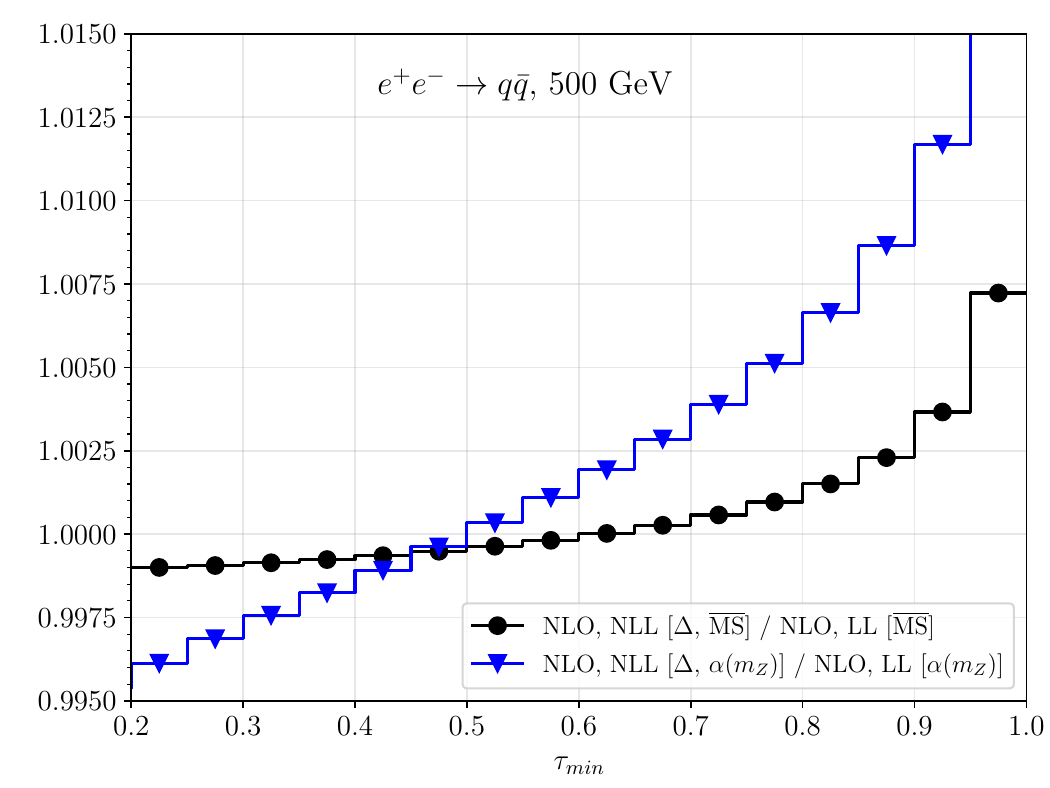}
$\phantom{a}$
  \includegraphics[width=0.47\textwidth]{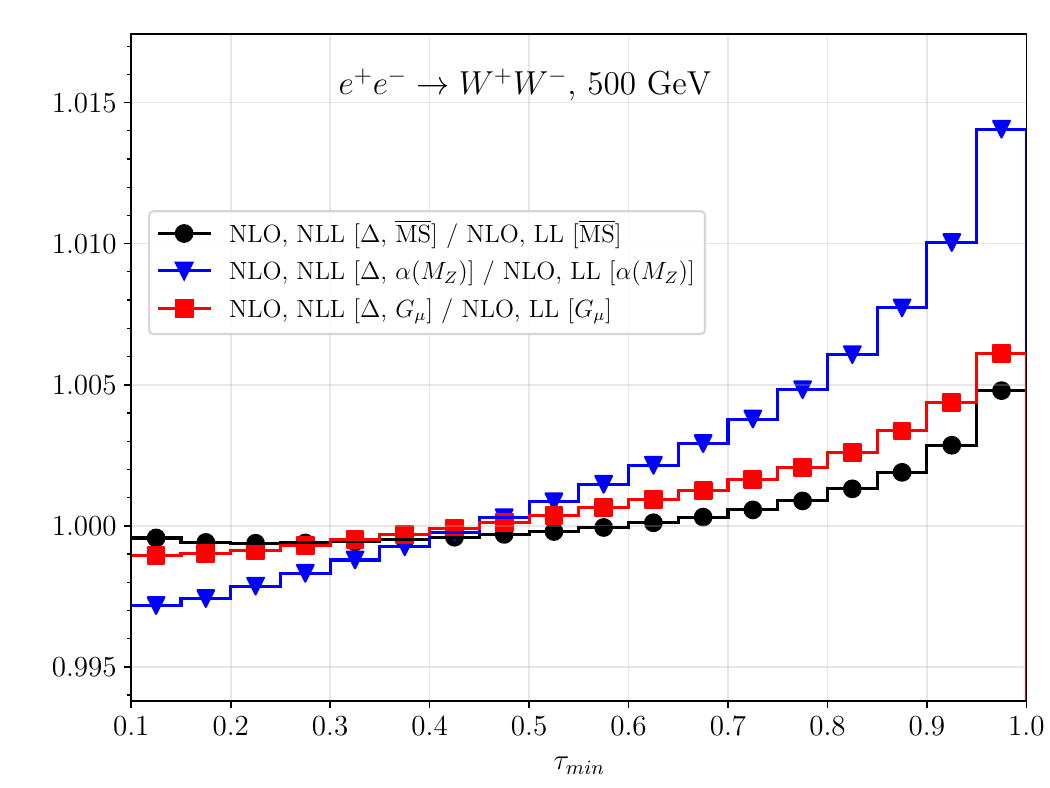}
\caption{\label{fig:NLLoLL1} 
Ratios of NLO cross sections computed with NLL-$\Delta$ PDFs over those
computed with LL PDFs, for different choices of renormalisation scheme.
Left panel: $q\bq$ production; right panel: $W^+W^-$ production.
}
  \end{center}
\end{figure}
\begin{figure}[t!]
  \begin{center}
  \includegraphics[width=0.47\textwidth]{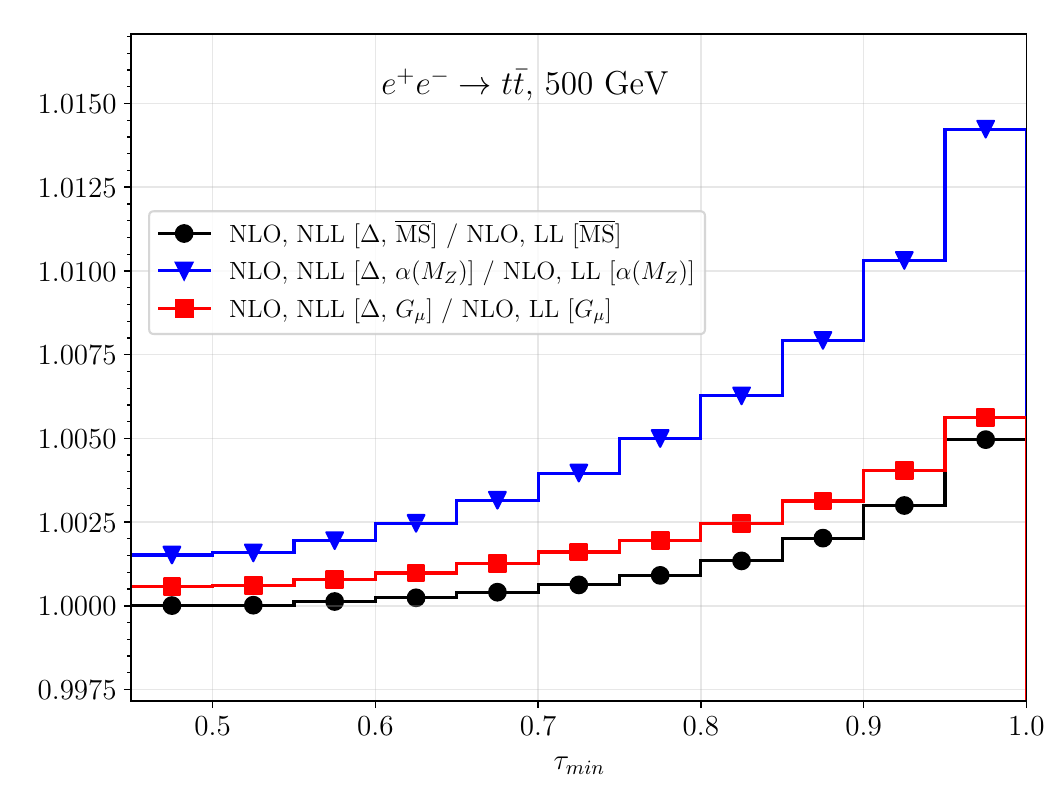}
$\phantom{a}$
  \includegraphics[width=0.47\textwidth]{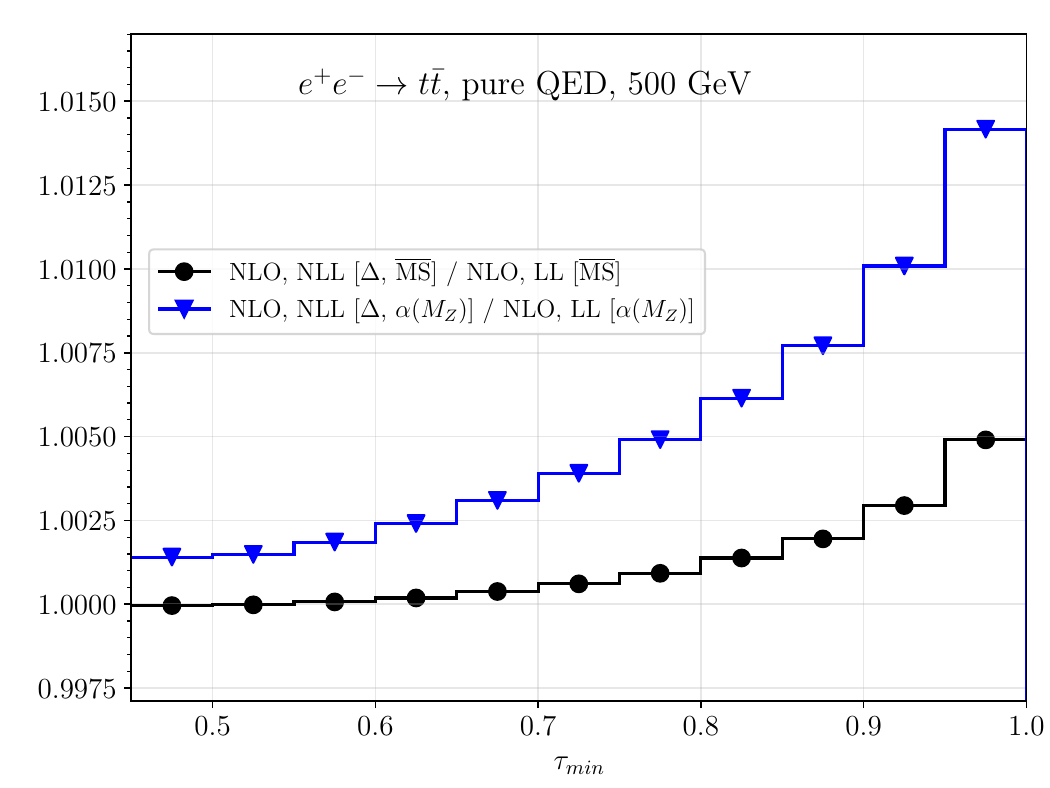}
\caption{\label{fig:NLLoLL2} 
As in fig.~\ref{fig:NLLoLL1}, for $t\bt$ production in the full SM
(left panel) and in QED (right panel).
}
  \end{center}
\end{figure}
We start by discussing the impact of the inclusion of the NLL terms in the 
PDFs. We do so by plotting, in figs.~\ref{fig:NLLoLL1} and~\ref{fig:NLLoLL2},
the ratio of the cross section of eq.~(\ref{Ixsec}) computed (at the NLO) 
with the NLO+NLL PDFs defined in the $\Delta$ factorisation scheme over the 
same quantity computed with LO+LL PDFs (for the latter ones, either the running 
or the collinear scheme is employed -- see the comment at the very end of 
sect.~\ref{sec:res0}). We do so for different values of $\tau_{min}$, 
the results of each of which are represented as bin entries
in a histogram\footnote{The ranges in $\tau_{min}$ cover all of the 
kinematically-accessible values; the leftmost bin that we include contains 
the threshold value \mbox{$\tau_{min}=4m_p^2/s$}, except in the case of $q\bq$ 
production (whose threshold value $\tau_{min}=0$ coincides with a singularity
of the matrix elements), where we consider \mbox{$\tau_{min}\ge 0.2$}.}. 
We carry out the computations in the three renormalisation schemes 
considered in this paper, using the same scheme in the numerator
and denominator: $\MSb$ (black curves overlaid with circles), $\aem(\mZ)$ 
(blue curves overlaid with triangles), and $G_\mu$ (red curves overlaid 
with boxes). Figure~\ref{fig:NLLoLL1} presents the predictions for $q\bq$ 
production (left panel) and $W^+W^-$ production (right panel), whereas 
fig.~\ref{fig:NLLoLL2} is relevant to $t\bt$ production, in the full SM 
(left panel) and in QED (right panel).

There are a couple of immediate conclusions that can be drawn from the
inspection of the figures. Firstly, the relative impact of the NLL 
contributions can be much larger than the typical precision targets 
at future $\epem$ colliders, and depends on both the process and the 
kinematical region one considers (since the histograms are not flat);
and, secondly, the dependence on the renormalisation scheme is
significant (conversely, we shall show in sect.~\ref{sect:resfacren}
that the one stemming from the factorisation scheme is much smaller,
which is the reason why we could concentrate here on $\Delta$-scheme
results). As far as the former aspect is concerned, it is representative
of a process- and observable-dependent pattern\footnote{For each process, 
we have computed several differential and cumulative observables, and studied
them in the same manner as what is done here for that of eq.~(\ref{Ixsec}).}
that renders it impossible to account for NLL PDF effects in some
``universal'' manner (e.g., with the multiplication of LL-accurate 
results by an overall factor). Thus, the key conclusion is the
following: while the assessment of the relevance of NLL PDF effects 
depends on the specific applications one pursues (in particular,
the observable one considers and the accuracy with which this is expected
to be determined experimentally), one should expect them to be 
phenomenologically important in high-energy $\epem$ collisions,
and thus regard NLL-accurate PDFs as the default choice for
precision studies in that context.

\subsection{Factorisation- and renormalisation-scheme dependences
\label{sect:resfacren}}
In this section we consider the dependence of the observable of 
eq.~(\ref{Ixsec}) upon the choice of the factorisation and the 
renormalisation schemes. We first point out that these two dependencies
{\em may} be seen as being of a different nature, in spite of the fact
that they both induce differences that are beyond the accuracy one is
working at (thus, in our case, the differences are of NNLO). In particular, 
it is often the case that a definite renormalisation scheme is chosen because
it is thought to be particularly apt at correctly capturing dominant effects 
of perturbative orders higher than those included in the computation one 
is performing (e.g., the $G_\mu$ scheme for processes that involve $W$'s
and $Z$'s, and no photons).
This viewpoint is of course legitimate, but its validity diminishes
with the ability to carry out computations of increasingly-high 
perturbative accuracy; in such a situation, it is more sensible to
regard the differences in predictions stemming from different renormalisation
schemes as a theoretical systematics. Conversely, one observes that a 
factorisation scheme is not defined in relation to some physical property, 
as is the case for (most of) the renormalisation scheme(s): it is a purely
theoretical artifact, in that it defines the finite part of the residue
in the subtraction of a collinear singularity. As such, the differences
between the predictions obtained with different factorisation schemes
are almost by definition a theoretical systematics, although some schemes 
can be better than others in terms of giving predictions more in line with 
higher order calculations\footnote{In the language of the FKS 
subtraction that is used here, where factorisation
schemes are defined by the choices of the $K_{ij}(z)$ functions, it
is particularly easy to see how the cancellation of the effects they
induce occurs in perturbation theory -- see e.g.~eq.~(\ref{tmp16})
and the comments that follow it.}. And yet, when increasing the 
perturbative accuracy of the computation such a systematics may become
the dominant source of uncertainty, and one may want to find theoretical
motivations for a definite choice of the factorisation scheme. While we
do not adopt this attitude here (also in view of the fact that we work 
at the NLO), we point out that the $\MSb$ and $\Delta$ schemes are
dramatically different in the $z\simeq 1$ (i.e.~the soft) region, and
this has some practical consequences.

\begin{figure}[t!]
  \begin{center}
  \includegraphics[width=0.47\textwidth]{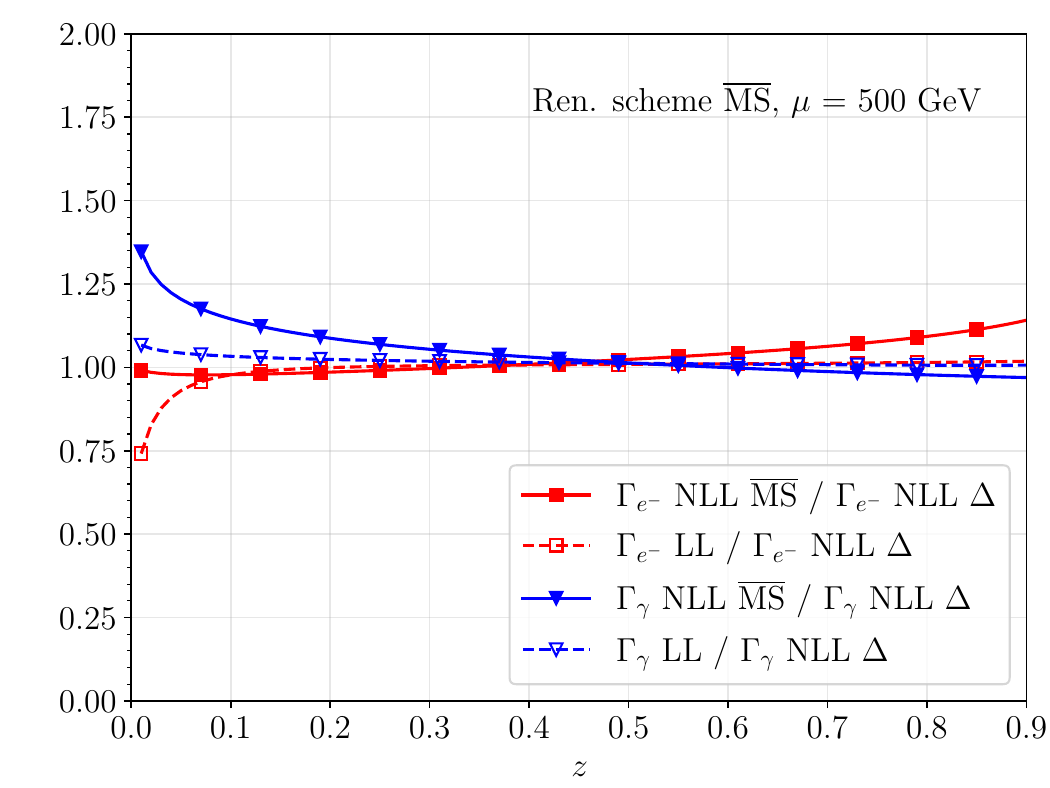}
$\phantom{a}$
  \includegraphics[width=0.47\textwidth]{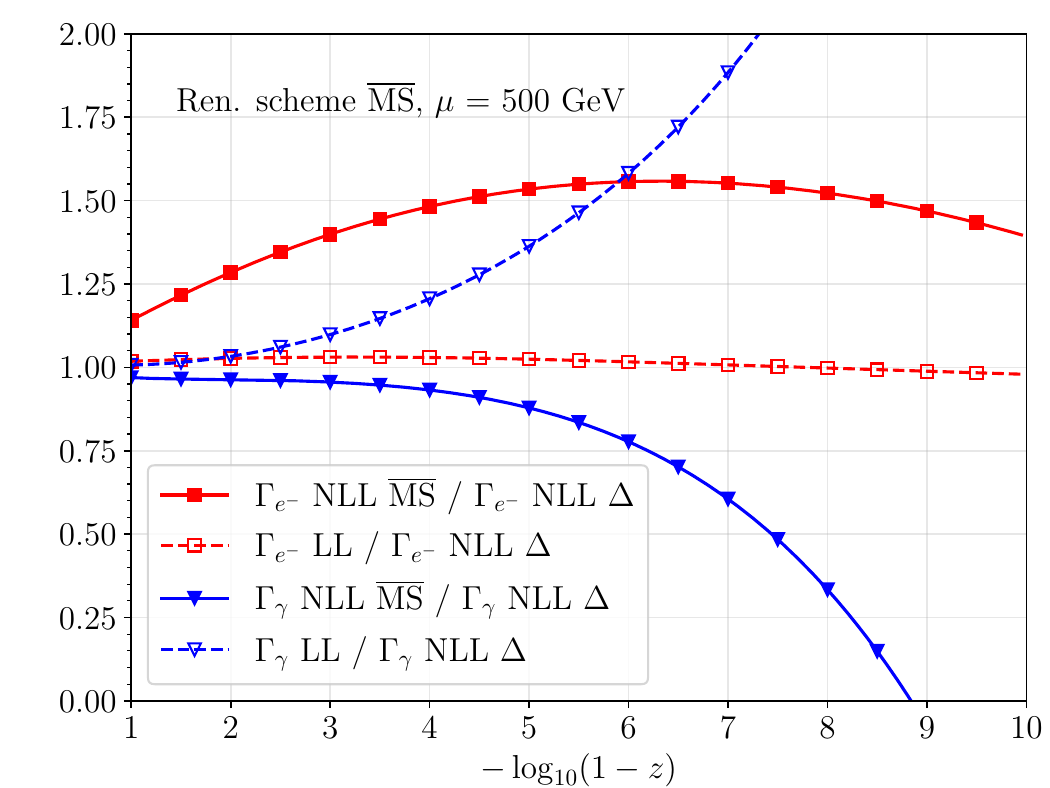}
\caption{\label{fig:PDFrat} 
Ratios of NLL $\MSb$ PDFs (solid curves) and LL PDFs (dashed curves)
over NLL $\Delta$ PDFs, for an electron (red curves overlaid with boxes) 
and a photon (blue curves overlaid with triangles); all of the PDFs are 
computed in the $\MSb$ renormalisation scheme. The ratios are shown for 
small- and intermediate-$z$ (left panel), and for $z\simeq 1$ (right panel).
}
  \end{center}
\end{figure}
In order to further the previous point, we must bear in mind that
while physical predictions are factorisation-scheme dependent only
beyond the perturbative accuracy one is working at, this is not true
for either the PDFs or the short-distance cross sections. In the 
case of the PDFs this is apparent from fig.~\ref{fig:PDFrat}. There,
we show the ratios of the NLL PDFs computed in the $\MSb$ factorisation
scheme over those computed in the $\Delta$ scheme -- for both, the $\MSb$
renormalisation scheme is adopted to be definite (the results in other
renormalisation schemes are totally analogous). The results for the 
electron (red solid curves overlaid with boxes) and photon
(blue solid curves overlaid with triangles) PDFs are presented,
in the small- and intermediate $z$ region (left panel), as well as
for $z\simeq 1$ (right panel). The ratios are extremely large ($\ord(1)$
deviations w.r.t.~one) in the large-$z$ region, which is particularly 
significant for the electron, since that region gives by far the dominant 
contribution to physical observables. For comparison, analogous ratios where 
the numerators are the LL-accurate PDFs (dashed lines overlaid with boxes and 
triangles for the electron and the photon, respectively) show deviations 
from one only of approximately $\ord(3\%)$ in the case of the electron
(except for very small $z$ values). In other words, NLL PDFs defined in 
the $\Delta$ scheme are quite similar to the LL ones, while very large 
differences are seen in the case of the $\MSb$ scheme.

\begin{figure}[t!]
  \begin{center}
  \includegraphics[width=0.47\textwidth]{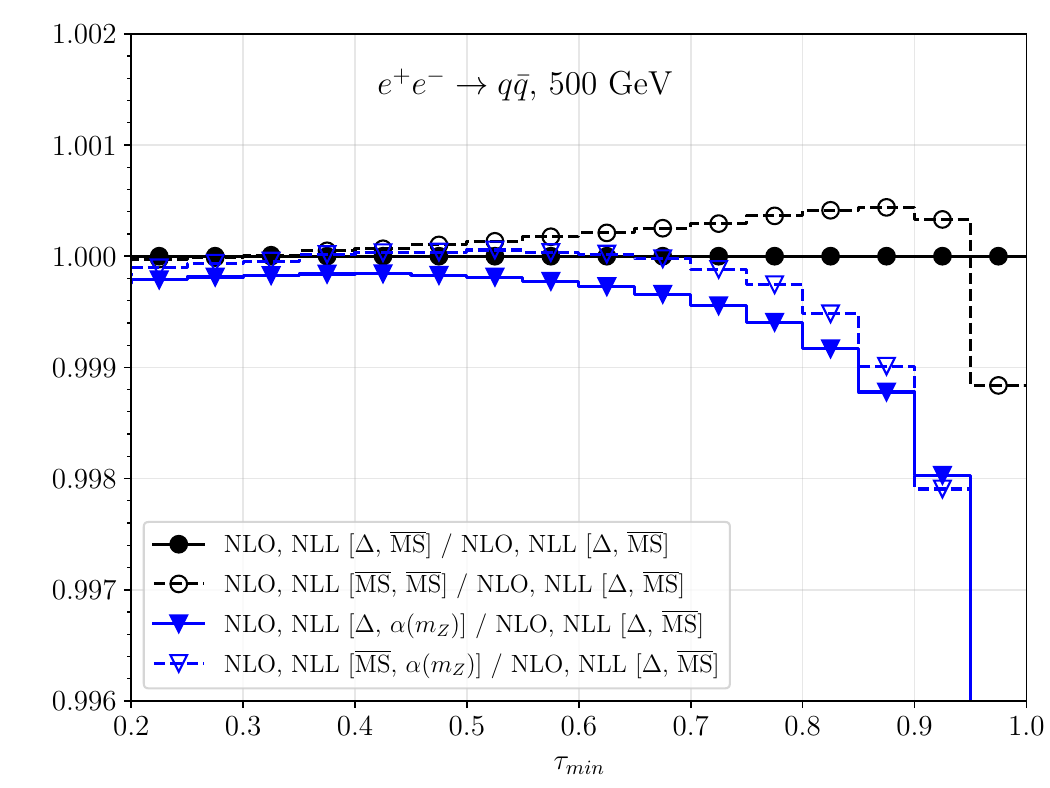}
$\phantom{a}$
  \includegraphics[width=0.47\textwidth]{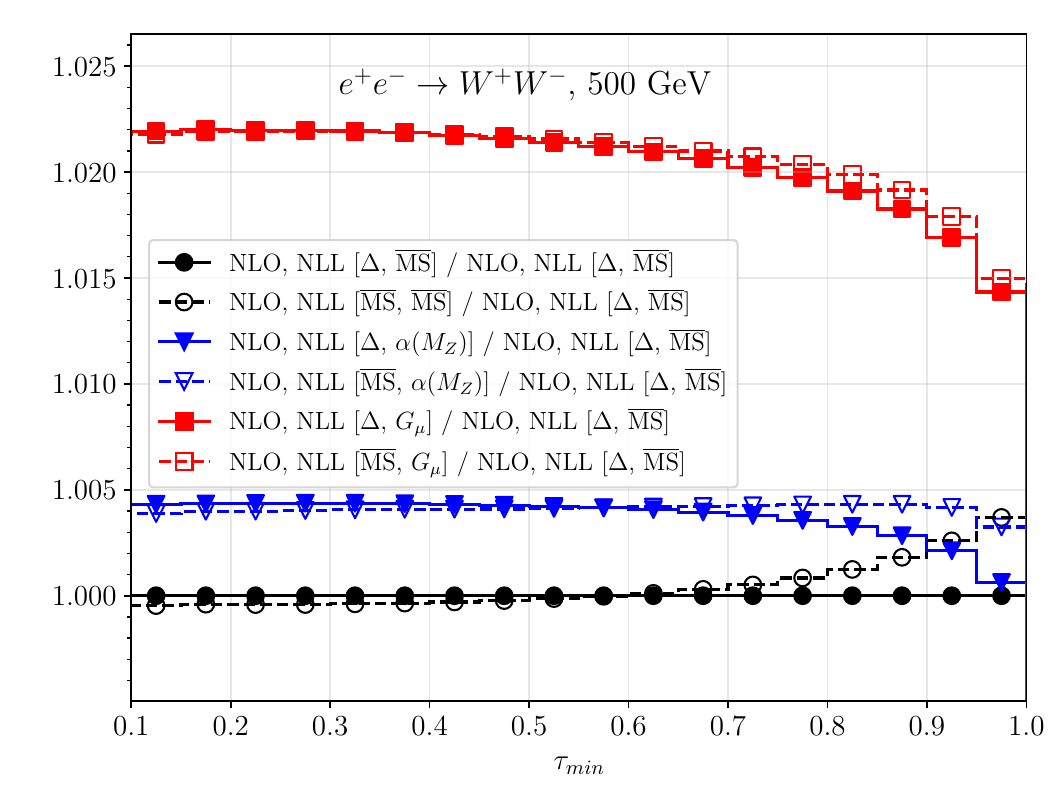}
\caption{\label{fig:facren1}
Ratios of cross sections computed with NLL PDFs for all possible
combinations of renormalisation and factorisation schemes, over
those computed with NLL-$\Delta$ PDFs in the $\MSb$ renormalisation
scheme. Left panel: $q\bq$ production; right panel: $W^+W^-$ production.
}
  \end{center}
\end{figure}
\begin{figure}[t!]
  \begin{center}
  \includegraphics[width=0.47\textwidth]{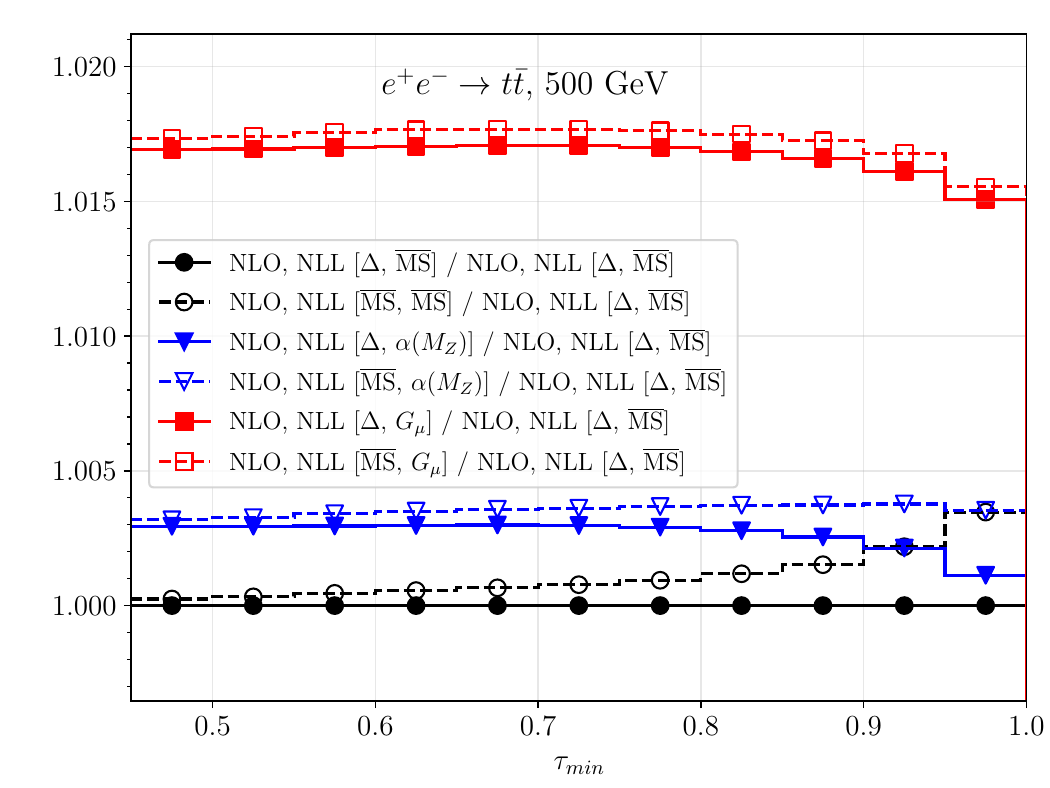}
$\phantom{a}$
  \includegraphics[width=0.47\textwidth]{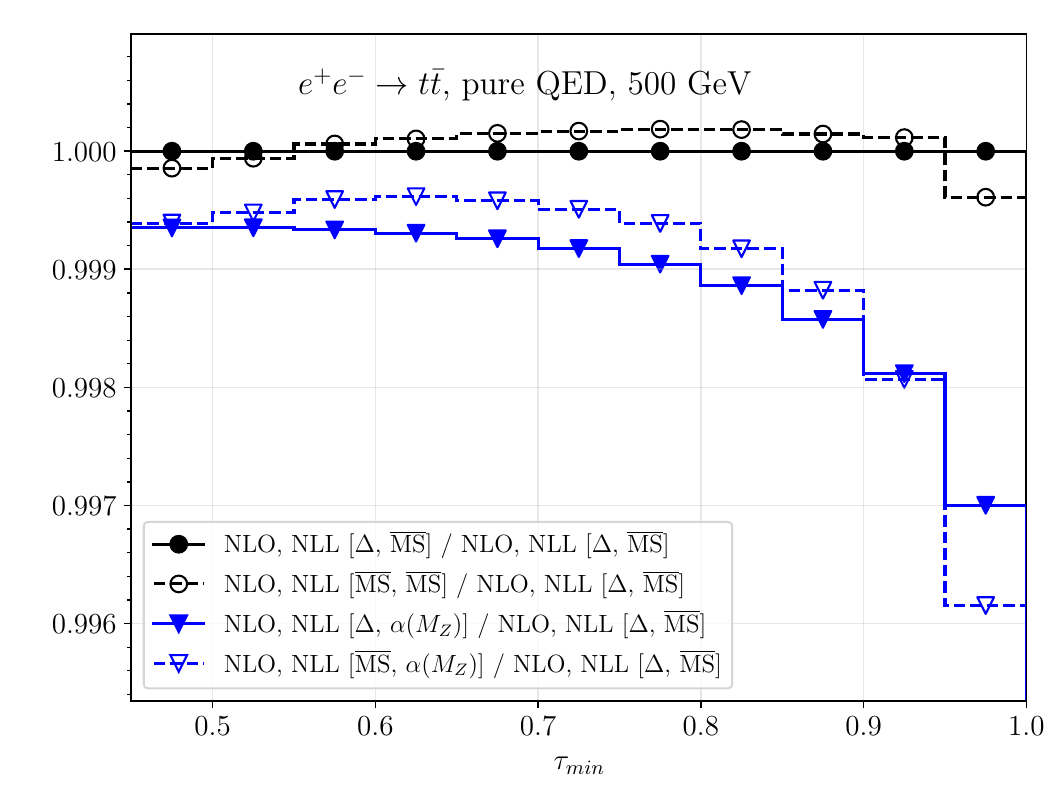}
\caption{\label{fig:facren2} 
As in fig.~\ref{fig:facren1}, for $t\bt$ production in the full SM
(left panel) and in QED (right panel).
}
  \end{center}
\end{figure}
\begin{figure}[t!]
  \begin{center}
  \includegraphics[width=0.47\textwidth]{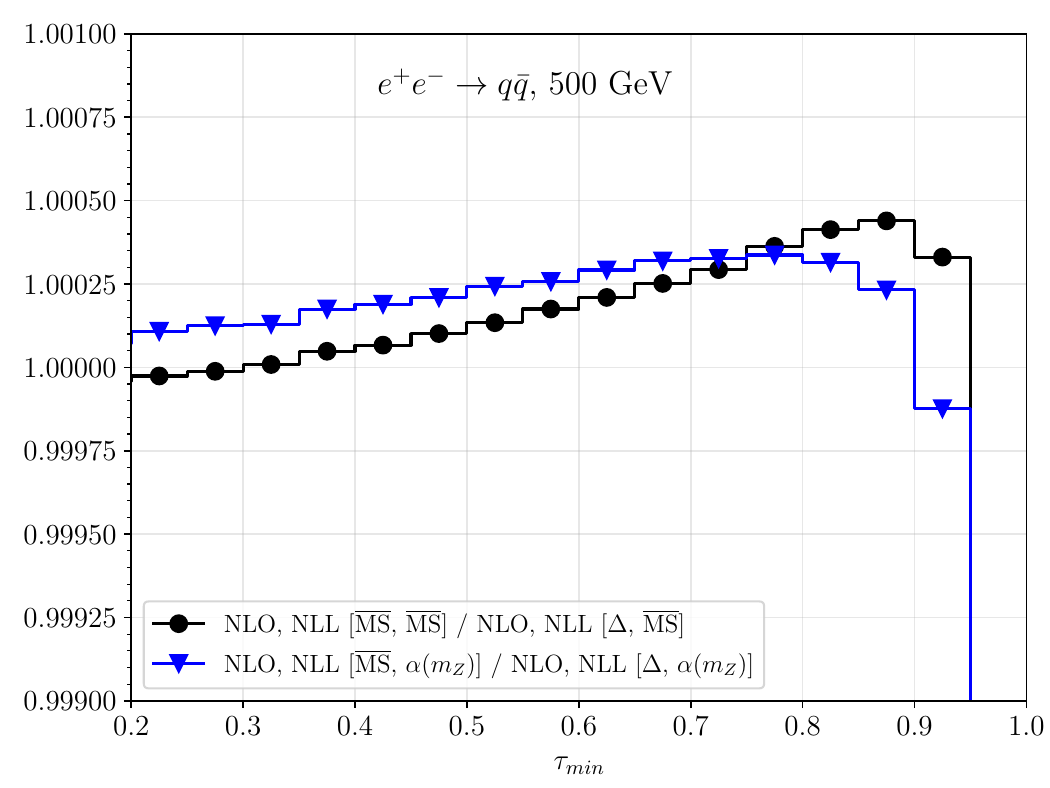}
$\phantom{a}$
  \includegraphics[width=0.47\textwidth]{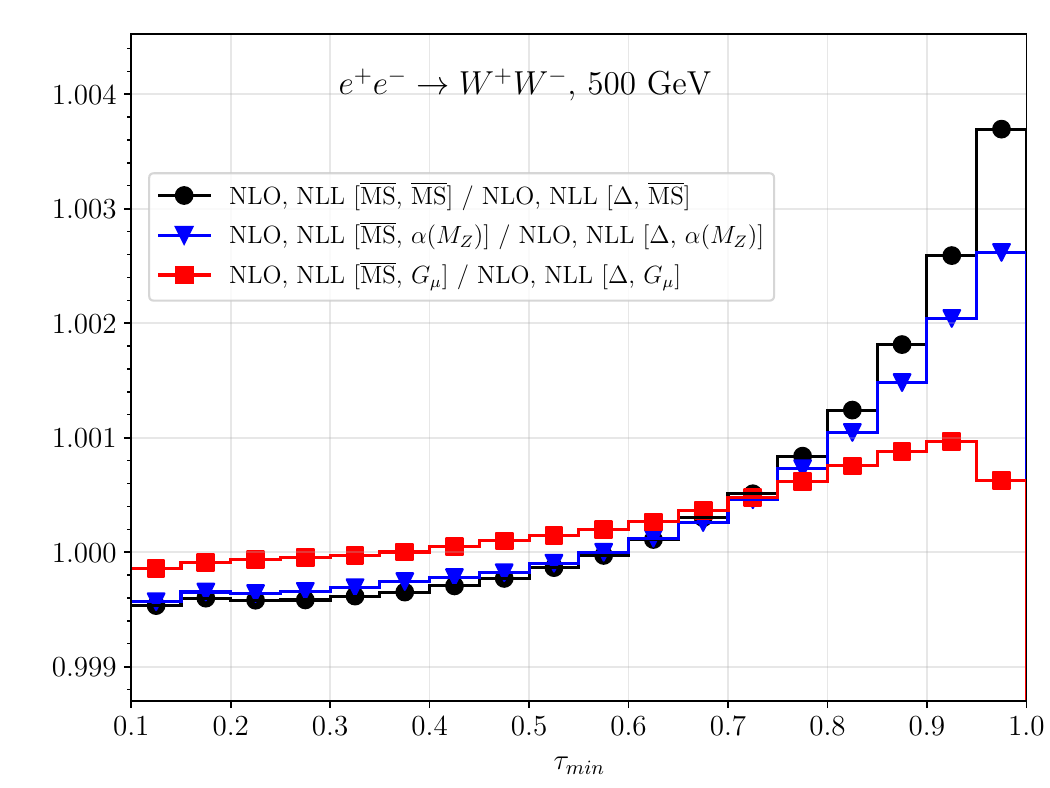}
\caption{\label{fig:fac1} 
Ratios of cross sections computed with NLL-$\MSb$ PDFs over those
computed with NLL-$\Delta$ PDFs, for different choices of renormalisation 
scheme. Left panel: $q\bq$ production; right panel: $W^+W^-$ production.
}
  \end{center}
\end{figure}
\begin{figure}[t!]
  \begin{center}
  \includegraphics[width=0.47\textwidth]{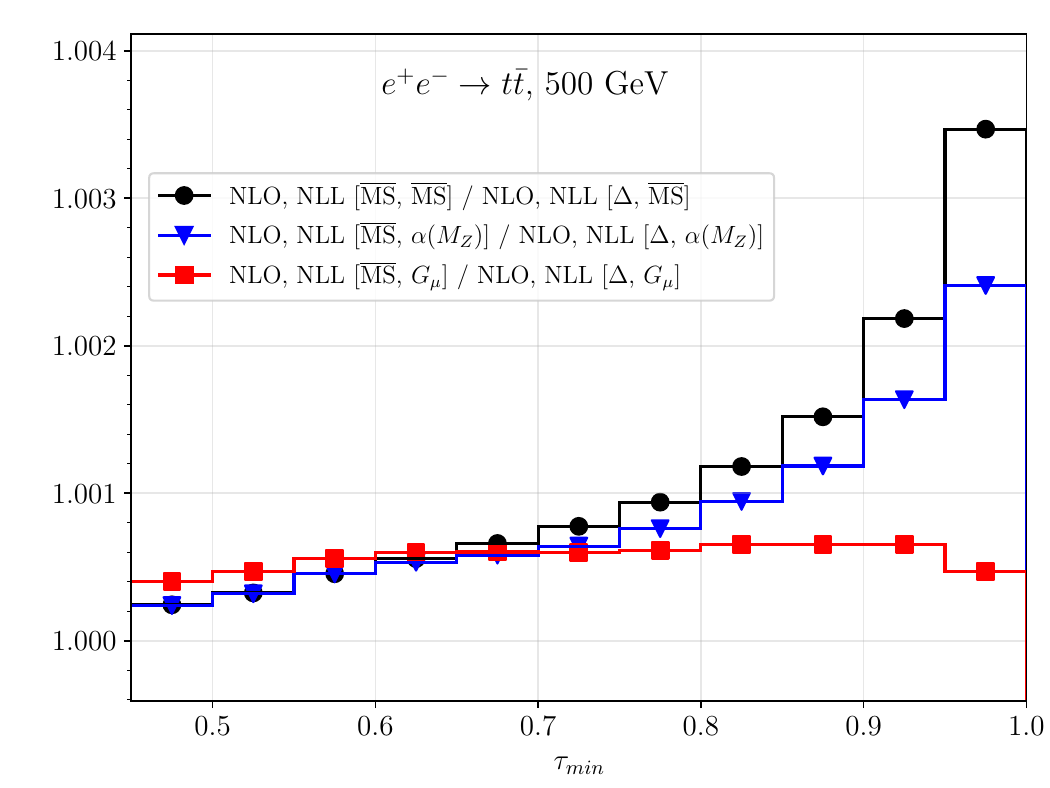}
$\phantom{a}$
  \includegraphics[width=0.47\textwidth]{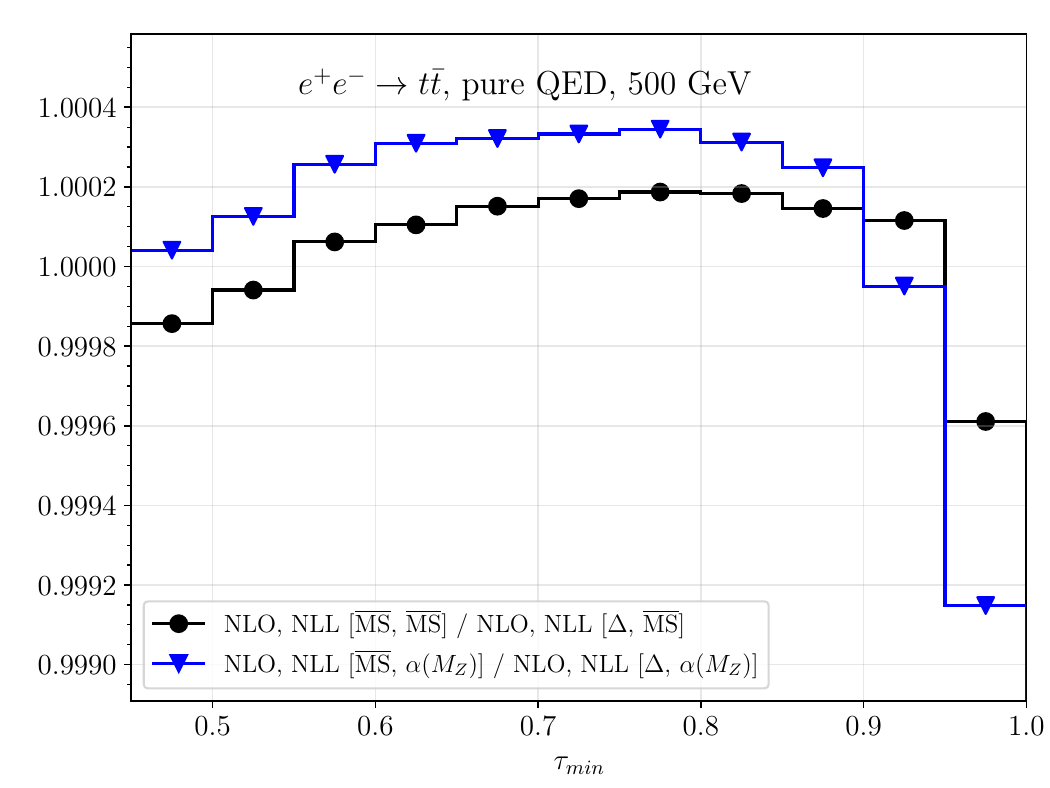}
\caption{\label{fig:fac2} 
As in fig.~\ref{fig:fac1}, for $t\bt$ production in the full SM
(left panel) and in QED (right panel).
}
  \end{center}
\end{figure}

Given the significant differences between the PDFs defined in the $\MSb$
and $\Delta$ schemes, it is remarkable how well the predictions that stem 
from them agree with each other at the level of observables. This is
shown in figs.~\ref{fig:facren1}--\ref{fig:fac2}, which we now comment
in some detail. In fig.~\ref{fig:facren1} (relevant to $q\bq$ and $W^+W^-$ 
production) and fig.~\ref{fig:facren2} (relevant to $t\bt$ production in
the full SM and in QED) we plot the ratios of the NLO results obtained with
all of the six combinations of renormalisation and factorisation schemes
(four for pure-QED processes, the $G_\mu$ scheme being not relevant there),
over those obtained by using PDFs defined in the $\Delta$ factorisation
scheme and in the $\MSb$ renormalisation scheme. The three solid histograms
are those associated with using, in the numerators, PDFs defined in the 
$\Delta$ factorisation scheme and the three renormalisation schemes -- $\MSb$ 
(black curves overlaid with circles), $\aem(\mZ)$ (blue curves overlaid with 
triangles), and $G_\mu$ (red curves overlaid with boxes). The three dashed 
histograms, that employ the same patterns as their solid counterparts, are 
obtained by using in the numerators PDFs defined in the $\MSb$ factorisation 
scheme. Thus, for a given colour/symbol (i.e.~a renormalisation-scheme choice) 
the differences between the solid and dashed histograms measure the 
factorisation-scheme dependences; whereas for a given pattern (solid or
dashed, i.e.~a factorisation-scheme choice) the differences among the three
colours/symbols measure the renormalisation-scheme dependences. 

The message
that emerges in a clear manner is that the renormalisation-scheme dependence
is significantly larger than the factorisation-scheme one\footnote{One 
sees a (process-dependent) breakdown of this pattern when $\tau_{min}\to 1$; 
we remind the reader that fixed-order predictions lose validity in that 
region, owing to the emergence of soft unresummed logarithms.}; this is 
true independently of the process considered. Note that this behaviour is 
consistent with the general observations made at the beginning of this
section regarding the different nature of theoretical systematics stemming
from factorisation- and renormalisation-scheme choices. With that being 
said, we remark that the renormalisation-scheme dependence is mostly
(the exception being again the large-$\tau_{min}$ region) a normalisation
effect. This is related to the observation, made immediately before
sect.~\ref{sec:Rsch}, that the parameters that control the electron PDF 
shape at $z\to 1$ ($\xi_1^{(k)}$ and $\hat{\xi}_1^{(k)}$) differ, across the
three UV schemes considered here, by terms of $\ord(\aem^3)$, which are thus
proven to be not significant numerically. From the previous discussion, 
however, we expect shape, and not only normalisation, differences to become 
more relevant with increasing c.m.~energy. We have also verified that shape 
differences are already present at \mbox{$\sqrt{s}=500$~GeV} if one naively 
switches the running of $\aem$ off in $\MSb$; this underscores our findings 
that such an unphysical case induces differences of $\ord(\aem^2)$ in the
$\xi_1^{(k)}$ and $\hat{\xi}_1^{(k)}$ parameters w.r.t.~the actual $\MSb$ 
ones, and confirms that with NLL PDFs simply ignoring the running of $\aem$
in $\MSb$ is not a viable option for physics simulations: a fixed-$\aem$ 
renormalisation scheme, such as $\aem(\mZ)$ and $G_\mu$, must be used
instead.

Another key conclusion from figs.~\ref{fig:facren1} and~\ref{fig:facren2} 
is that these figures show definitely what has been anticipated above, namely 
that the very significant differences between the NLL PDFs defined in 
different factorisation schemes do {\em not} result in large differences at 
the level of observables: there is a large compensation mechanism at play 
between the PDFs and the short-distance cross sections.

In order to document the latter remark in a more quantitative manner,
in fig.~\ref{fig:fac1} (relevant to $q\bq$ and $W^+W^-$ production) and 
fig.~\ref{fig:fac2} (relevant to $t\bt$ production in the full SM and in QED) 
we present the ratios for our observable obtained by using PDFs defined in
the $\MSb$ ($\Delta$) factorisation scheme in the numerator (denominator).
We do so separately for the three renormalisation schemes, i.e.~$\MSb$ 
(black curves overlaid with circles), $\aem(\mZ)$ (blue curves overlaid 
with triangles), and $G_\mu$ (red curves overlaid with boxes). Thus,
for any given colour/symbol, these histograms correspond to the ratios
of the dashed over solid histograms with the same patterns that appear
in figs.~\ref{fig:facren1} and~\ref{fig:facren2}. We conclude that the
$\ord(1)$ differences between different factorisation schemes observed
in the PDFs result in \mbox{$\ord(10^{-4}-10^{-3})$} differences for
this observable. While this conclusion holds true irrespective
of the process one considers, we point out that the best agreement between
the results in the two factorisation schemes is observed for $q\bq$ and $t\bt$ 
production computed in QED. This is because these processes have a K-factor
which is closer to one than that of the full-SM ones -- this is 
shown in figs.~\ref{fig:Kfac1} and~\ref{fig:Kfac2} (we stress that K-factors
are unphysical quantities; the definition we have adopted emphasises the
role of the matrix elements, and is largely independent of the PDF choice;
see appendix~\ref{sec:LOxNLL} for more details). The farther away from one
the K-factor, the larger the relative impact of matrix elements of NLO 
($\ord(\aem^3)$ here) w.r.t.~LO ($\ord(\aem^2)$ here) ones; and NLO 
matrix elements, when convoluted with PDFs, induce a factorisation-scheme 
dependence of NNLO. Thus, such scheme-dependent NNLO terms are larger for 
processes with K-factors that differ from one by larger amounts.

A second conclusion to be drawn from figs.~\ref{fig:fac1} and~\ref{fig:fac2}
can be reached by considering the cancellation of the factorisation-scheme
dependence at the NLO. It turns out that in computations performed in $\MSb$ 
double-logarithmic terms appear in both the PDFs and in the short-distance
cross sections (see the discussion in appendix~\ref{sec:LOxNLL}), which 
mutually cancel in the convolution between these quantities. Conversely,
such terms are simply not present in the $\Delta$ scheme. The implication
is that, from the numerical point of view, computations carried out in the
$\MSb$ scheme require a much larger amount of CPU time w.r.t.~those performed
in the $\Delta$ scheme, in order to obtain the same statistical accuracy as
the latter ones -- this larger CPU footprint is clearly due to the loss of 
precision that cancellations among large terms entail. In fact, by looking 
at fig.~\ref{fig:PDFrat}, and in particular at the fact that NLL PDFs defined 
in the $\Delta$ ($\MSb$) factorisation scheme are very close (very different) 
from the LL ones, one heuristically understands that $\Delta$-based NLO+NLL 
computations are expected, from a statistical viewpoint, to behave similarly 
to their NLO+LL analogues.

\subsection{Impact of photon-induced contributions\label{sect:gammaproc}}
As was mentioned at the beginning of sect.~\ref{sec:res}, the results
of sects.~\ref{sect:resNLL} and~\ref{sect:resfacren} have been obtained
by keeping only the contributions due to the $\epem$ partonic channels,
which are expected to be largely dominant. Still, the master collinear
factorisation formula, eq.~(\ref{master0}), in principle requires an
incoherent sum (indices $i$ and $j$ on the r.h.s.) to be performed 
over all possible partonic channels. Given that our PDFs include 
$\Nl=3$ leptons, $\Nu+\Nd=5$ quarks, and the photon, there are up to
\mbox{$(\Nl+\Nu+\Nd+1)^2$} channels to be considered. While in QED
(at variance with QCD) the well-defined perturbative expansion of the
PDFs allows one to formally establish an $\aem$-based hierarchy among
the contributions due to the above partonic channels, this can be misleading
sometimes; for example, this happens when all-order effects significantly 
modify the PDFs w.r.t.~their expressions obtained with a perturbative 
expansion at some fixed order.

\begin{figure}[t!]
  \begin{center}
  \includegraphics[width=0.47\textwidth]{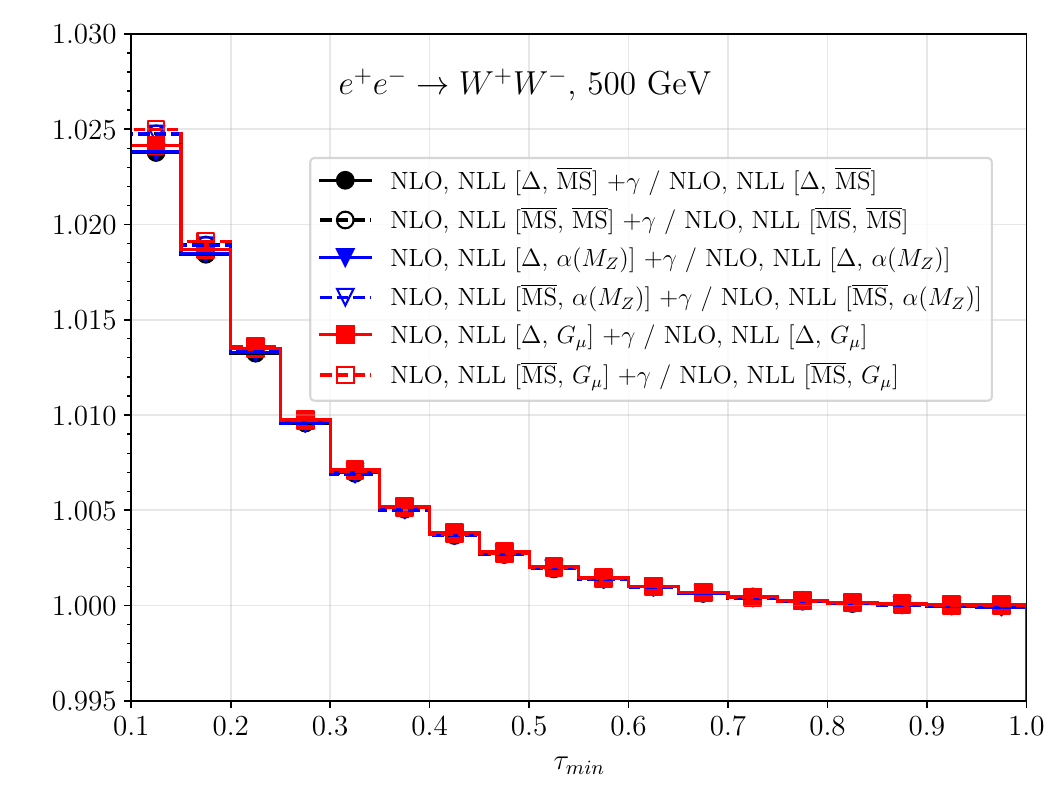}
$\phantom{a}$
  \includegraphics[width=0.47\textwidth]{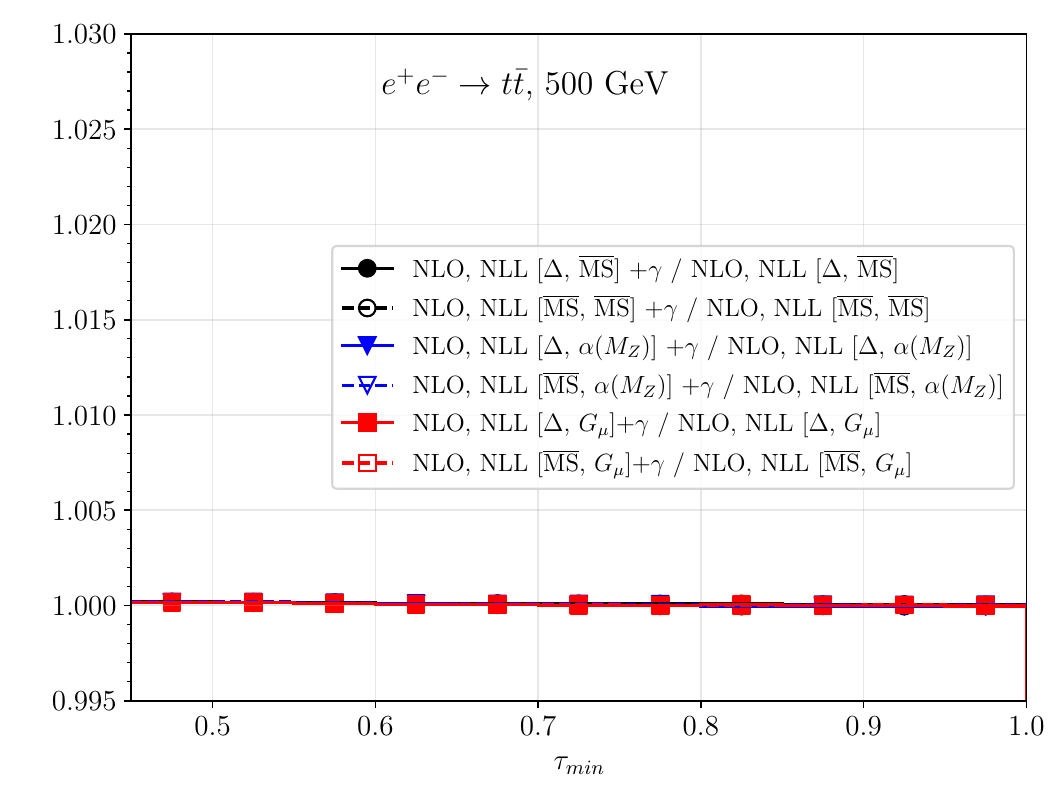}
\caption{\label{fig:gamma} 
Ratios of cross sections computed by including photon-initiated partonic
contributions over their counterparts computed by discarding such 
contributions, for all possible combinations of renormalisation and 
factorisation schemes. Left panel: $W^+W^-$ production; right panel:
$t\bt$ production in the full SM.
}
  \end{center}
\end{figure}
The most interesting case is that of the photon, since its PDF is only
suppressed by one power of $\aem$ w.r.t.~that of the electron (all of the
other partons have an $\ord(\aem^2)$ suppression). A preliminary discussion
on this case has already been given in the final part of 
sect.~\ref{sec:res0}; here, we aim to study the impact of photon-initiated
partonic channels in a couple of explicit cases, namely for $W^+W^-$ and
$t\bt$ production in the full SM. In particular, we compare the results
one obtains by retaining only the $\epem$ partonic channels with those
obtained by including photon-initiated ones as well -- at the LO for the
two processes considered here, these are eqs.~(\ref{LOeeprocWWtt})
and~(\ref{LOggprocWWtt})\footnote{We stress that at the NLO the partonic
structure is richer -- our results include contributions from all of
the possible photon/electron/positron partonic combinations, 
e.g.~\mbox{$\lp\gamma\to W^+W^-\lp$} 
and~\mbox{$\gamma\gamma\to W^+W^-\gamma$}.}. Our predictions are presented in 
fig.~\ref{fig:gamma}, in the forms of ratios of cross sections obtained
with all electron- and photon-initiated partonic channels over those relevant 
to electron-only channels. All of the six combinations of factorisation-
and renormalisation-scheme choices have been considered. The difference
between the two processes is striking (this is emphasised graphically by
the choice of the same range on the $y$ axis for the two panels of
the figure): while for $t\bt$ production the relative impact of the 
photonic channels is of $\ord(10^{-4})$, i.e.~within 
factorisation- and renormalisation-scheme 
uncertainty, for $W^+W^-$ production at small $\tau_{min}$ (i.e.~when 
the cross section approaches its fully-inclusive value) it is of $\ord(1\%)$, 
larger than any theoretical systematics at this order: it thus represents 
a physical effect. We note that, at the fully differential level, 
in regions dominated by small pair invariant masses, the photon-induced
contributions can actually be in excess of 50\% of the total. Needless to say, 
a key point here is the process dependence of the results: channels different 
from the $\epem$ one may or may not give sizable contributions, with a 
definite answer to be obtained only with specific running conditions and 
selection cuts. It is therefore important that the PDFs have the ability 
to include all partonic channels prescribed by the factorisation theorem.

Before concluding this section, a couple of general remarks are in order.
Firstly, we remind the reader that the photon PDF is {\em not} equal
to the Weizsaecker-Williams function~\cite{vonWeizsacker:1934nji,
Williams:1934ad} (WW henceforth); while at $\ord(\aem)$ these two 
quantities are relatively close to each other (but do not coincide -- see 
e.g.~ref.~\cite{Frixione:2019lga}), this is not the case for the all-order 
PDF vs the WW function. 
This may induce visible discrepancies between predictions obtained with 
the photon PDF and the WW function. We also point out that PDFs automatically 
include a unitarity condition: in other words, when summing over all possible
branching types that underpin PDF evolution, the number of incoming
particles (the electron in this case) is conserved, so that the fraction
of electron- vs photon-initiated partonic processes is the one correctly
determined by QED. This is not the case if the LL electron PDF (in particular
if evolved purely as a non-singlet) and the WW function are employed (as is 
often done in the context of N$^k$LO+LL simulations), since they separately 
implement a unitary constraint; thus, an appropriate rescaling of the 
respective contributions must be envisaged.

Secondly, we note that in the context of an YFS-based approach all
partonic processes that are not $\epem$-initiated enter (necessarily
at orders at least one higher than the Born's) via the IR-finite 
residues. This implies that if photon-induced (or any other parton
type) partonic processes give sizable contributions in a 
collinear-factorisation description owing predominantly to the all-order 
structure of the PDFs, such contributions cannot be reliably predicted 
by an YFS-based approach, unless the relevant collinear logarithms
in the residues are resummed to all orders.

\subsection{Simulations with beamstrahlung\label{sect:beamstr}}
In this section we consider the impact of beam-dynamics effects, which
we identify with beamstrahlung, on the observable of eq.~(\ref{Ixsec}).
Beamstrahlung is parametrised as indicated in eq.~(\ref{Bee}), so that
the only initial-state particles we consider are electrons and positrons.
As far as the function \mbox{${\cal B}_{\epem}(\yp,\ym)$} is concerned,
we use the form associated with a $500$-GeV collider with an ILC-type 
configuration as is given in ref.~\cite{Frixione:2021zdp}; the interested 
reader is urged to check that paper for further details. Here, we limit 
ourselves to pointing out that in our setup we define effective 
``beamstrahlung+ISR'' PDFs, obtained by convoluting once and for all 
(i.e.~prior to any physics runs) the beamstrahlung functions with the 
ISR PDFs, so that at runtime the number of integration variables is the 
same as that in the case where beamstrahlung is ignored; this also 
implies that the numerical behaviour of the two environments is
essentially identical.

\begin{figure}[t!]
  \begin{center}
  \includegraphics[width=0.47\textwidth]{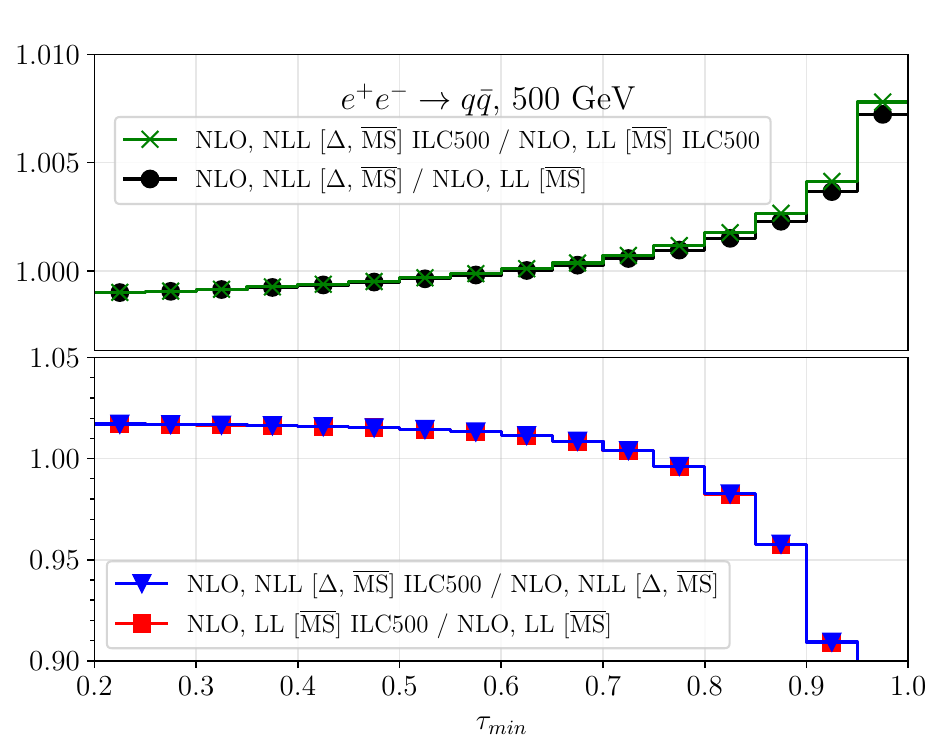}
$\phantom{a}$
  \includegraphics[width=0.47\textwidth]{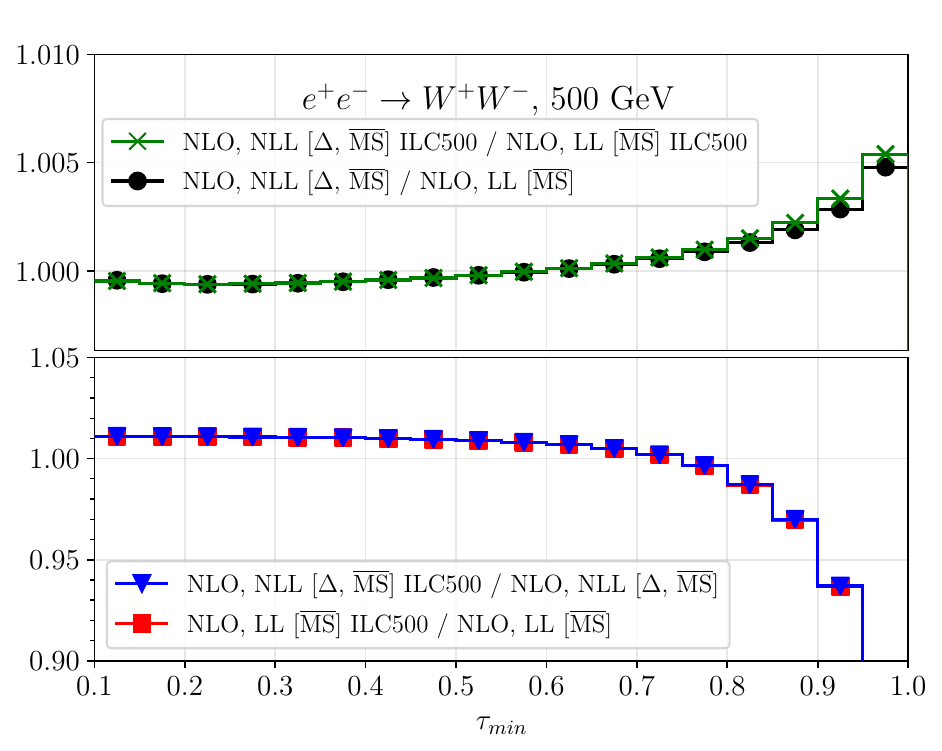}
\caption{\label{fig:beams1} 
Results of simulations that include beamstrahlung effects at ILC500, and 
their comparisons with those obtained without beamstrahlung. See the text 
for details. Left panel: $q\bq$ production; right panel: $W^+W^-$ production.
}
  \end{center}
\end{figure}

\begin{figure}[t!]
  \begin{center}
  \includegraphics[width=0.47\textwidth]{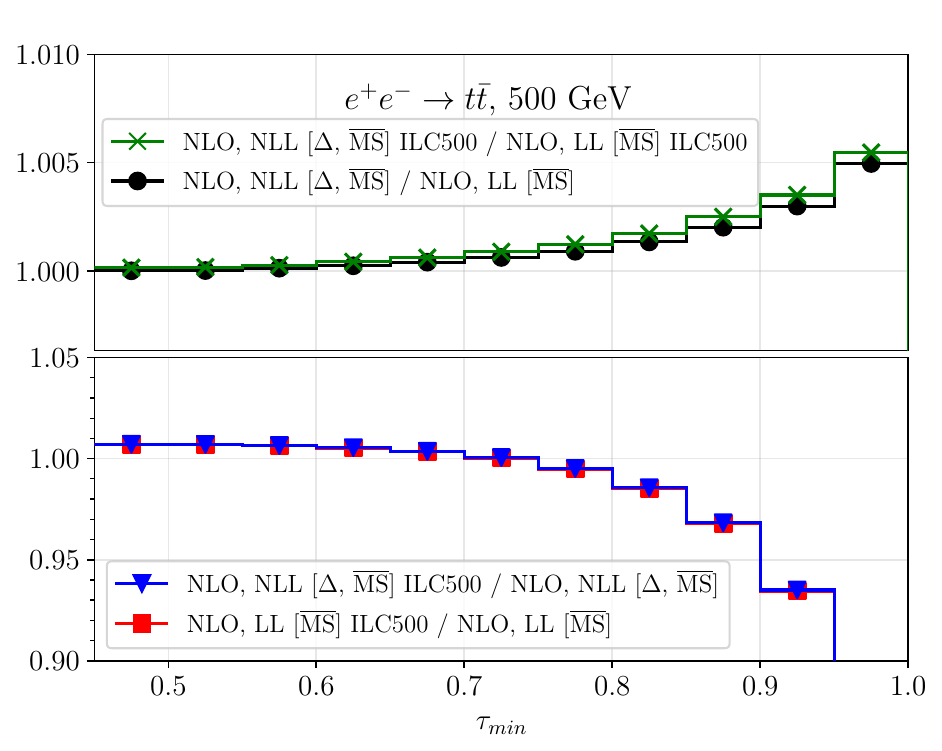}
$\phantom{a}$
  \includegraphics[width=0.47\textwidth]{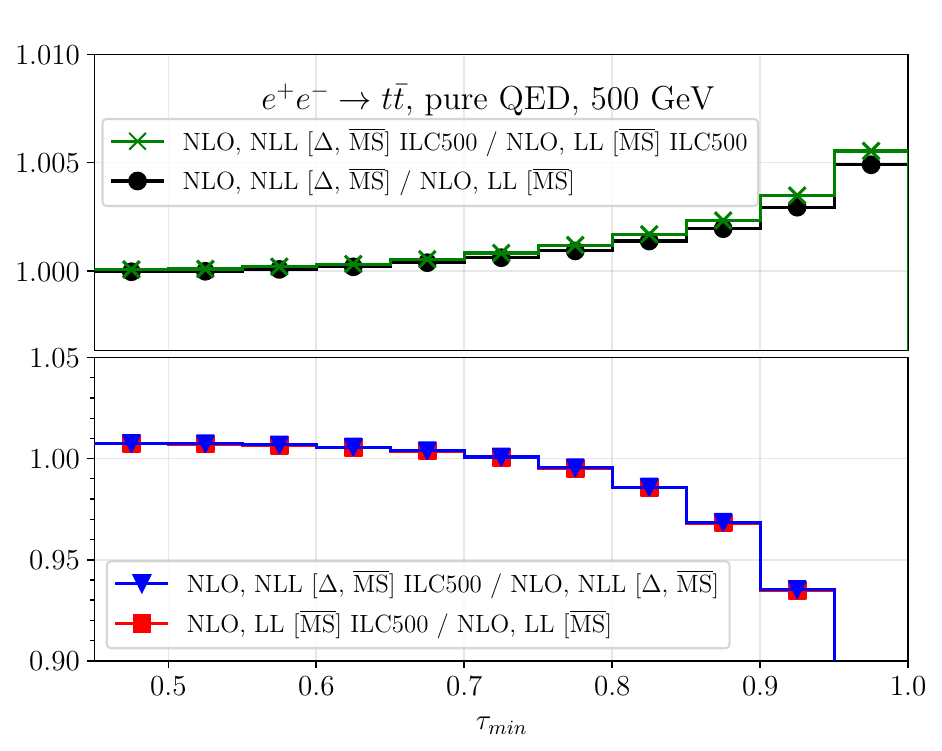}
\caption{\label{fig:beams2} 
As in fig.~\ref{fig:beams1}, for $t\bt$ production in the full SM
(left panel) and in QED (right panel).
}
  \end{center}
\end{figure}

Figure~\ref{fig:beams1} presents the predictions for $q\bq$ 
production (left panel) and $W^+W^-$ production (right panel), whereas 
fig.~\ref{fig:beams2} is relevant to $t\bt$ production, in the full SM 
(left panel) and in QED (right panel); the renormalisation scheme is
chosen to be $\MSb$, and NLO+NLL PDFs are defined in the $\Delta$ 
factorisation scheme. All panels have the same layout, namely: in the 
lower-half frame, we show the ratio of the cross section obtained
by including beamstrahlung effects over that obtained by neglecting them,
in the case of NLO+NLL PDFs (blue curves overlaid with triangles) 
and of LO+LL PDFs (red curves overlaid with boxes). In the upper-half
frame, we show the ratio of the cross section obtained with NLO+NLL
PDFs over that obtained with LO+LL PDFs, with (green curves overlaid 
with crosses) and without (black curves overlaid with circles) beamstrahlung
effects -- thus, the latter curves are exactly the same as those that
appear in figs.~\ref{fig:NLLoLL1} and~\ref{fig:NLLoLL2}, and they are
reported here for ease of comparison with their beamstrahlung-based
counterparts. We can conclude what follows: for the configuration 
considered here (ILC-type), beamstrahlung effects have a clearly visible
impact (up to $\ord(10\%)$), which is observable-dependent and local.
However, they affect in an almost identical manner the predictions stemming
from LO+LL and NLO+NLL PDFs (the red and blue curves are very close to each
other, and so are the black and green ones); this observation is valid for
all of the production processes. The above implies that the conclusions drawn 
in sects.~\ref{sect:resNLL} and~\ref{sect:resfacren} apply to simulations
with beamstrahlung as well.

\section{Conclusions\label{sec:conc}}
This paper builds upon the results for the electron QED PDFs of 
refs.~\cite{Frixione:2019lga,Bertone:2019hks,Frixione:2021wzh},
and it features both theoretical and phenomenology material.
From a theoretical viewpoint, the single-fermion-family UV-$\MSb$ treatment 
of refs.~\cite{Frixione:2019lga,Bertone:2019hks,Frixione:2021wzh} 
is extended to include the evolution with multiple fermion families and 
their mass thresholds, and to give one the possibility to choose among 
three different UV-renormalisation schemes ($\MSb$, $\aem(\mZ)$, and $G_\mu$);
for each of them, it is possible to adopt either the 
$\MSb$~\cite{Bertone:2019hks} or the $\Delta$~\cite{Frixione:2021wzh} 
factorisation scheme. From a phenomenology viewpoint, we have presented 
here for the first time in the literature fixed-order (LO and NLO)
predictions for actual observables based on NLO+NLL PDFs. We have done 
so by exploiting the automated framework of \aNLOs, thus improving the
accuracy of the treatment of ISR in $\epem$ collisions w.r.t.~that available 
in its previous public version~\cite{Frixione:2021zdp}, that was limited
to LO+LL effects. An implication of this improvement is that now
\aNLOs\ can compute NLO EW corrections for processes with massless 
initial-state leptons, which was the only typology of NLO computations
not publicly doable thus far. We also remark that the simulations of
beamstrahlung effects introduced in ref.~\cite{Frixione:2021zdp} 
remains viable in conjunction with NLO+NLL PDFs, and performs numerically
as well as the LO+LL-based one.

The theoretical novelties presented here have different underlying
motivations. The necessity of evolving PDFs with multiple fermion families
stems from a requirement of consistency with the evolution of $\aem$ which,
if performed with a single fermion or without properly accounting for
mass thresholds, would give a poor description of the data. Conversely,
the possibility of choosing among different factorisation and renormalisation
schemes gives one sufficient flexibility for high-energy simulations,
regardless of whether a definite choice is made depending on the 
process and the running conditions under consideration, or whether 
theoretical systematics must be fully assessed. We note that which
of these two options is adopted need not be the same for the factorisation
and renormalisation schemes. In particular, we stress that for increasingly
large energies the usage of fixed-$\aem$ UV schemes such as $\aem(\mZ)$ and 
$G_\mu$ may become questionable.

As far as phenomenology is concerned, our main conclusions are the following.
At the level of observables, the NLL effects in the PDFs have an impact
which is local (i.e.~it depends on both the observable and its kinematical
range), in both shape and size. It is thus impossible to account for it
in a predictive and overall manner in the context of simulations based
on LL-accurate PDFs. The differences between results obtained with different 
factorisation schemes at a given renormalisation scheme are generally much 
smaller than those between results obtained with different renormalisation 
schemes at a given factorisation scheme. For the inclusive observable we
have studied here, the relative factorisation-scheme dependence is of
\mbox{$\ord(10^{-4}-10^{-3})$}; should this figure, and its analogue
for other observables, become comparable to or larger than the expected
experimental accuracy, we note that there is some evidence that the $\Delta$
factorisation scheme is a better choice than $\MSb$, since it has, w.r.t.~the 
latter: a better behaviour in the soft region; a form closer in shape and 
size to the LL PDF (for the electron); and a better numerical stability in
cross section computations. Finally, the approach we follow, based on 
collinear factorisation, renders it straightforward to include contributions
due to partonic processes that are not $\epem$-initiated. As an example,
we have considered the photon-induced contributions to $W^+W^-$
and $t\bt$ production, and found them to be very significant for the former
process at small pair invariant masses. This is yet another example of
an effect whose impact is local and process-dependent; it also shows
why NLL-accurate evolution is to be preferred to an LL one, since it is
only in the former case that electron PDFs associated with non-electron
partons can unambiguously be defined. 

The electron PDFs derived in this paper include all of the ingredients
that are necessary for a systematic study of high-energy $\epem$ 
production processes which is both phenomenologically viable and 
of higher accuracy in its ISR treatment w.r.t.~what has been done
so far in the literature. Theoretically, the only item that remains
to be addressed at the NLL is the possible inclusion of resummed soft 
non-collinear logarithms into the PDFs, which we leave to a future
work. While this is conceptually interesting, we note that since the NLL PDFs
defined here do exponentiate soft logarithms which are also collinear,
when working at the NLO (or at the LO with the prescription introduced
in appendix~\ref{sec:LOxNLL}) the effects not taken into account are
at least of NNLO, and logarithmically subleading.

We finally remark that, together with this paper, we release both
the code that returns the evolved electron PDFs (\eMELA), and a new
public version of \aNLOs, that can be used to reproduce the results
presented here.

\section*{Acknowledgements}
SF is particularly indebted to G.~Degrassi and A.~Vogt for a few 
illuminating discussions. The help of D.~de~Florian, A.~Denner, S.~Forte, 
M.~Greco, V.~Hirschi, E.~Laenen, D.~Pagani, J.~Reuter, H-S.~Shao, and B.~Ward 
at various stages of this work is also gratefully acknowledged.
VB is supported by the European Union's Horizon 2020 research and
innovation programme under grant agreement STRONG 2020 - No 824093.
SF thanks the CERN TH Division for hospitality during the course
of this work.
MZ is supported by the ``Programma per Giovani Ricercatori Rita Levi
Montalcini'' granted by the Italian Ministero dell'Universit\`a e 
della Ricerca (MUR).
XZ is supported by the Italian Ministry of Research (MUR) under grant
PRIN 20172LNEEZ.

\appendix
\section{NLO cross sections with LO+LL PDFs\label{sec:NLOxLL}}
Let us suppose that the parton-level cross sections\footnote{Particle
and parton indices are understood throughout this appendix, in order
to have a leaner notation; they play obvious roles, and the reader
will have no difficulty in reinstating them if necessary.} $d\hsig$ that
enter eq.~(\ref{master0}) are computed at $\ord(\aem^{b+p})$; with this,
we understand that their Born contributions are of $\ord(\aem^b)$, and the 
overall perturbative accuracy is thus N$^p$LO. If the PDFs which such
cross sections are convoluted with in eq.~(\ref{master0}) are
N$^q$LO accurate (the logarithmic accuracy is irrelevant in this
argument, and is omitted), then the accuracy of the predicted
particle-level cross sections $d\sigma$ is N$^{\min(p,q)}$LO,
which follows trivially from a series expansion of the r.h.s.~of 
eq.~(\ref{master0}).

This is inconvenient, since typically $q<p$ -- in particular, all of
the phenomenological predictions published so far, with the exception
of the NLO+NLL ones presented in this paper, have used LO-accurate
PDFs (i.e.~$q=0$). This issue can be addressed by supplementing the 
partonic cross sections with a compensating contribution that
features terms of \mbox{$\ord(\aem^{q+1},\ldots\aem^p)$}, 
constructed (by using the perturbative expansion of the PDFs)
so that the short-distance N$^p$LO accuracy is preserved
by the convolution integral. 

In this appendix we consider the case of NLO cross sections ($p=1$)
and LO PDFs ($q=0$). We show that the compensating contribution can
be written in a universal (i.e.~process- and observable-independent) 
manner, and that it can be viewed (although improperly) as defining
a PDF-specific factorisation scheme. In order to do that, we re-write
eq.~(\ref{master0}) symbolically as follows:
\beq
d\sigma=\Gamma^{(K)}\star\Gamma^{(K)}\star d\hsig^{(K)}\,,
\label{factsimp}
\eeq
where both the partonic cross sections and the PDFs are NLO-accurate for
the time being. On the r.h.s.~of eq.~(\ref{factsimp}) we have included 
an upper index $(K)$ to remind one explicitly that the corresponding
quantities are factorisation-scheme dependent; order-by-order in
perturbation theory, particle-level cross sections are factorisation-scheme 
independent, hence no index $(K)$ appears on the l.h.s.. Using the
conventions of eq.~(\ref{PDFex}) for the perturbative coefficients of
the series expansion of the PDFs, and their analogues for the cross 
sections, eq.~(\ref{factsimp}) leads to:
\beqn
d\sigma^{[0]}&=&d\hsig^{[0]}\,,
\label{fct0}
\\
d\sigma^{[1]}&=&d\hsig^{(K)[1]}+d\delta^{(K)[1]}\,,
\label{fct1}
\\
d\delta^{(K)[1]}&=&\Gamma^{(K)[1]}\star\Gamma^{[0]}\star d\hsig^{[0]}+
\Gamma^{[0]}\star\Gamma^{(K)[1]}\star d\hsig^{[0]}
\label{fctdel}
\\*&=&
\Gamma^{(K)[1]}\star d\hsig^{[0]}+
\Gamma^{(K)[1]}\star d\hsig^{[0]}\,.
\label{fctdel2}
\eeqn
Equation~(\ref{fctdel2}) follows from eq.~(\ref{fctdel}) because of
eq.~(\ref{G0sol}). Furthermore, $\Gamma^{(K)[1]}$ is given by the
r.h.s.~of eq.~(\ref{G1sol2}) (for the electron, which is the only
relevant case since we shall eventually focus on LO PDFs) with $\muz\to\mu$ 
there. By using the explicit formulae for the partonic cross sections of 
the FKS formalism~\cite{Frixione:1995ms,Frixione:1997np} one has:
\beqn
d\hsig^{(K)[1]}&=&d\hsig^{(K=0)[1]}-d\kappa^{[1]}\,,
\label{dhsKvsMSb}
\\
d\kappa^{[1]}&=&
K\star d\hsig^{[0]}+
K\star d\hsig^{[0]}\,.
\label{dkappa}
\eeqn
This shows that the l.h.s.~of eq.~(\ref{fct1}) is indeed 
factorisation-scheme independent, since from eq.~(\ref{G1sol2}):
\beq
\Gamma^{(K)[1]}=\Gamma^{(K=0)[1]}+K\,.
\eeq
We remind the reader that by setting $K=0$ one works in the $\MSb$
factorisation scheme.

If we now convolute the NLO partonic cross sections with LO PDFs $\GammaLO$,
i.e.~we write
\beq
d\sigma=\GammaLO\star\GammaLO\star d\hsig^{(K)}\,,
\label{factsimpLO}
\eeq
instead of eq.~(\ref{factsimp}), the analogues of eqs.~(\ref{fct1})
and~(\ref{fctdel2}) read:
\beqn
d\sigma^{[1]}&=&d\hsig^{(K)[1]}+d\delta_{\rm LO}^{[1]}\,,
\label{fct1LO}
\\
d\delta_{\rm LO}^{[1]}&=&
\GammaLO^{[1]}\star d\hsig^{[0]}+
\GammaLO^{[1]}\star d\hsig^{[0]}\,,
\label{fctdel2LO}
\eeqn
respectively. It is clear that, in general, the l.h.s.'s of 
eqs.~(\ref{fct1}) and~(\ref{fct1LO}) are not equal to each other.
However, one can {\em impose} them to be so, by equating their 
r.h.s.'s thus:
\beq
d\hsig^{(K=0)[1]}+d\delta^{(K=0)[1]}=
d\hsig^{(K)[1]}+d\delta_{\rm LO}^{[1]}\,,
\label{KeqLL}
\eeq
where we have exploited the factorisation-scheme independence of
eq.~(\ref{fct1}) to choose $K=0$ there. Equation~(\ref{KeqLL}) is
then solved for $K$ (we denote the solution by $K_{\rm LO}$), leading to:
\beq
K_{\rm LO}=\GammaLO^{[1]}-\Gamma^{(K=0)[1]}\,.
\label{Kcond}
\eeq
When this function is used in eq.~(\ref{dkappa}) (i.e.~by
setting $K=K_{\rm LO}$ there), $d\kappa^{[1]}$ plays
the role of the compensating contribution to the partonic cross sections
(eq.~(\ref{dhsKvsMSb})) which has been introduced in the discussion at
the beginning of this appendix. We point out that the choice of the
$\MSb$ factorisation scheme in the second term on the r.h.s.~of 
eq.~(\ref{Kcond}) is dictated by simplicity. Still, another scheme 
could be chosen, but this would entail using it in the first term
on the r.h.s.~of eq.~(\ref{dhsKvsMSb}), since the property of scheme
independence of the final result must be preserved.

Lest eq.~(\ref{Kcond}) generate some misunderstanding, we stress that
with LO PDFs the definition of a factorisation scheme does not make sense.
However, the framework provided by the scheme-change functions $K$ in 
the context of the FKS subtraction is very convenient for computing 
the compensating contribution that allows one to obtain NLO-accurate
particle cross sections.

In order to be explicit, we now compute the $K_{\rm LO}$ functions
for different LO PDFs. We write the generic functional form of the
latter as follows:
\beq
\GammaLO(z)=
\frac{\exp\left(3\beta_S/4-\gamma_E\beta_E\right)}
{\Gamma\left(1+\beta_E\right)}\,
\beta_E(1-z)^{\beta_E-1}
-\half\beta_H(1+z)
+\ord(\aem^2)\,.
\label{GLLsol}
\eeq
We point out that both the $\ord(\aem^2)$ and $\ord(\aem^3)$ terms on the 
r.h.s.~of eq.~(\ref{GLLsol}) are explicitly known~\cite{Skrzypek:1990qs,
Skrzypek:1992vk,Cacciari:1992pz,Bertone:2019hks}, but are not necessary
to obtain the results that follow. Any choice of the parameters $\beta_E$, 
$\beta_S$, and $\beta_H$ in eq.~(\ref{GLLsol}) is customarily (and 
unfortunately) called a ``scheme''; here, we shall consider the following 
ones:
\begin{itemize}
\item Beta scheme:
\beq
\beta_E=\beta_S=\beta_H=\ee^2\beta\,.
\label{bschdef}
\eeq
\item Eta scheme:
\beq
\beta_E=\beta_S=\ee^2\beta\,,\;\;\;\;\;\;\;\;
\beta_H=\ee^2\eta\,.
\label{eschdef}
\eeq
\item Mixed scheme:
\beq
\beta_E=\ee^2\beta\,,\;\;\;\;\;\;\;\;
\beta_S=\beta_H=\ee^2\eta\,.
\label{mschdef}
\eeq
\item Collinear scheme:
\beq
\beta_E=\beta_S=\beta_H=\ee^2\eta_0\,.
\label{collschdef}
\eeq
\item Running scheme:
\beq
\beta_E=\beta_S=\beta_H=2\ee^2 t\,.
\label{rschdef}
\eeq
\end{itemize}
We have used the quantities:
\beqn
&&\eta=\frac{\aem}{\pi}\log\frac{\mu^2}{m^2}\,,\;\;\;\;\;\;\;\;
\beta=\frac{\aem}{\pi}\left(\log\frac{\mu^2}{m^2}-1\right)\,,
\label{Leb}
\\*
&&\eta_0=\frac{\aem}{\pi}\log\frac{\mu^2}{\mu_0^2}\,,\;\;\;\;\;\;\;\;
t=\frac{1}{2\pi b_0}\log\frac{\aem(\mu)}{\aem(\mu_0)}=
\frac{\aem}{2\pi}\,\log\frac{\mu^2}{\mu_0^2}+
\ord(\aem^2)\,.
\label{Leb2}
\eeqn
The beta, eta, and mixed schemes are by now standard (see e.g.~appendix A.1 
of ref.~\cite{Beenakker:1996kt}), while what we have called here collinear 
and running schemes are less so -- we use them in the forms introduced in
ref.~\cite{Bertone:2019hks}, given there by eqs.~(5.46) and (5.66), 
respectively, extended as is explained in sects.~\ref{sec:evol} 
and~\ref{sec:UV} to account for multiple fermion families and their
thresholds (for previous single-fermion LL-accurate solutions with running 
$\aem$, see e.g.~ref.~\cite{Gribov:1972ri,Dokshitzer:1977sg,Nicrosini:1986sm}). 
Note that in the latter two schemes we have kept the dependence on the 
starting scale $\mu_0$, rather than setting $\mu_0=m$ as was done for the 
phenomenological applications of sect.~\ref{sec:res}. In a strict LO+LL 
evolution, the $\mu_0$ dependence is beyond accuracy, but when convoluting 
with NLO cross sections some care is required, as we shall soon see.

We express the results for the $K_{\rm LO}$ functions in the various LO
schemes considered in eqs.~(\ref{bschdef})--(\ref{rschdef}) by using the 
following functional form (which is the most general at this perturbative 
order):
\beq
K_{\rm LO}(z)/\ee^2=K_\delta\delta(1-z)+K_+(z)\pdistr{1-z}{+}
+K_{\rm L}(z)\lppdistr{1-z}{+}+K_{reg}(z)\,.
\label{Kcan}
\eeq
The explicit computation of the r.h.s.~of eq.~(\ref{Kcond}) leads then
to the following results:
\begin{itemize}
\item Beta scheme:
\beq
K_\delta=-\frac{7}{2}\,,\;\;\;
K_+(z)=0\,,\;\;\;
K_{\rm L}(z)=2(1+z^2)\,,\;\;\;
K_{reg}(z)=0\,.
\label{Ksolb}
\eeq
\item Eta scheme:
\beq
K_\delta=-\frac{7}{2}\,,\;\;\;
K_+(z)=0\,,\;\;\;
K_{\rm L}(z)=2(1+z^2)\,,\;\;\;
K_{reg}(z)=-(1+z)\,.
\label{Ksole}
\eeq
\item Mixed scheme:
\beq
K_\delta=-2\,,\;\;\;
K_+(z)=0\,,\;\;\;
K_{\rm L}(z)=2(1+z^2)\,,\;\;\;
K_{reg}(z)=-(1+z)\,.
\label{Ksolm}
\eeq
\item Collinear scheme:
\beq
K_\delta=-2-\frac{3}{2}L_0\,,\;\;\;
K_+(z)=(1-L_0)(1+z^2)\,,\;\;\;
K_{\rm L}(z)=2(1+z^2)\,,\;\;\;
K_{reg}(z)=0\,.
\label{Ksolc}
\eeq
\item Running scheme:
\beq
{\rm as~in~eq~(\ref{Ksolc})}\,.
\label{Ksolr}
\eeq
\end{itemize}
Note that the results of eqs.~(\ref{Ksolb})--(\ref{Ksolr}) are specific 
to the $\MSb$ scheme (since eqs.~(\ref{dhsKvsMSb}) and~(\ref{Kcond}) 
use $K=0$), and must thus be used in conjunction not only with the 
appropriate LO+LL PDFs, but also with NLO short distance cross sections 
calculated in $\MSb$. Equivalent results appropriate for other 
factorisation schemes (i.e.~with $K\ne 0$) can easily be derived.

In eq.~(\ref{Ksolc}) we have defined:
\beq
L_0=\log\frac{\mu_0^2}{m^2}\,.
\label{Lzdef}
\eeq
The dependence on $\mu_0$ in the $K_{\rm LO}$ function defined by
eq.~(\ref{Ksolc}) cancels at relative $\ord(\aem)$ against that
of $\GammaLO$ in the collinear and running schemes, so that the
NLO particle cross section remains independent of $\mu_0$, as
is expected.

\section{LO cross sections with (N)LO+(N)LL PDFs\label{sec:LOxNLL}}
In appendix~\ref{sec:NLOxLL} we have shown how LO+LL PDFs must be
used in conjunction with NLO short-distance cross sections without 
spoiling the accuracy of the latter. Here we consider the complementary
question, namely: given LO-accurate cross sections (i.e.~of $\ord(\aem^b)$), 
which PDFs should they be convoluted with (i.e.~only LO+LL, or NLO+NLL as 
well)? We note that this is not an academic question: the vast majority of 
simulations relevant to BSM physics at future $\epem$ colliders still rely 
on LO cross sections.

From a formal viewpoint, at the level of observables the accuracy of
predictions stemming from LO cross sections is LO, irrespective of 
whether LO+LL or NLO+NLL PDFs are used. However, numerically
the choice of PDFs can have a much larger impact than simple perturbative
considerations suggest. In order to see this, let us first make the
obvious observation that in LO-based simulations all contributions
factor out the $\ord(\aem^b)$ Born cross section\footnote{With some
abuse of language, we shall say that such contributions are ``proportional''
to the Born, understanding a convolution.}. In a proper NLO-accurate
computation, contributions proportional to the Born emerge from three
different sources: 
\begin{itemize}
\item[{\em a)}] the $\ord(\aem)$ term in the expansion of the PDFs; 
\item[{\em b)}] the degenerate $(n+1)$-body contributions to the 
short-distance cross sections; 
\item[{\em c)}] the virtual, soft-, and collinear-reminder contributions 
to the  short-distance cross sections.
\end{itemize}
We point out that, while items {\em b)} and {\em c)} are formulated in
terms of quantities that appear in the FKS subtraction formalism, their
analogues exist in any scheme that allows one to compute NLO results.
Leaving aside the quantities in {\em c)} for the time being, and noting
that we are only interested in the electron channel (since we shall 
eventually work at the LO), the kernel relevant to item {\em b)} can be
read e.g.~from eq.~(4.88) of ref.~\cite{Frixione:2019lga}, and re-written
as follows:
\beqn
{\cal K}_{ee}/\ee^2&=&\left[\frac{1+z^2}{1-z}\left(
\log\frac{s}{\mu^2}+2\log(1-z)\right)\right]_+
+(1-z)
\nonumber\\*&&
+\left(\frac{7}{2}-\frac{3}{2}\log\frac{s}{\mu^2}\right)\delta(1-z)
-K_{ee}(z)\,.
\label{calKFKS}
\eeqn
Adding the r.h.s.~of eq.~(\ref{calKFKS}) to the $\ord(\aem)$ term in 
the expansion of the PDFs (item {\em a)} above) leads to:
\beqn
&&\left(\Gamma_e^{(K)[1]}+{\cal K}_{ee}\right)/\ee^2=
\left(\GammaLO^{[1]}+{\cal K}_{ee}\right)/\ee^2=
\nonumber\\*&&\phantom{aaa}
\left(\frac{1+z^2}{1-z}\right)_+\left(\log\frac{s}{m^2}-1\right)
+(1-z)
+\left(\frac{7}{2}-\frac{3}{2}\log\frac{s}{\mu^2}\right)\delta(1-z)\,.
\label{tmp16}
\eeqn
A few comments on eq.~(\ref{tmp16}) are in order. Firstly, as the notation
suggests this result is relevant to both NLO+NLL PDFs and LO+LL PDFs. For
this to happen, it is crucial that at the LO the $K_{\rm LO}$ functions 
derived in appendix~\ref{sec:NLOxLL} be used in \mbox{${\cal K}_{ee}$};
conversely, with NLO+NLL PDFs the dependence 
on the $K$ functions (i.e., on the factorisation scheme)
disappears, as it should by construction. Secondly, at both the LO and 
the NLO the double-logarithmic term (present in both ${\cal K}_{ee}$ and
the $\ord(\aem)$ term in the PDFs) also drops out; this is consistent
with the usage of the PDFs in the factorisation theorem (eq.~(\ref{master0})),
whose l.h.s.~cannot have double-logarithmic terms of this kind given
its nature of a massive-electron cross section. Thirdly, the term
proportional to $\delta(1-z)$ has kinematically the same form as those
from item {\em c)} above, and will naturally combine with them; in particular,
this will imply the cancellation of the dependence upon $\mu$.

We can now go back to the original problem posed in this appendix, namely
that of the convolution of the PDFs with LO cross sections. In this case, 
only the contribution due to {\em a)} is relevant; with NLO+NLL PDFs, this 
implies a dependence of the result on the factorisation scheme (through the
$K$ functions) which, among other things, may induce an $\ord(\aem^{b+1})$ 
double-logarithmic term of the kind mentioned above (to be specific,
this happens with $\MSb$ PDFs, but does {\em not} happen with those
defined in the $\Delta$ scheme). Equation~(\ref{tmp16}) then suggests 
a universal (i.e., process-independent) way to address this issue. The
idea is the following: when performing an LO-based computation, one adds an
$\ord(\aem^{b+1})$ term that has the same form as the degenerate $(n+1)$-body 
contribution; such a term is defined by means of the following kernel:
\beq
{\cal K}_{{\rm LO},ee}/\ee^2=\left(\frac{1+z^2}{1-z}\right)_+
\log\frac{s}{\mu^2}+K_{ee}^{{\rm ref}}(z)-K_{ee}(z)\,.
\label{calKLO}
\eeq
Here, $K_{ee}^{{\rm ref}}$ is a scheme-change function associated with
a fictitious ``reference'' scheme, that we determine as follows.
Firstly, we require that when such a function is used in eq.~(\ref{calKFKS})
it cancels exactly all of the $\mu$-independent terms for $z<1$. This gives:
\beq
K_{ee}^{{\rm ref}}(z)=2\,\frac{1+z^2}{1-z}\log(1-z)+(1-z)\,,
\;\;\;\;\;\;\;
z<1\,.
\label{Krefmo}
\eeq
Secondly, we extend the result of eq.~(\ref{Krefmo}) to all $z$'s
by adding an endpoint contribution proportional to $\delta(1-z)$, whose 
coefficient we compute by turning the functions in eq.~(\ref{Krefmo})
into plus distributions, and by requiring that the integral in \mbox{$[0,1]$}
of $K_{ee}^{{\rm ref}}(z)$ vanishes (this is equivalent to imposing a 
charge-conservation condition, see ref.~\cite{Frixione:2019lga}). Thus:
\beq
K_{ee}^{{\rm ref}}(z)=2\left(\frac{1+z^2}{1-z}\log(1-z)\right)_+ +(1-z)
-\half\,\delta(1-z)\,.
\label{Kref}
\eeq
Therefore, the inclusion of the term induced by the kernel of 
eq.~(\ref{calKLO}) is so that at $\ord(\aem^{b+1})$ in an LO-based
calculation one obtains a contribution:
\beqn
&&\left(\Gamma_e^{(K)[1]}+{\cal K}_{{\rm LO},ee}\right)/\ee^2=
\left(\GammaLO^{[1]}+{\cal K}_{{\rm LO},ee}\right)/\ee^2=
\nonumber\\*&&\phantom{aaa}
\left(\frac{1+z^2}{1-z}\right)_+\left(\log\frac{s}{m^2}-1\right)
+(1-z)-\half\,\delta(1-z)\,.
\label{KKfacLO}
\eeqn
For comparison, e.g.~in the Beta scheme one has:
\beq
\GammaLO^{[1]}/\ee^2=
\left(\frac{1+z^2}{1-z}\right)_+\left(\log\frac{\mu^2}{m^2}-1\right).
\label{GLO1beta}
\eeq
Equation~(\ref{KKfacLO}) shows that by following the procedure advocated 
here in the convolution of the LO cross sections with NLO+NLL PDFs there 
are no residual double-logarithmic terms at $\ord(\aem^{b+1})$, and 
no factorisation-scheme dependence. Conversely, when LO+LL PDFs are
employed, the behaviour is quite analogous to that of a standard
LO computation, with differences emerging solely from the inclusion of
process-independent contributions (which also imply that the dependence
upon the scale $\mu$ is of $\ord(\aem^{b+2})$ and not of $\ord(\aem^{b+1})$ 
as in eq.~(\ref{GLO1beta})\footnote{This happens because the $\mu$ dependence
of the non-$\delta(1-z)$ terms of eq.~(\ref{calKLO}) is the same as that
of eq.~(\ref{calKFKS}). As far as the $\delta(1-z)$ terms are concerned,
it has already been observed that the $\mu$ dependence in eq.~(\ref{calKFKS})
is cancelled by that of the soft and collinear reminders. Although these are 
universal, for simplicity we have chosen to neglect them, and therefore we 
did not include a $\mu$-dependent $\delta(1-z)$ term in eq.~(\ref{calKLO}),
bypassing the problem by imposing a charge-conservation condition.}).

Clearly, there is ample freedom in the choice of the kernel of
eq.~(\ref{calKLO}). We stress again that the contribution it induces
in the cross section is of $\ord(\aem^{b+1})$, and therefore formally beyond 
accuracy in the context of an LO-accurate computation. However, this
procedure has at least a couple of positive features. Firstly, it prevents
numerical results from becoming pathological, so that the coefficients
of the perturbative expansion are well-behaved regardless of whether
one employs LO+LL or NLO+NLL PDFs. And secondly, by making universal terms
of $\ord(\aem^{b+1})$ independent of the PDF choice it emphasises the role
of matrix elements (as opposed to PDFs) in the comparison between LO and
NLO results. In other words, in this way one expects K-factors computed by 
including the convolution with PDFs to be quite similar to those obtained 
without such a convolution.

\begin{figure}[t!]
  \begin{center}
  \includegraphics[width=0.47\textwidth]{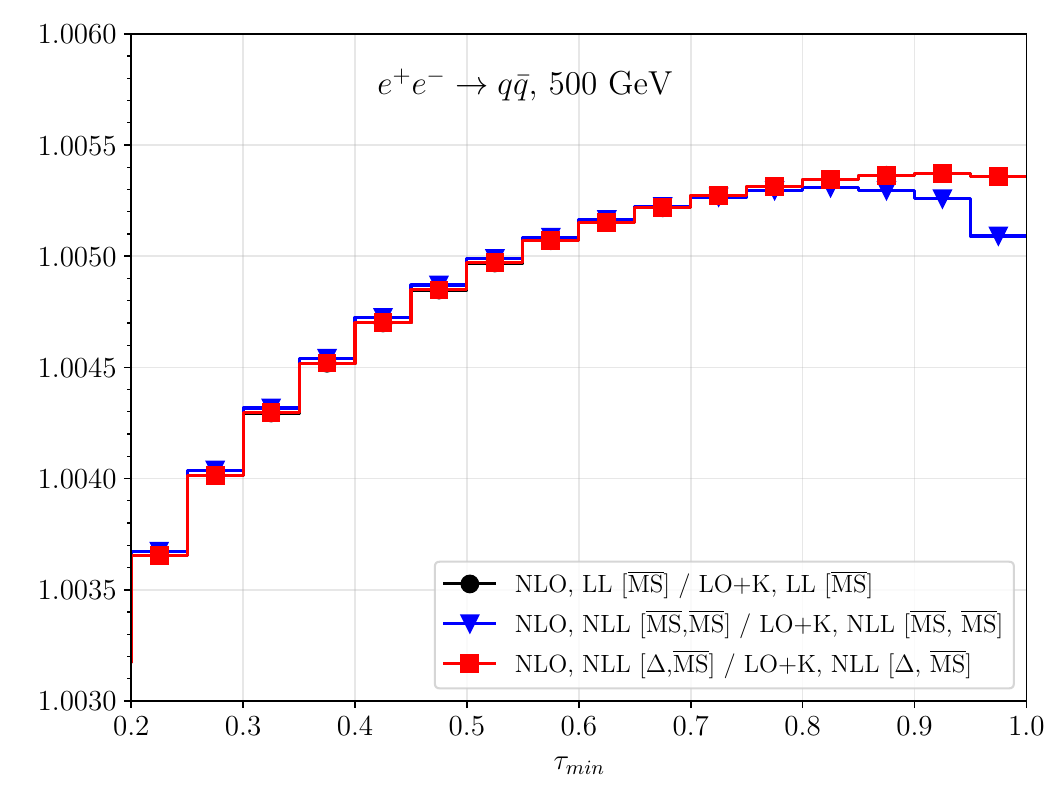}
$\phantom{a}$
  \includegraphics[width=0.47\textwidth]{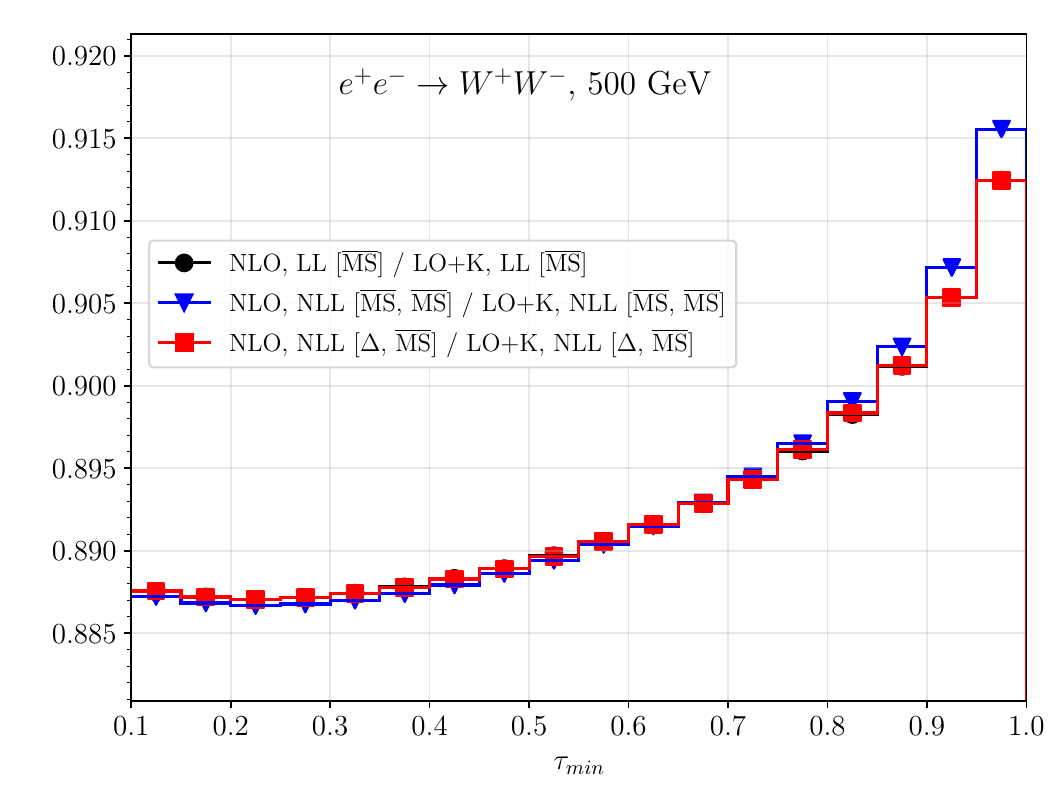}
\caption{\label{fig:Kfac1} 
K-factors computed by including the contribution stemming from
eq.~(\ref{calKLO}) in the LO cross sections, in the $\MSb$ renormalisation
scheme for both NLL PDFs (in the $\MSb$ and $\Delta$ factorisation schemes)
and LL PDFs.
Left panel: $q\bq$ production; right panel: $W^+W^-$ production.
}
  \end{center}
\end{figure}
\begin{figure}[t!]
  \begin{center}
  \includegraphics[width=0.47\textwidth]{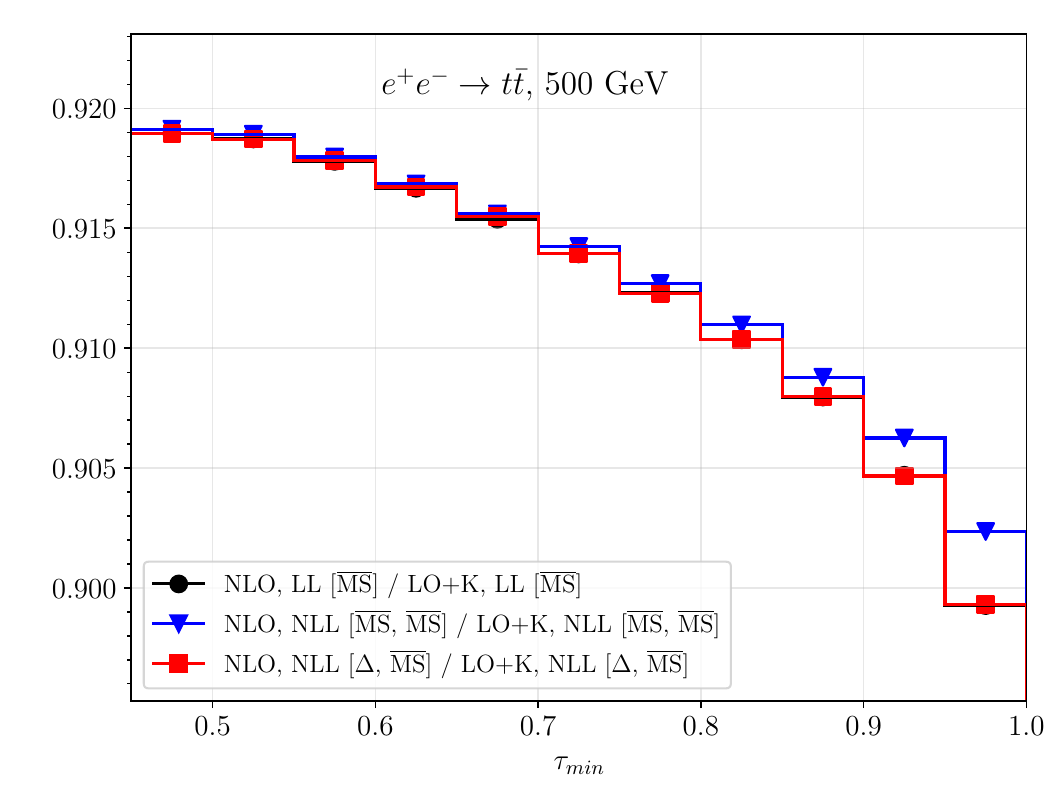}
$\phantom{a}$
  \includegraphics[width=0.47\textwidth]{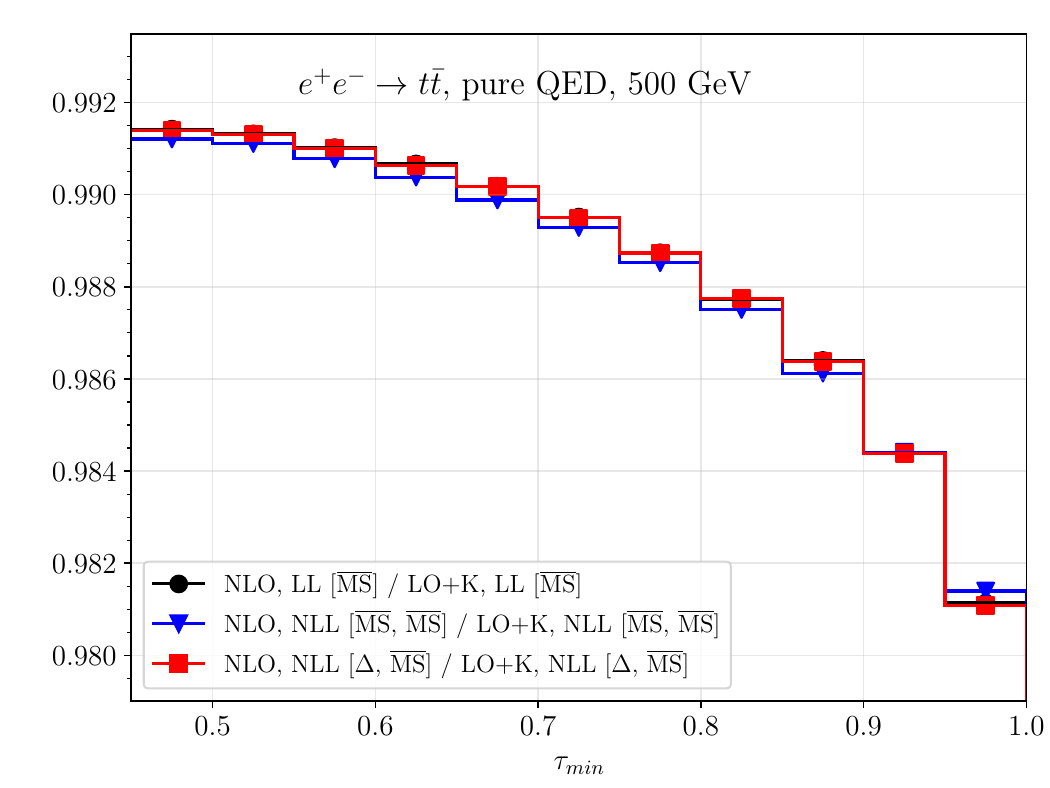}
\caption{\label{fig:Kfac2} 
As in fig.~\ref{fig:Kfac1}, for $t\bt$ production in the full SM
(left panel) and in QED (right panel).
}
  \end{center}
\end{figure}
\begin{figure}[t!]
  \begin{center}
  \includegraphics[width=0.47\textwidth]{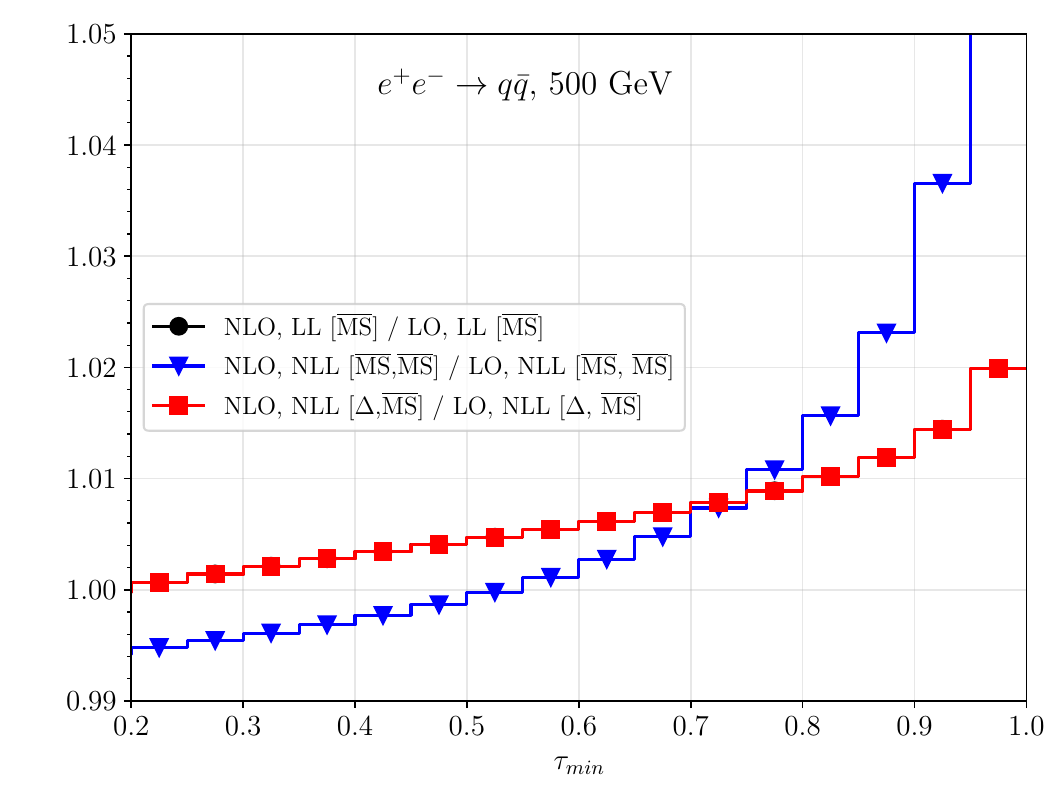}
$\phantom{a}$
  \includegraphics[width=0.47\textwidth]{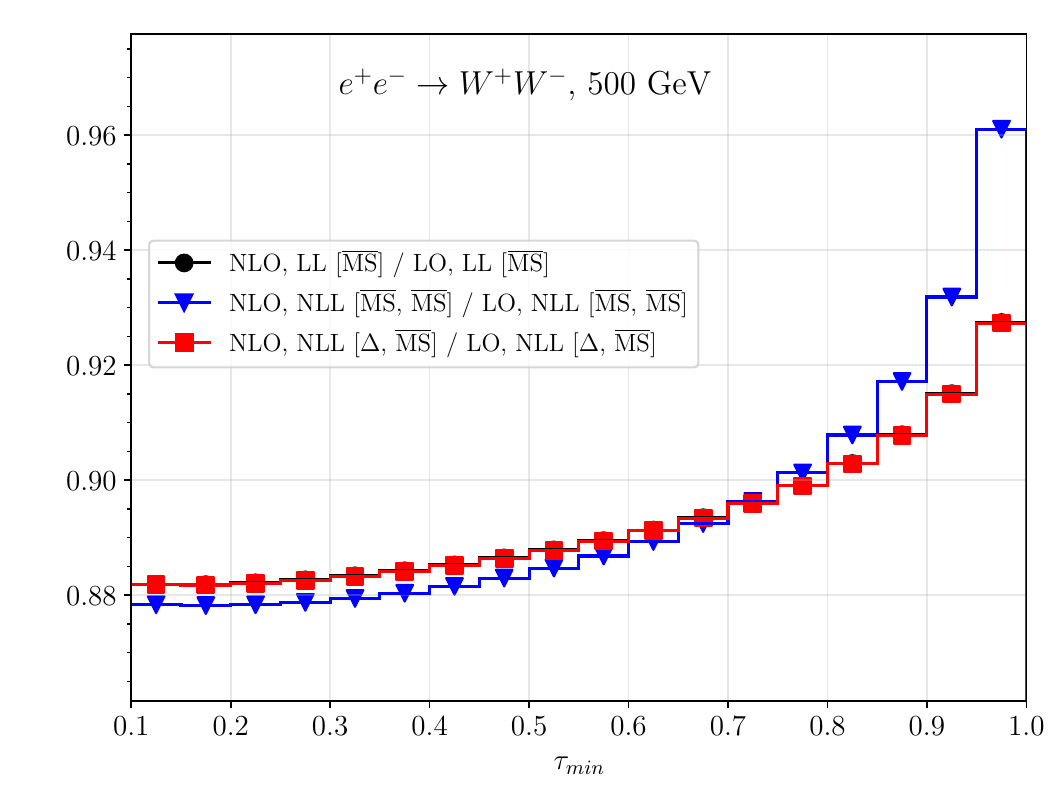}
\caption{\label{fig:Kfac1s} 
As in fig.~\ref{fig:Kfac1}, with K-factors computed by {\em not} including 
the contribution stemming from eq.~(\ref{calKLO}) in the LO cross sections.
Left panel: $q\bq$ production; right panel: $W^+W^-$ production.
}
  \end{center}
\end{figure}
\begin{figure}[t!]
  \begin{center}
  \includegraphics[width=0.47\textwidth]{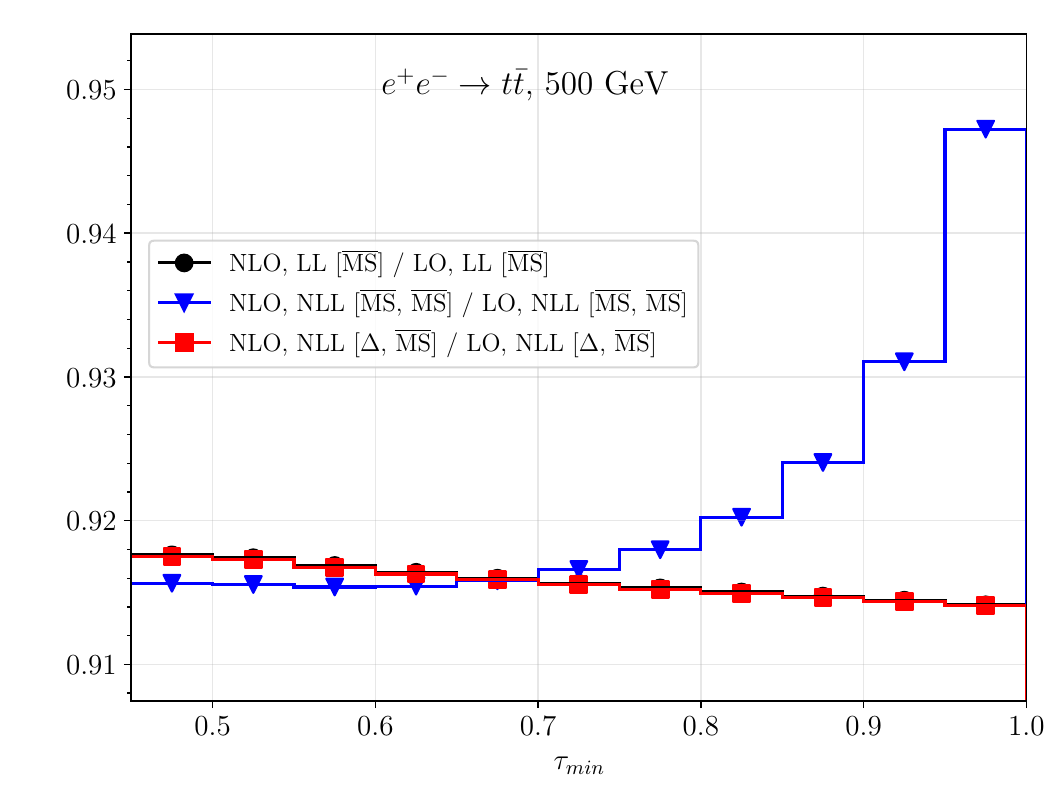}
$\phantom{a}$
  \includegraphics[width=0.47\textwidth]{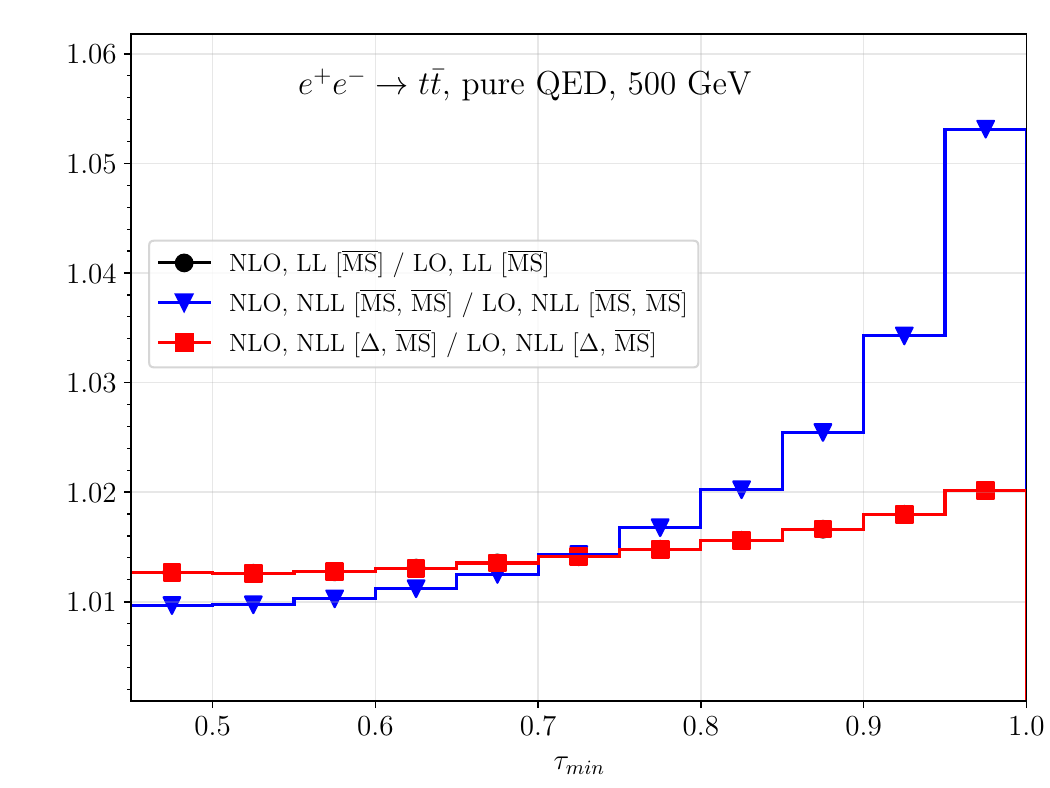}
\caption{\label{fig:Kfac2s} 
As in fig.~\ref{fig:Kfac1s}, for $t\bt$ production in the full SM
(left panel) and in QED (right panel).
}
  \end{center}
\end{figure}

In order to see the impact of the procedure for the computation of
LO-accurate cross sections discussed here, we present the results for 
the K-factors obtained by employing LO+LL (black curves overlaid with circles)
and NLO+NLL PDFs (in the $\MSb$ (blue curves overlaid with triangles)
and $\Delta$ (red curves overlaid with boxes) schemes) {\em with} 
(figs.~\ref{fig:Kfac1} and~\ref{fig:Kfac2}) and {\em without} 
(figs.~\ref{fig:Kfac1s} and~\ref{fig:Kfac2s}) the inclusion of the 
contribution stemming from eq.~(\ref{calKLO}). In order to be definite,
we restrict ourselves to working with the $\MSb$ renormalisation scheme.

Before proceeding to commenting the results, we make two observations.
Firstly, significant differences between K-factors computed with different
prescriptions and/or choices of factorisation scheme are predominantly 
induced by the denominators (i.e.~by the LO cross sections), since we have 
seen in sects.~\ref{sect:resNLL} and~\ref{sect:resfacren} that NLO results 
are all relatively close to each other under the same conditions. Secondly,
regardless of their definitions K-factors remain unphysical quantities,
that cannot be used to draw conclusions about physics issues.

The main implication of figs.~\ref{fig:Kfac1}--\ref{fig:Kfac2s} is
that by using eq.~(\ref{calKLO}) the K-factors computed with the
three different PDF choices are close to each other, while with the
standard definition those obtained by using NLO+NLL $\MSb$-defined PDFs
are clear outliers. The analytical formulae presented above show that
this is due to the presence, in the LO-accurate results computed in the
standard manner with such PDFs, of double-logarithmic terms that must not
feature at observable level on the l.h.s.~of the factorisation formula,
and do so on the r.h.s.~owing only to their being beyond accuracy. Conversely,
and regardless of the specific definition adopted for the K-factor, results 
obtained with LO+LL PDFs are quite similar to those obtained with NLO+NLL 
$\Delta$-defined PDFs (in fact, with the $y$ ranges used in the plots,
they are essentially indistinguishable from one another). This is consistent 
with the observation already made about fig.~\ref{fig:PDFrat}, that the 
electron densities stemming from these two PDFs are quite similar in shape 
and size.

We can further observe that, in the case of LO+LL PDFs and NLO+NLL ones with
the $\Delta$ factorisation scheme, K-factors computed with eq.~(\ref{calKLO})
or without it are close to each other, with the quality of such an agreement
decreasing with $\tau_{min}$. This is not surprising, in view of the fact 
that K-factors are meant to quantify inclusive properties, and the level 
of inclusiveness decreases with $\tau_{min}$ (to the extent that at
$\tau_{min}\simeq 1$ one exposes the presence of soft logarithms).
Furthermore, the K-factors relevant to the processes computed in QED
are closer to one than their full-SM counterparts -- this is the origin
of the behaviour underlined in sect.~\ref{sect:resfacren}, namely that
the former processes have a smaller factorisation-scheme dependence
w.r.t.~the latter ones.

\section{NLO EW corrections with massless initial-state leptons
\label{sec:NLOxsec}}
In this appendix, we document the changes to the \aNLOs\ code that have
occurred for the computation of NLO EW corrections at lepton colliders
to become possible. We remind the reader that \aNLOs\ has originally
automated NLO QCD corrections~\cite{Alwall:2014hca}; this has later
been extended~\cite{Frederix:2018nkq} to the case of NLO EW (and mixed,
i.e.~the simultaneous perturbative expansion in two coupling constants)
corrections. Technically, the NLO implementations of 
refs.~\cite{Alwall:2014hca,Frederix:2018nkq} are based on the assumption 
that the initial-state partons that initiate the hard process are not
monochromatic, but rather have non-trivial energy spectra, which are 
given by the PDFs. In view of the fact that the emphasis of
refs.~\cite{Alwall:2014hca,Frederix:2018nkq} is hadronic physics,
such PDFs has been taken so far to be the hadronic ones. Thus, in order
to extend the NLO-EW-correction capabilities of \aNLOs\ to $\epem$
collisions, a precondition is the implementation of electron PDFs,
which pose a challenging numerical problem in view of their
integrable-singularity behaviour at $z\to 1$. This issue has been
addressed in ref.~\cite{Frixione:2021zdp}, whose solution is however
limited to cross sections which are LO-accurate in $\aem$.

In summary: given what has been achieved for EW corrections in 
ref.~\cite{Frederix:2018nkq} (complete automation of NLO matrix elements,
and of NLO cross sections in hadronic collisions) and 
ref.~\cite{Frixione:2021zdp} (inclusion in \aNLOs\ of electron
PDFs), in the present work we have addressed the only remaining missing
item, namely the automation of the computations of NLO-EW cross sections
in the presence of electron PDFs. Here, the key issue (which sets this
case apart from its LO-EW counterpart solved in ref.~\cite{Frixione:2021zdp})
is the efficiency of the phase-space integration of locally-subtracted 
real-emission matrix elements, which is severely degraded in the presence 
of electron PDFs, and specifically because of that of the electron/positron 
(i.e.~$\PDF{\lpm}{\lpm}$).

The nature of the problem is the following: in \aNLOs\, the cancellation 
of soft and collinear divergences relies on the FKS subtraction 
formalism~\cite{Frixione:1995ms,Frixione:1997np} and on its automation
(see ref.~\cite{Frederix:2009yq}). FKS first achieves a simplification
of the singularity structure of the matrix elements by effectively 
partitioning the phase space into sectors, identified by two parton indices 
$i$ (the FKS parton) and $j$ (its sister), so that in each of them at most 
one soft (when $E_i\to 0$) and one collinear (when $\theta_{ij}\to 0$)
singularity occurs. If parton $j$ belongs to the initial state, the 
way in which the real-emission kinematics is generated in \aNLOs\ is 
based on the so-called event projection (see e.g.~refs.~\cite{Frixione:2002ik,
Frederix:2011ss}): the sum of all final-state momenta bar that of parton $i$ 
has the same invariant mass and rapidity as the sum of the final-state momenta 
in the underlying Born kinematics. This procedure, that requires initial-state
partons to be non-monochromatic, is a necessary feature in order to be 
able to match NLO computations with current {\em hadronic} parton showers, 
as it mimics what the latter do in the case of initial-state backward 
evolution. It is not mandatory, but greatly improves the integration
efficiency, in the case of fixed-order computations. One of the implications
of event projection is that it leads to the fact that real-emission matrix
elements and their subtraction terms are associated with different Bjorken 
$x$'s. While this is not a problem in the case of hadronic collisions
(since the PDFs are slowly decreasing functions), in the case of $\epem$ 
collisions (with the PDFs sharply increasing functions) it creates serious 
efficiency issues.

In order to cope with this problem, in \aNLOs\ a new way of generating
the real-emission kinematics has been introduced, and applied to those FKS 
sectors relevant to the cases where the (massless) FKS sister $j$ is in the 
initial state and the underlying branching is a QED one. More in detail,
we proceed as was anticipated in sect.~6 of ref.~\cite{Frixione:2022ofv},
namely:
\begin{enumerate}
\item
The four-momentum $k_i$ of the FKS parton $i$ is generated by means of 
the variables:
\beq
\xi_i=\frac{2E_i}{\sqrt{\hat{s}}}\,,\quad
y_{ij}=\cos\theta_{ij}\,,\quad
\varphi_{ij}\,.
\eeq
These correspond to the rescaled (w.r.t.~the c.m.~partonic energy 
$\sqrt{\hat{s}}$) energy of the FKS parton, its polar angle w.r.t.~the FKS 
sister $j$, and an azimuthal angle $\varphi_{ij}$, respectively, all of which 
are defined in the partonic c.m.~frame. These variables are in a one-to-one 
correspondence with some integration random numbers (generally with 
adaptive sampling). 
\item
The kinematics of the other $n$ final-state momenta 
\mbox{$\{\bar k_l\}_{l=1,n+1}^{l\ne i}$}
is first generated in their c.m.~frame, knowing that their total invariant 
mass squared is:
\beq
\left(\sum_{\stackrel{l=1}{l\ne i}}^{n+1}\bar k_l\right)^2 = 
(1-\xi_i) \, \hat s\,.
\eeq
\item 
Finally, the momenta \mbox{$\{\bar k_l\}_{l=1,n+1}^{l\ne i}$} are boosted 
in the partonic c.m.~frame
\beq
k_l = \mathbf B \bar k_l
\eeq
with the boost $\mathbf B$ along the three-direction of $k_i$,
and so that total momentum conservation is achieved, namely:
\beq
\sum_{l=1}^{n+1} k_l =  \left(\sqrt{\hat s}, \vec 0 \right)\,.
\eeq
\end{enumerate}
In this way, a stable and efficient evaluation of NLO EW corrections 
can be performed also for lepton collisions within \aNLOs.

We point out that, since the generation of momenta outlined above does not
rely on event projection, it is also incompatible with an MC@NLO-type
matching for initial-state QED emissions, {\em assuming} that for these
kinds of branchings Parton Shower Monte Carlos work precisely in the
same way as they do in hadronic collisions. This assumption is not
necessarily correct, but is in any case useful to remind one that an
NLO-accurate matrix element-Monte Carlo matching for $\epem$ collisions
constitutes an open problem, whose solution is highly desirable.

\section{Synopsis of previous results on NLO+NLL PDFs\label{sec:pap123}}
In view of the systematic usage of the results of
refs.~\cite{Frixione:2019lga,Bertone:2019hks,Frixione:2021wzh} made
in this paper, we give here some additional information about them,
which complement the discussion in sect.~\ref{sec:elem}. While this 
renders the current work essentially self-contained, for a fuller
understanding of the issues involved the reader is urged to consult
the original publications.

We start by pointing out again that refs.~\cite{Frixione:2019lga,
Bertone:2019hks,Frixione:2021wzh} are based on considering a single
lepton family and on working in the $\MSb$ renormalisation scheme;
both of these limitations have been lifted in this paper. Then:
\begin{itemize}
\item 
In ref.~\cite{Frixione:2019lga} the NLO (i.e.~$\ord(\aem)$)
initial conditions are computed, for both PDFs and fragmentation 
functions (FFs) and for any combination of particle (i.e.~the object that
branches) and parton (i.e.~the object that emerges from the branching and
enters the short-distance cross section)\footnote{This applies to spacelike
branchings (i.e.~to PDFs); the role of particle and parton is reversed in
timelike branchings (i.e.~for FFs).}.
Such NLO initial conditions are factorisation-scale dependent, and are
meant to be imposed at a scale $\mu=\muz\sim m$ in order to start the
evolution. Furthermore, at any generic value of $\mu$ they coincide
by construction with the $\ord(\aem)$ expansions of the respective
PDFs and FFs. In the case of the PDFs, they are reported here
in eqs.~(\ref{G0sol})--(\ref{Gpossol2}).
\item 
In ref.~\cite{Bertone:2019hks} the NLO initial conditions computed
in ref.~\cite{Frixione:2019lga} are employed to obtain the NLL-evolved
PDFs for an electron particle (in other words, PDFs for a photon particle
and FFs are not considered, but can be dealt with in a pretty analogous
manner). Ref.~\cite{Bertone:2019hks} works in the $\MSb$ factorisation
scheme, and achieves the sought-for NLL-evolved PDFs in three different
ways: {\em a)} analytically to all orders in $\aem$ for $z\to 1$ (this
is called the asymptotic solution); {\em b)} analytically up to 
$\ord(\aem^3)$ for $z<1$ (this is called the recursive solution);
{\em c)} numerically for any $z\lesssim 1-10^{-\kappa}$, with 
$\kappa\simeq 10$ (this is called the numerical solution). The asymptotic
and recursive solutions are then matched to each other additively,
and found to agree extremely well with the numerical solution for all
of the $z$ values where the latter is reliable. While this provides one
with a powerful self-consistency check, in practical applications it is
more convenient to employ the numerical solution for $z\lesssim 1$,
and to switch to the asymptotic one when $z\to 1$. The availability of 
the latter in an analytic form is crucial in particular for the electron 
parton, in view of the fact that its PDFs has a power-like integrable 
divergence at $z=1$.
\item 
In ref.~\cite{Frixione:2021wzh} the NLO initial conditions computed
in ref.~\cite{Frixione:2019lga} are employed to obtain the NLL-evolved
PDFs for an electron particle by working in a DIS-inspired factorisation
scheme, called $\Delta$, expected to be better behaved in the $z\sim 1$
region w.r.t.~the $\MSb$ factorisation scheme. Given the final remark of 
the previous bullet point, at variance with ref.~\cite{Bertone:2019hks} 
only the asymptotic solution has been considered. More in detail, two
different solutions, called $\Delta_1$ and $\Delta_2$, have been obtained 
in ref.~\cite{Frixione:2021wzh}, which differ from each other in the
fact that certain effects in the running of $\aem$ are kept to all orders
in the former, while are limited to $\ord(\aem^2)$ in the latter. As it turns
out, the $\Delta_1$ solution is vastly superior. Therefore, the $\Delta_2$
one has not been considered any further, and in order to simplify the notation
the $\Delta$-scheme results of this paper coincide with those stemming
from the $\Delta_1$ solution of ref.~\cite{Frixione:2021wzh}.
\end{itemize}
As it has been pointed out in sects.~\ref{sec:evol} and~\ref{sec:UV},
the functional forms of the asymptotic solutions of 
refs.~\cite{Bertone:2019hks,Frixione:2021wzh} are unchanged in the
case of an evolution with multiple fermion families\footnote{Note that
PDFs for partons different from the electron and the photon vanish for 
$z\to 1$.}, and only the 
parametric replacements of eq.~(\ref{replxi}) are necessary (plus
$t\to t_k$ in the case of the photon PDFs). For the reader's convenience, 
we report below such functional forms, which are implicitly employed in 
sects.~\ref{sec:evol} and~\ref{sec:UV} of this paper.
\begin{itemize}
\item
Electron parton, $\MSb$ factorisation scheme. This is eq.~(5.63) of 
ref.~\cite{Bertone:2019hks}:
\beqn
\!\!\!\!\!\Gamma_{\lm}^{(\MSb)}(z,\mu)&\stackrel{z\to 1}{\longrightarrow}&
\frac{e^{-\gE\xi_1}e^{\hat{\xi}_1}}{\Gamma(1+\xi_1)}\,
\xi_1(1-z)^{-1+\xi_1}
\label{NLLsol3run}
\\*&&\phantom{aaa}\times
\Bigg\{1+\frac{\aem(\mu_0)}{\pi}\Bigg[\left(\log\frac{\mu_0^2}{m^2}-1\right)
\left(A(\xi_1)+\frac{3}{4}\right)-2B(\xi_1)+\frac{7}{4}
\nonumber\\*&&\phantom{\times aaa1+\frac{\aem}{\pi}\Bigg[}\;
+\left(\log\frac{\mu_0^2}{m^2}-1-2A(\xi_1)\right)\log(1-z)
-\log^2(1-z)\Bigg]\Bigg\}\,.
\nonumber
\eeqn
\item
Photon parton, $\MSb$ factorisation scheme. This is eq.~(B.87)\footnote{Where
use is made of eq.~(B.25) of ref.~\cite{Bertone:2019hks} to obtain a 
numerically-equivalent form which is more convenient in practical 
applications.} of ref.~\cite{Bertone:2019hks}:
\beqn
\Gamma_{\gamma}^{(\MSb)}(z,\mu)&\stackrel{z\to 1}{\longrightarrow}&
\frac{t\,\aemz^2}{\aemmu}\,\frac{3}{2\pi\xi_1}\,\log(1-z)
-\frac{t\,\aemz^3}{\aemmu}\,\frac{1}{2\pi^2\xi_1}\,\log^3(1-z)\,.
\phantom{aaa}
\label{ePDFgaasy}
\eeqn
\item
Electron parton, $\Delta$ factorisation scheme. This is eq.~(4.40) of 
ref.~\cite{Frixione:2021wzh}:
\beqn
\Gamma_{\lm}^{(\Delta)}(z,\mu)&\stackrel{z\to 1}{\longrightarrow}&
\frac{e^{-\gE\xi_1}e^{\hat{\xi}_1}}{\Gamma(1+\xi_1)}\,
\xi_1(1-z)^{-1+\xi_1}
\label{NLLsol8Del}
\\*&&\times
\left[\left(1+\frac{3\aem(\mu_0)}{4\pi}L_0\right)
\sum_{p=0}^{\infty} {\cal S}_{1,p}(z)
-\frac{\aem(\mu_0)}{\pi}L_0\sum_{p=0}^{\infty} {\cal S}_{2,p}(z)\right].
\nonumber
\eeqn
\item
Photon parton, $\Delta$ factorisation scheme. This is eq.~(5.50) of 
ref.~\cite{Frixione:2021wzh}:
\beqn
\Gamma_\gamma^{(\Delta)}(z,\mu)&\stackrel{z\to 1}{\longrightarrow}&
\frac{1}{2\pi}\,\frac{\aem^2(\mu_0)}{\aem(\mu)}\,\frac{1+(1-z)^2}{z}\,L_0
+\frac{1}{2\pi\xi_1}\,\frac{t\,\aem^2(\mu_0)}{\aem(\mu)}\,L_0
\nonumber\\*&&
-\frac{t\,\aem(\mu)}{2\pi\xi_1}\,
\frac{e^{-\gE\xi_1}e^{\hat{\xi}_1}}
{\Gamma\left(1+\xi_1\right)}\,(1-z)^{\xi_1}\,L_0\,.
\label{gaNLLsol4runQED2}
\eeqn
\end{itemize}
In these equations, several secondary quantities are employed. $L_0$ has
been defined in eq.~(\ref{Lzdef}). We also have the identities:
\beqn
d_1(\xi)&=&-A(\xi)\,,
\label{dores}
\\
d_2(\xi)&=&2B(\xi)-\frac{\pi^2}{6}\,,
\label{dtres}
\eeqn
with:
\beqn
d_k(\xi)&=&\frac{1}{{\cal G}_d(\xi,0)}
\left.\frac{\partial^k{\cal G}_d(\xi,\delta)}{\partial\delta^k}
\right|_{\delta=0}\,,
\label{dffbygen}
\\
{\cal G}_d(\xi,\delta)&=&
\frac{e^{-\gE(\xi-\delta)}}{\Gamma(\xi-\delta)}\,.
\label{dffgenfun}
\eeqn
The functions ${\cal S}_{i,p}(z)$ that appear on the r.h.s.~of
eq.~(\ref{NLLsol8Del}) are:
\beqn
{\cal S}_{1,p}(z)&=&\delta_{p0}\,\frac{\aem(\mu)}{\aem(\mu_0)}
+\frac{\aem(\mu)-\aem(\mu_0)}{\pi}\,
\frac{\hat{{\cal S}}_{1,p}(z)}{D(z)^{p+1}}\,,
\label{cS10def}
\\
{\cal S}_{2,p}(z)&=&\frac{\aem(\mu)}{\aem(\mu_0)}
\Big(-\delta_{p0}\log(1-z)+\delta_{p1}d_1(\xi_1)\Big)
+\frac{\aem(\mu)-\aem(\mu_0)}{\pi}\,
\frac{\hat{{\cal S}}_{2,p}(z)}{D(z)^{p+1}}\,,\phantom{aaaa}
\label{cS20def}
\eeqn
and have the following properties:
\beqn
\frac{\hat{{\cal S}}_{1,p}(z)}{D(z)^{p+1}}
&\stackrel{z\to 1}{\longrightarrow}&\log^{-2-p}(1-z)\;
\longrightarrow\;0\,,
\label{S1psupp}
\\
\frac{\hat{{\cal S}}_{2,p}(z)}{D(z)^{p+1}}
&\stackrel{z\to 1}{\longrightarrow}&\log^{-1-p}(1-z)\;
\longrightarrow\;0\,,
\label{S2psupp}
\eeqn
where
\beq
D(z)=1+\frac{\aem(\mu_0)}{\pi}\,\log^2(1-z)\,.
\label{Ddendef}
\eeq
The functions above can be computed recursively for any $p$, starting
from their definitions given in eqs.~(4.18), (4.29), and~(4.30) of
ref.~\cite{Frixione:2021wzh}, but owing to eqs.~(\ref{S1psupp}) 
and~(\ref{S2psupp}) only the lowest $p$ values are needed in numerical 
computations. From ref.~\cite{Frixione:2021wzh}:
\beqn
&&\hat{{\cal S}}_{1,0}(z)=-\frac{\pi}{\aem(\mu_0)}\,,
\label{hS10}
\\
&&\hat{{\cal S}}_{1,1}(z)=\big(1-2d_1(\xi_1)\big)\log(1-z)\,,
\label{hS11}
\\
&&\hat{{\cal S}}_{1,2}(z)=\frac{\aem(\mu_0)}{\pi}\left(C-1
+3d_1(\xi_1)-3d_2(\xi_1)\right)\log^2(1-z)
\nonumber
\\*&&\phantom{\hat{{\cal S}}_{1,2}(z)=}
+C-d_1(\xi_1)+d_2(\xi_1)\,,
\label{hS12}
\\*
&&\hat{{\cal S}}_{2,0}(z)=\frac{\pi}{\aem(\mu_0)}\log(1-z)\,,
\label{hS20}
\\
&&\hat{{\cal S}}_{2,1}(z)=-\big(1-d_1(\xi_1)\big)\log^2(1-z)
-\frac{\pi}{\aem(\mu_0)}\,d_1(\xi_1)\,,
\label{hS21}
\\
&&\hat{{\cal S}}_{2,2}(z)=-\frac{\aem(\mu_0)}{\pi}\left(C-1
+2d_1(\xi_1)-d_2(\xi_1)\right)\log^3(1-z)
\nonumber
\\*&&\phantom{\hat{{\cal S}}_{2,2}(z)=}
-\Big(C-2d_1(\xi_1)+3d_2(\xi_1)\Big)\log(1-z)\,,
\label{hS22}
\eeqn
and:
\beq
C=\frac{\pi^2}{6}-1\,.
\label{Cconst}
\eeq
In the numerical implementation these contributions have actually been 
included up to \mbox{$p=5$}; we refrain from reporting here the corresponding 
analytical expressions, which are cumbersome without being particularly 
illuminating.

\phantomsection
\addcontentsline{toc}{section}{References}
\bibliographystyle{JHEP}
\bibliography{eexsecs}

\end{document}